\documentclass[epj,nopacs]{svjour}
\usepackage{fixltx2e}
\usepackage{graphicx}
\usepackage{amsmath}
\usepackage{amssymb}
\usepackage{mathptmx}
\usepackage{helvet}
\usepackage{eurosym}
\RequirePackage{lineno}
\DeclareFontFamily{U}{eur}{\skewchar\font'177}
\DeclareFontShape{U}{eur}{m}{n}{
  <5> <6> <7> <8> <9> gen * eurm
  <10> <10.95> <12> <14.4> <17.28> <20.74> <24.88> eurm10
  }{}
\DeclareFontShape{U}{eur}{b}{n}{
  <5> <6> <7> <8> <9> gen * eurb
  <10> <10.95> <12> <14.4> <17.28> <20.74> <24.88> eurb10
  }{}
\DeclareSymbolFont{EUr}{U}{eur}{m}{n}
\DeclareSymbolFont{EUb}{U}{eur}{b}{n}
\DeclareMathSymbol{\upmu}{\mathord}{EUr}{"16}%
\def\mm{\ifmmode  {\mathrm{\ mm}}\else \textrm{~mm}\fi}
\def\ns{\ifmmode  {\mathrm{\ ns}}\else \textrm{~ns}\fi}
\def\mrad{\ifmmode  {\mathrm{\ mrad}}\else \textrm{~mrad}\fi}
\def\mum{\ifmmode  {\mathrm{\ \upmu m}}\else \textrm{~\upmu m}\fi}
\def\murad{\ifmmode  {\mathrm{\ \upmu rad}}\else \textrm{~\upmu rad}\fi}
\newlength{\capindent}
\newlength{\capwidth}
\newlength{\figwidth}
\newcommand{\icaption}[2][!*!,!]{\hspace*{\capindent}%
  \begin{minipage}{\capwidth}
    \ifthenelse{\equal{#1}{!*!,!}}%
      {\caption{#2}}%
      {\caption[#1]{#2}}
      \vspace*{3mm}
  \end{minipage}}
\usepackage{geometry}
\begin{document}
%
%\begin{titlepage}
%\vspace*{-6mm}
%\includegraphics[width=3cm]{plots/atlas_logo1.eps} \hfill
%\begin{minipage}[b]{7cm}
%\begin{center}
%\mbox{\Huge \bf ATLAS}
%\end{center}
%\begin{center}
%\mydocversion
%\end{center}
%\begin{center}
%\thedate
%\end{center}
%\end{minipage}
%\hfill \includegraphics[width=3cm]{plots/cern_bw.eps}

%\vspace{1cm}

\title{\bf Commissioning of the ATLAS Muon Spectrometer with Cosmic Rays}
%%%%%\vspace{1.cm}
%\author{\it Draft 1$^{*)}$ See following pages for the list of authors}
\hugehead
\author{% ATLAS Collaboration author list for 21-FEB-2010
% Data extracted on 21-May-2010 for paperid 21
%\documentclass[11pt]{article}
%\usepackage{a4wide}\begin{document}
\begin{flushleft}
{\Large The ATLAS Collaboration}

\bigskip

G.~Aad$^{\rm 48}$,
B.~Abbott$^{\rm 111}$,
J.~Abdallah$^{\rm 11}$,
A.A.~Abdelalim$^{\rm 49}$,
A.~Abdesselam$^{\rm 118}$,
O.~Abdinov$^{\rm 10}$,
B.~Abi$^{\rm 112}$,
M.~Abolins$^{\rm 88}$,
H.~Abramowicz$^{\rm 152}$,
H.~Abreu$^{\rm 115}$
% ...
\bigskip

% ...
$^{10}$ Institute of Physics, Azerbaijan Academy of Sciences, H. Javid Avenue 33, AZ 143 Baku, Azerbaijan\\
$^{11}$ Institut de F\'isica d'Altes Energies, IFAE, Edifici Cn, Universitat Aut\`onoma  de Barcelona,  ES - 08193 Bellaterra (Barcelona), Spain\\
% ...
$^{48}$ Albert-Ludwigs-Universit\"{a}t, Fakult\"{a}t f\"{u}r Mathematik und Physik, Hermann-Herder Str. 3, D - 79104 Freiburg i.Br., Germany\\
$^{49}$ Universit\'e de Gen\`eve, Section de Physique, 24 rue Ernest Ansermet, CH - 1211 Geneve 4, Switzerland\\
% ...
$^{88}$ Michigan State University, Department of Physics and Astronomy, High Energy Physics Group, East Lansing, MI 48824-2320, United States of America\\
% ...
$^{111}$ University of Oklahoma, Homer L. Dodge Department of Physics and Astronomy, 440 West Brooks, Room 100, Norman, OK 73019-0225, United States of America\\
$^{112}$ Oklahoma State University, Department of Physics, 145 Physical Sciences Building, Stillwater, OK 74078-3072, United States of America\\
% ...
$^{115}$ LAL, Univ. Paris-Sud, IN2P3/CNRS, Orsay, France\\
% ...
$^{118}$ Oxford University, Department of Physics, Denys Wilkinson Building, Keble Road, Oxford OX1 3RH, United Kingdom\\
% ...
$^{152}$ Tel Aviv University, Raymond and Beverly Sackler School of Physics and Astronomy, Ramat Aviv, IL - Tel Aviv 69978, Israel\\
% ...
\end{flushleft}

%\end{document}
}
\date{Received: date / Revised version: date}

\abstract{
 The ATLAS detector at the Large Hadron Collider has
collected several hundred million cosmic ray events during 2008 and
2009. These data were used to commission the Muon Spectrometer and
to study the performance of the trigger and tracking chambers, their
alignment, the detector control system, the data acquisition and the
analysis programs. We present the performance in the relevant parameters 
that determine the quality of the muon measurement.
We discuss the single element efficiency,
resolution and noise rates, the calibration method of the detector
response and of the alignment system, the track reconstruction
efficiency and the momentum measurement. The results show that the
detector is close to the design performance and that the Muon
Spectrometer is ready to detect muons produced in high
energy proton-proton collisions.
}

\maketitle

%\vspace{1.cm}

%tml - General note:  Sometimes "end-cap" is "end-cap" and "barrel" is "barrel".  Need to pick one.
% If we want caps, then I suggest "End-Cap" instead of "end-cap".

%\begin{abstract}
%\noindent The ATLAS detector at the Large Hadron Collider has
%collected several hundred million cosmic ray events during 2008 and
%2009. These data were used to commission the Muon Spectrometer and
%to study the performance of the trigger and tracking chambers, their
%alignment, the detector control system, the data acquisition and the
%analysis programs. We present the performance in the relevant parameters 
%that determine the quality of the muon measurement.
%We discuss the single element efficiency,
%resolution and noise rates, the calibration method of the detector
%response and of the alignment system, the track reconstruction
%efficiency and the momentum measurement. The results show that the
%detector is close to the design performance and that the Muon
%Spectrometer is ready to detect muons produced in high
%energy proton-proton collisions.
%\end{abstract}

%\begin{center}
%{\it (To be submitted to Nuclear Instruments and Methods)}
%\end{center}

%\end{titlepage}

%%\newpage
%{\small \tableofcontents}
%\small 
\tableofcontents
%%%\newpage

%\setpagewiselinenumbers
%\modulolinenumbers[5]
%\linenumbers

\section{The ATLAS Muon Spectrometer}
\label{introduction}
\begin{figure*}
\begin{center}
\includegraphics[width=12cm,height=12cm]{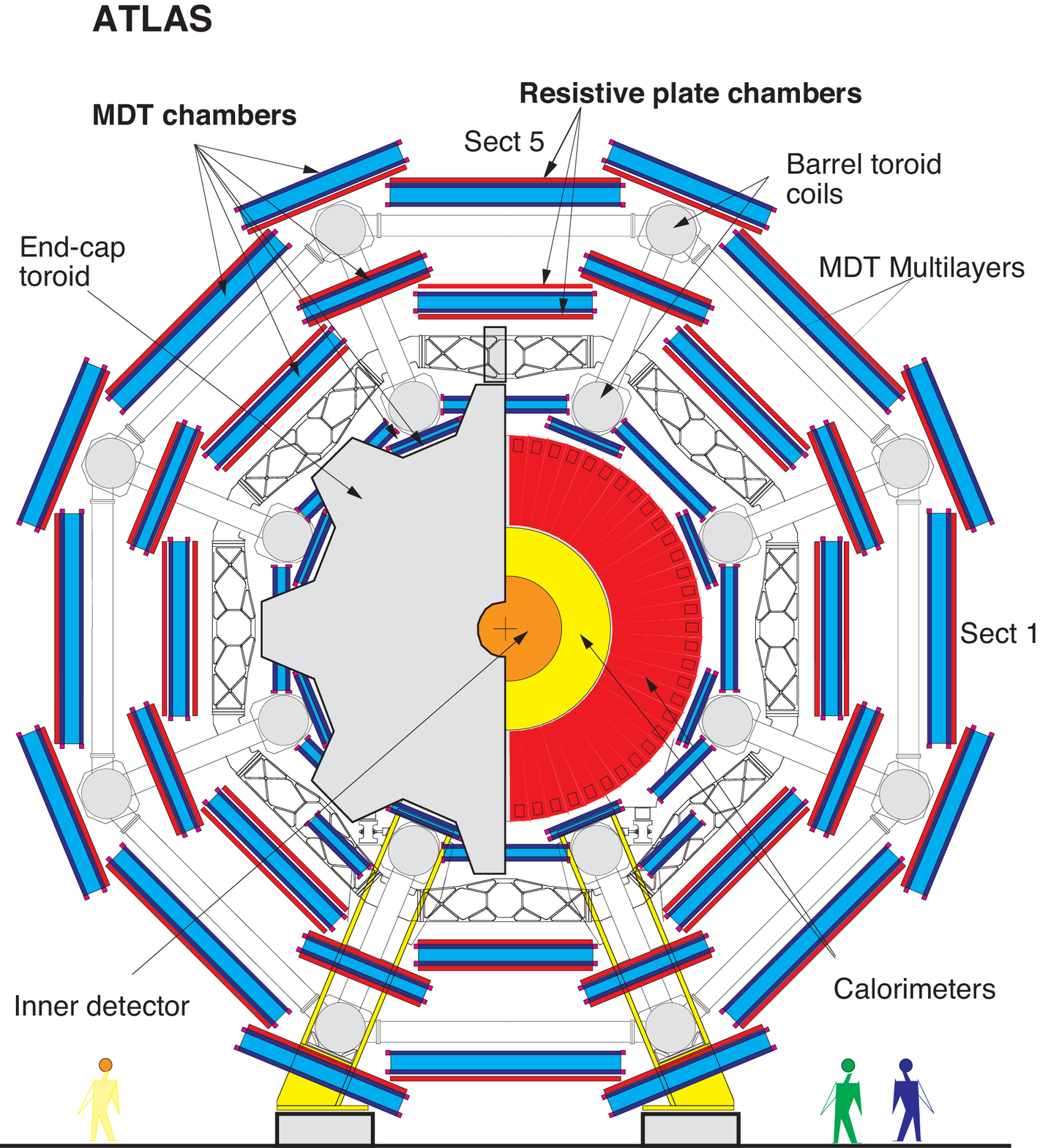}
%\vspace*{5cm}       % Give the correct figure height in cm
%\vspace{0.5cm}
\includegraphics[width=12cm]{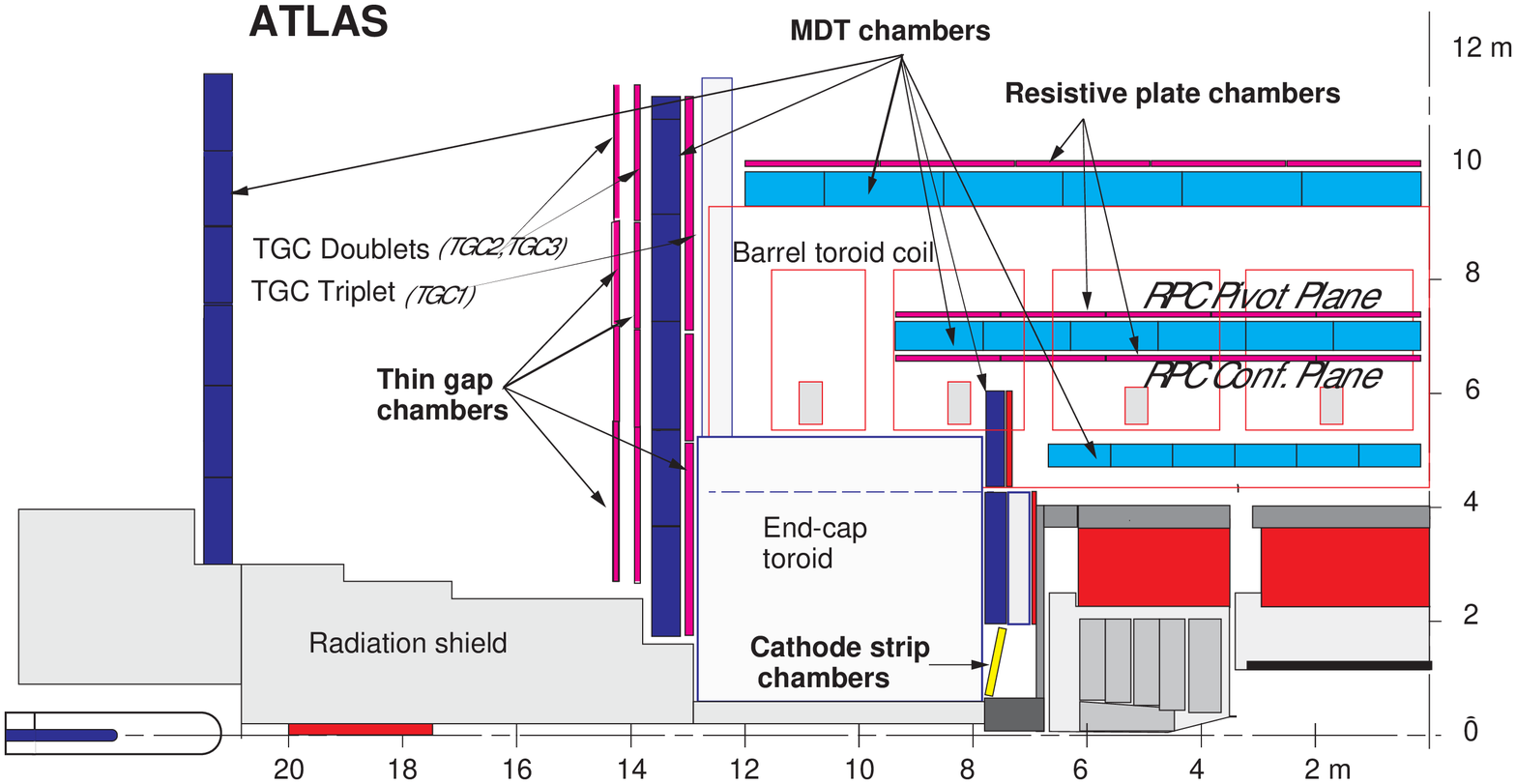}
%\includegraphics[width=12cm,height=12cm]{newplots/muonspectroXY.eps}
%\vspace*{5cm}       % Give the correct figure height in cm
%\vspace{0.5cm}
%\includegraphics[width=12cm]{newplots/muonspectroRZ.eps}
%\vspace*{1.5cm}       % Give the correct figure height in cm
\end{center}
\caption{Schematic view of the muon spectrometer in the $x$-$y$ (top) and $z$-$y$ (bottom) projections. Inner, Middle and Outer chamber stations are denoted BI, BM, BO in the barrel and EI, EM, EO in the end-cap.}
\label{fig:muon-spect}
\end{figure*}

%{\bf Editor: F. Cerutti}\\
%{\bf few changes here -}
The ATLAS Muon Spectrometer (MS in the following) is designed to
provide a standalone measurement of the muon momentum with an
 uncertainty in the transverse momentum varying from 3$\%$ at 100 GeV to about 10$\%$
at 1 TeV, and to provide a trigger for muons with varying transverse momentum thresholds down to a few GeV. 
A detailed description of the
muon spectrometer and of its expected performance can be found in
~\cite{MuonTDR}~\cite{ATLASdet}~\cite{CSCpaper}. Here only a brief
overview is given. The muon momentum is determined by measuring the track
curvature in a toroidal magnetic field. The muon trajectory is
always normal to the main component of the magnetic field so that
the transverse momentum resolution is roughly independent of $\eta$
over the whole acceptance. The magnetic field is provided by three
toroids, one in the ``barrel" ($ | \eta  | < 1.1 $) and one for each
``end-cap" ($1.1 <  | \eta |  <  2.7$), with a field integral
between 2 and 8 Tm. The muon curvature is measured by means of
%tml - The word "chamber" is used in various ways, sometimes as a generic term
% as in "trigger chambers" and sometimes as a very specific collection of
%MDT tubes that are referred to as a "chamber".  This is confusing to the reader
%and moreover, the latter term (MDT chamber) is never defined.
three precision chamber stations positioned along its trajectory. In order
to meet the required precision each muon station should provide a
measurement on the muon trajectory with an accuracy of 50 $\mu$m. In
Figure~\ref{fig:muon-spect} a schematic view of the muon
spectrometer   \footnote   {The ATLAS reference system is a cartesian right-handed coordinate system, with
the nominal collision point at the origin. The positive x-axis is defined as pointing from the collision point to
the centre of the LHC ring and the positive y-axis points upwards while the z-axis is tangent to the beam direction at the collision point. The azimuthal angle $\phi$ 
is measured around the beam axis, 
and the polar angle $\theta$
is the angle measured with respect to the z-axis. The pseudorapidity is defined as $\eta$=-ln tan$\theta$/2.} is given.

%\begin{figure}[!htb]
%\begin{center}
%\includegraphics[width=10cm,height=10cm]{newplots/muonspectroXY.eps}
%\vspace{0.5cm}
%\includegraphics[width=12cm]{newplots/muonspectroRZ.eps}
%\end{center}
%\caption{Schematic view of the muon spectrometer in the $x$-$y$ (top) and $z$-$y$ (bottom) projections. Here $z$ is the beam axis,
%$x$ the horizontal axis and $y$ the vertical axis. Inner, Middle and Outer chamber stations are denoted BI, BM, BO in the barrel and EI, EM, %EO in the end-cap.}
%\label{fig:muon-spect}
%\end{figure}

For most of the acceptance Monitored Drift Tube (MDT) chambers are deployed~\cite{ATLASdet}. The coordinate
in the plane perpendicular to the wires, measured by the MDT, is referred to as
the precision, or bending coordinate, being mainly perpendicular to
the direction of the toroidal field. In the end-cap inner region,
for $| \eta |  <  2.0$,  Cathode Strip Chambers
(CSC)~\cite{ATLASdet} are used because of their capability to cope with higher background rates.\par
%Small paragraph with the MDT and CSC characteristics.
The MDT chambers are composed of two MultiLayers (ML) made of three or four layers of tubes. Each tube is 30 mm in diameter and has an anode wire of 50 $\mu$m diameter. The gas  mixture used is 93\% Ar and 7\% CO$_2$ with a small admixture of water vapour,  the drift velocity is not saturated and the total drift time is about 700 ns. The space resolution attainable with a single tube is about 80$\mu$m, measured in a test beam ~\cite{h8a}, \cite{h8b}. The CSC chambers are multiwire proportional chambers with cathode strip read out. The cathode planes are equipped with orthogonal strips and the precision coordinate is obtained measuring the charge induced on the strips making the charge interpolation between neighbouring strips. Typical resolution obtained with this read-out scheme is about 50 $\mu$m.\par

\begin{sloppypar}
%tml - The concept of "trigger chambers" has not been introduced.  Need a few words in the intro
%section that explains that the trigger is provided by dedicated detector elements, separate from
%the MDTs.
The trigger  system of the MS is  based on two different chamber technologies: 
Resistive Plate Chambers (RPC)~\cite{ATLASdet} instrument the barrel
region while  Thin Gap Chambers (TGC)~\cite{ATLASdet} are used in
the higher background environment of the end-cap regions. Two RPC
chambers are attached to the middle barrel chambers providing a
low-$p_T$ trigger. A high-$p_T$ trigger is provided by the RPC
modules installed on the outer barrel chambers in combination with the low $p_T$  signal provided by the middle chambers. The RPCs also provide
the coordinate along the MDT wires that is not measured by the MDT
chambers. 
%The RPCs have a typical strip pitch of about 3~cm.
\end{sloppypar}

Similarly in the end-cap two TGC doublets and one triplet are
installed close to the middle station and provide the low-$p_T$ and
high-$p_T$ trigger signals. The TGCs also measure the coordinate of
the muons in the direction parallel to the MDT wires. This
coordinate is referred to as the second, or non-bending coordinate.
For this purpose TGC chambers are also installed close to the MDTs in the inner layer of the end-cap (EI).\par
Some MS naming conventions adopted in this paper are introduced
here. The MS is divided in the $x$-$y$-plane (also referred to as
$\phi$-plane) in 16 sectors: Sector 5 being the upper most
and Sector 13 the lower most.
In both barrel and end-cap regions the MS is divided into 8 `Large'
sectors (odd numbered sectors) and 8 `Small' sectors (even numbered sectors),
determined by their coverage in $\phi$. The muon stations are named
`Inner', `Middle', and `Outer', according to the distance from the
Interaction Point (IP).  The three stations for the barrel are
denoted BI, BM, and BO, and for the End-Cap EI, EM, and EO,
respectively. Along the $z$ axis, the MS is divided into two sides,
called side A (positive $z$) and C (negative $z$).

As a complementary source of information, two
publications~\cite{h8a}, \cite{h8b} on a detector system-test 
with a high momentum muon beam can be
consulted.

%tml- The intro sentence here about the Fall 2008, seems superceded by the intro to the
% section ("Data sample").  I suggest removing it and just beginning with "This paper
%is organized as follows.  The cosmic ray trigger is described in Section..."
Beginning in September 2008 the ATLAS detector was operated continuously up to November 2008 and then for different periods starting from Spring 2009. The first beams were circulated in the LHC machine in
September 2008 but no beam-beam collisions were delivered. During these
periods, the ATLAS detector collected mainly cosmic ray data. All muon detector technologies were included in the run with the exception of CSCs for which the Read Out chain was still not yet commissioned and therefore they are not included in the results presented in this paper.\par
The analyzed data samples and
the reconstruction software are described in Section~\ref{Samples}.
The cosmic ray trigger is described in Section~\ref{Trigger}. 
Studies of data quality, calibration, and alignment  are presented in
Sections~\ref{DQA},~\ref{Calibration},~\ref{Performance},~\ref{Alignment} respectively,while studies on tracking
performance are presented in sections~\ref{Segments}
and~\ref{Tracks}. The results are summarized in Section~\ref{Conclusions}.

%%%%%%   START   KEVIN    ROSY %%%%%%%%%%%%%%%%%%%%%%%%%%%%%%%
\section{Data sample and reconstruction software}
%{\bf Editors: Rosy, Kevin and Niels}\\
\label{Samples}
\subsection{Data sample}
\begin{sloppypar}
In preparation for LHC collisions, the ATLAS detector has acquired
several hundred million cosmic ray events during several run periods
in 2008 and 2009. The analysis of a subset of data corresponding to
about 60 M events is presented here. These runs allowed
commissioning the ATLAS experiment, the trigger, the data
acquisition, the various detectors and the reconstruction software.
Most of the cosmic rays reach the underground detectors via the two big shafts.
They have incident angles close to the vertical axis and they
are mainly triggered by the RPCs. The selected runs, together with
the status of the magnetic field in the MS and the number of
collected events for the different trigger streams, are listed in
Table~\ref{runlist}.
\end{sloppypar}

\begin{table}[htb]
\centering
\begin{tabular}{| c | c | c | c | c |}
\hline\noalign{\smallskip}
Run & Trigger & B-field  & N of Evts & Period\\
\hline \hline
91060  & RPC & Off & 17 M  & Fall 08\\
\hline
91060 & TGC  & Off & 0.2 M & Fall 08\\
\hline
89106 & TGC & Off & 0.4 M  & Fall 08\\
\hline
89403 & TGC & Off & 0.4 M & Fall 08\\
\hline
91803 & TGC & {\bf On} &  50 K  & Fall 08\\
\hline
91890 & RPC & {\bf On} &  16 M  & Fall 08\\
\hline
113860 & RPC & Off &  \ 6 M & Spring 09\\
\hline
121080 & RPC & {\bf On} &  21 M & Summ 09\\
\hline
\end{tabular}
\caption{List of analyzed data runs together with the corresponding
trigger stream, statistics and status of the MS magnetic field. All
runs were collected in Fall 2008, with the exceptions of run 113860
collected in Spring 2009, and run 121080 in Summer 2009.}
\label{runlist}

\end{table}

\subsection{Muon reconstruction software}
The data were processed using the complete ATLAS software
chain~\cite{ATHENA}: data decoding, data preparation (which includes
calibration and alignment), and track reconstruction. 
%The first two items are discussed in some detail in Section~\ref{Calibration}
%and~\ref{Alignment}. 
Muon reconstruction has been handled by two
independent packages, namely {\it Moore}~\cite{MOORE} and {\it
Muonboy}~\cite{Mboy}. The two reconstruction algorithms are similar
in design but differ in some details. The general strategy is to
reconstruct muon trajectories both at the local (individual
chamber), as well as at the global (spectrometer), level. The
%tml - Here's one of the places where the definition of a "chamber" is needed:
trajectories reconstructed in individual chambers can be
approximated as straight lines over a short distance where bending
has little effect and are therefore fit to track {\it segments}. Full
tracks are formed by combining segments from multiple chambers.

Prompt muons produced in proton-proton collisions have trajectories
that point back to the Interaction Point (IP). Moreover they are synchronous with the
collision since all the detector front end electronics are
synchronized with the LHC bunch crossing frequency of 40 MHz. In
contrast, cosmic ray muons are ``non-pointing" and are asynchronous
with the detector clock: they have an additional 25 ns jitter with
respect to the clock selected by the trigger. In addition, during
commissioning the different trigger detectors were not timed with
sufficient precision, leading to variations in timing depending on the
region of the detector that originated the trigger. A further
difficulty in track reconstruction was due to the lack of precise
alignment of the muon detectors during this commissioning phase, 
as described in Section~\ref{Alignment}.

The reconstruction algorithms were adapted for these ``cosmic ray"
conditions as described below. Both programs were modified by
relaxing the standard tracking requirements and implementing a
procedure to accommodate the cosmic ray timing conditions. The
tolerance for hit association to form track segments and the
%tml - Changed "hit error function.." below because it's not the "function" that's increased.
%hit error function were increased.
uncertainty associated with each hit position were increased.
 Moreover, a
procedure called {\it global $t_{0}$ refit} (G$t_{0}$--refit) was
developed in both reconstruction algorithms to compensate for the 25
ns time jitter and the imprecise trigger timing. The aim of this procedure is to determine with
better precision the time when the cosmic ray crossed the detector
by introducing a free global timing  parameter (g$t_{0}$) in the
segment reconstruction. The implementation of the G$t_{0}$--refit in
the two reconstruction algorithms is briefly described below while
the results are presented in Section~\ref{Calibration}.

\subsection{Muonboy track reconstruction}
The strategy of the $\it{Muonboy}$ reconstruction algorithm can be summarized in four main steps:
\begin{itemize}
 \item identification of Regions Of Activity (ROA) in the muon system with the information provided by the RPC/TGC detectors;
 \item reconstruction of local segments in each muon station in the identified ROA;
 \item combination of segments of different muon stations to form muon track candidates using three-dimensional tracking in magnetic field;
 \item global track fit of the muon track candidates through the full system using individual hit information.
\end{itemize}
The topology of cosmic ray tracks is accommodated by relaxing the
Region Of Activity requirement of pointing in a projective geometry when
associating hits to form segments, or matching segments to form
tracks. Moreover, since cosmic ray events have low occupancy, looser
quality criteria were used for the selection of segments and tracks.

The {\it Muonboy} algorithm for the G$t_{0}$--refit consists of a
scan of different g$t_{0}$ values in steps of 10 ns, doing the full
segment reconstruction at each step. The g$t_{0}$ value giving the
%tml - a "parabolic fit" to what?  Apparently it's the quality factor.
% Why should it be a parabola?  Is it the log of a Gaussian quantity?
best reconstruction quality factor is kept and a parabolic fit is
performed using this best value and the two closer values along the
parabola. Then the g$t_{0}$ corresponding to the minimum of the
quality factor parabola is chosen. In order to obtain high
efficiency, the  accuracy requirement for the MDT single hit resolution is relaxed  by adding in
quadrature a 0.5 mm constant smearing to the intrinsic resolution
function (described in Section~\ref{MDTperf}). This smearing is
increased by additional 0.5 mm if the G$t_{0}$--refit 
%procedure described above 
fails. Moreover, a less demanding track quality
factor is required for tracks when hits are missing or are not
associated to the track.

\subsection{Moore track reconstruction}
The {\it Moore} reconstruction algorithm is built out of several distinct stages:
\begin{sloppypar}
\begin{itemize}
\item identification of global roads throughout the entire spectrometer using all muon detectors (MDT, CSC, RPC and TGC);
\item reconstruction of local segments in each muon station seeded by the identified global roads;
\item combination of segments of different muon stations to form muon track candidates;
\item global, three-dimensional, tracking and final track fit.
\end{itemize}
\end{sloppypar}
\begin{sloppypar}
Several modifications to the standard pattern recognition were made
to optimize the reconstruction of cosmic ray tracks. In the global
%Reference for "Hough transform"?
road finding step, a straight line {\it Hough transform} was used to
allow for non-pointing tracks. The cuts on distance and direction
between the road and the segment were relaxed. In the segment
finding no cuts were applied on the number of missing hits ($i.e.$
drift tubes that are expected to be crossed 
%by the segment 
but have no hits).
\end{sloppypar}

\begin{sloppypar}
The G$t_{0}$--refit consists in varying simultaneously the global
time offset (g$t_{0}$) for each segment reconstructed in a chamber.
Then all measured times of hits associated to the segment are
translated into drift radii after subtraction of the g$t_{0}$. The
g$t_{0}$ value that minimizes the sum in quadrature of the weighted
residuals (corresponding to the segment reconstruction $\chi^{2}$)
is selected. 
\end{sloppypar}

In this fit the MDT uncertainties are set to twice the
test-beam drift tube resolution. If the segment fit is not successful,
a straight line fit is performed assuming a constant 1 mm error.
Hits are removed  if their distance from the segment
is greater than 7$\sigma$. In the track fit the MDT errors are
enlarged to 2 mm to account for uncertainties in the alignment of
chamber stations.

%%%%%% END KEVIN ROSY %%%%%%%%%%%%%%%%%%%%%%%%%%%%%%%%%
%%%%%%%%%% New part from Filippo 1/1/2010 %%%%%%%%%%%%%%%%
\section{Trigger configuration during data taking}
\label{Trigger}
\begin{sloppypar}
%{\bf Editor: Masaya EC, Mimmo barrel }\\
A more detailed description of the trigger system can be found in~\cite{ATLASdet},\cite{LEVEL1-TDR}. 
Here only specific issues related to the 2008-2009 cosmic ray data taking are introduced.
The muon level-1 trigger is issued by the RPC in the barrel and by the TGC in the end-caps.
During cosmic ray data taking most of the statistics were collected using this trigger. 
Special trigger configurations were adopted with different geometries (eg non pointing to the IP) and different timing (e.g. delaying the triggers issued by the upper sectors in order to trigger only in the lower sectors to mimic particles coming from the IP)
when commissioning the muon trigger system itself or when selecting 
cosmic rays for commissioning the other ATLAS sub-systems.

In {\it beam-collision} configuration, the level-1 muon trigger selects pointing tracks with six different thresholds in transverse momentum and sends information to the Central-Trigger-Processor (CTP). 
The six thresholds, three low-$p_T$ and three high-$p_T$, do not distinguish between different detector regions, barrel or end-cap.
For cosmic rays, to help commissioning separately the two regions, it was chosen to assign three thresholds to the barrel and three to the end-cap.
\end{sloppypar}

\subsection{Barrel level-1 trigger}
%%%%%%%%%%%%%%%% Beginning New Part from Mimmo
%========================
The barrel trigger detectors are arranged in three stations each having a doublet of RPC layers at increasing distances from the IP. 
In each sector the first two stations are mechanically coupled to the BM MDT while the third is coupled with the BO MDT as shown in Figure~\ref{fig:muon-spect}.\par
\begin{sloppypar}
The trigger algorithm is steered by signals on the middle layers, named Pivot plane. When a hit is found on this plane, the low-$p_{T}$ trigger logic searches for hits in the inner layers, named Confirm plane, and requires a coincidence in time of three hits over the four layers in a pre-calculated cone. The width of this cone defines the $p_{T}$ threshold. 
%The trigger logic searches for a geometrical and time coincidence between the middle layers, named Pivot plane, and the other two, Confirm plane, requiring a majority coincidence of 3 out of 4 layers (2 in the pivot and 2 in the confirm). If a coincidence is found between 
%the pivot and the confirm planes, a low-$p_{T}$ trigger is issued.
If hits are also found in a pre-calculated cone of the outermost plane in coincidence with a low-$p_{T}$ trigger, a high-$p_{T}$ trigger is issued. Also in this case the $p_{T}$ threshold is defined by the width of the cone.
In addition to the $p_{T}$ requirement, the trigger logic also demands the track to be pointing towards the IP both in  $\phi$ and $\eta$.
In the cosmic ray runs only three of the six thresholds were used in the barrel and were defined  as {\sl MU0\_LOW}, {\sl MU0\_HIGH} and {\sl MU6}.
The two thresholds {\sl MU0\_LOW/HIGH} did not select a physical $p_T$ range; in fact, the {\sl MU0\_LOW} was triggered only by the time coincidence of 3 out of 4
hits without any pointing constraint and the {\sl MU0\_HIGH} was triggered by the coincidence of a {\sl MU0\_LOW} with at least a hit in the corresponding outer plane.
The threshold {\sl MU6} required not only a time coincidence but also an IP-pointing constraint in the $\phi$-projection only. 
%In the final configuration the IP-pointing constraint for MU6 will be required in both 
%$\phi$ and $\eta$ projections.
To emulate the timing expected for beam collisions, and to enhance the illumination of the Inner Detector (ID),
the cosmic ray trigger was issued mainly by the bottom sectors. This was  achieved by delaying the top sector trigger by 5 BC (125ns) preventing it from arriving first at the Central Trigger Processor (CTP) and thus forming the trigger.
\end{sloppypar}

In the fall 2008 data taking period, the timing of the low-$p_{T}$ trigger and the data read-out latencies were
still under commissioning. This had a large impact on the detector coverage.
The situation has largely improved for the runs taken in 2009 both in terms of detector coverage and in trigger timing as shown in 
Sections~\ref{DQA-RPC} and~\ref{RPCperf}.
%%%%%%%%%%%%%%%%%%%%%% End New Part from Mimmo

\subsection{End-cap level-1 trigger}
\begin{sloppypar}
The level-1 TGC trigger system provided three thresholds, 
named {\sl MU0\_TGC\_HALO}, {\sl MU0\_TGC} and {\sl MU6\_TGC}.
The trigger was issued by the coincidence between several TGC layers. The logic was based both on timing (BC identification) and geometry (pointing track). The main difference between the three trigger thresholds is related to the required number of layers and to the degree of pointing to the IP.
{\sl MU0\_TGC\_HALO} required a 3 out of 4 layer coincidence in the two outermost TGC stations, the so-called {\it Doublet chambers}, in both $\eta$ (bending) and $\phi$ (non-bending) projections and a pointing 
requirement within 20$^{\circ}$. 
{\sl MU0\_TGC} and {\sl MU6\_TGC} required in addition a 2 out of 3 layer coincidence in the 
TGC stations closer to the IP, the so-called {\it Triplet chambers}, in the $\eta$ projection only. The pointing requirement of {\sl MU0\_TGC} 
was of $\pm$10$^{\circ}$ degrees while for  {\sl MU6\_TGC} was of $\pm$5$^{\circ}$.

The trigger was timed for high-momentum muons coming from the IP.  
All the delays due to different time-of-flight and cable lengths
were properly set and cross-checked using a test pulse system achieving a relative timing within 4 ns.
For most of the cosmic run period, only the level-1 trigger generated from the TGC bottom sectors  was used.
This was chosen to ensure good timing of the trigger with the read-out of the ID, since cosmic muons triggered by the TGC bottom sectors and crossing the ID have a time-of-flight similar to muons produced in collisions.
\end{sloppypar}
%%%%%%%%%% New part from Filippo 1/1/2010 %%%%%%%%%%%%%%%%

\section{Data quality assessment}
\label{DQA}
\subsection{Introduction}
\label{DQAIntroduction}

The data quality assessment  consists of several software
algorithms working at different levels of the data taking. The
Detector Control System (DCS)~\cite{DAQ} is the first source of
information available during the operation of the detector. Here
information on the hardware status of the different sub-detectors
and on the settings of Low Voltage (LV) and High Voltage (HV) power
supplies and on the gas system is available. The DCS also receives
information from the Data Acquisition (DAQ)~\cite{DAQ} as soon as
problems during the read-out of a chamber appear.

The next  stage in the chain of data quality assessment is the
on-line monitoring. It receives input from the data acquisition
system running in a spectator mode. Once the data are decoded,
monitoring histograms are filled showing quantities related to the
detector operation. Part of the muon data selected by the level-1
trigger Region Of Interest (ROI) are transferred by the level-2
trigger processors to three dedicated computing farms (referred to
as {\it calibration centers}) to monitor and determine the
calibration parameters of the MS chambers. The larger event samples
available at the calibration centers allow the analysis of single
drift tube responses. The goal of the analysis at the calibration
centers is to provide drift tube and trigger chambers calibration constants and to give
general feedback on the detector operation within 24 hours, which is the time needed, at high luminosity,  to collect enough statistics 
to calculate new calibration constants.

On a longer time scale, using the full reconstructed ATLAS event information,
the off-line data monitoring provides the final
information on the data quality. At each  step a flag
summarizing the data quality at that level is stored in a database.

\subsection{MDT chambers}
\label{DQA-MDT}
\begin{figure*}[!tb]
\begin{center}
\includegraphics[width=12cm]{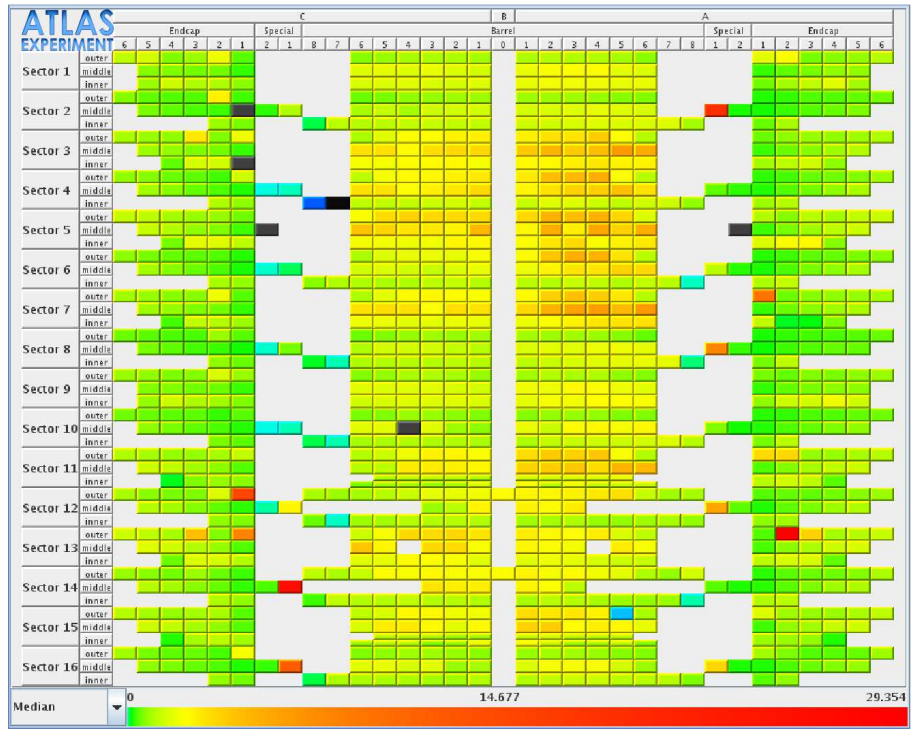}
%\vspace*{5cm}       % Give the correct figure height in cm
\end{center}
\caption{Screen shot of a monitoring application displaying the MDT
hit occupancy for all chambers. Each chamber is represented by a
small box. The color of the box is related to the average number of
raw hits per tube. The boxes are arranged in
an $\eta$-$\phi$ grid: a column represents an $\eta$ slice,
perpendicular to the beam axis; a row represents one of the sixteen $\phi$
sectors. Within each sector chambers of the Inner, Middle, Outer
ring are displayed separately.}
\label{fig:mdt-gnamon}
\end{figure*}

%{\bf Editor: Philipp - 2.5 pages}\\
%{\it Co-editors: Mauro, Jonas, Daniela, Gabriella, Joerg, Fabio, Alberto, Nectarios, Arely, Tony, Orin, Justin, James ....}\\
%%%%%%%% Text from Philipp ============================================
In the  fall 2008 period (e.g. Run 91060) only five out of 1110 MDT chambers were not included in the data taking. 
Of these five chambers, two were not yet connected to services and three had 
problems with the gas system.
Due to the cosmic ray illumination and the trigger coverage not all
chambers had sufficient event samples to determine the performance
of single drift tubes. 
The studies reported here were done at
different levels of detail, from chamber information down to single
drift tube information when the event samples were sufficient. 
%tml - It's not completely clear here that this is from online monitoring and
%not the calibration centers.
The data survey searched for problems of individual read-out
channels as well as of clusters corresponding to hardware related
groups of tubes. A screen shot of one of the online monitoring applications
used for the MDT chambers is shown in Figure~\ref{fig:mdt-gnamon}. Here the
average number of hits per tube for each MDT is represented in a
$\eta$-$\phi$ plot where the higher cosmic illumination on the top and bottom sectors (3-7, 11-15) 
compared to the vertical sectors (16- 2 and 8-10) is clearly seen as well as the larger illumination on the A side of the detector where the larger shaft is present.
The five chambers not included in the data
acquisition are marked as dark gray boxes. Two more chambers are
visible with very low statistics due to problems with the HV
supplies.
%The reason for this was the switched off HV supply for those chambers.
For  32 MDT chambers one of the two multi-layers was disconnected from HV.
%(each MDT station is divided in two drift tube multi-layers as shown in 
%Figure~\ref{fig:muon-spect}).

%\begin{figure}[!htb]
%\begin{center}
%\includegraphics[width=12cm]{newplots/mdt_run91060_rpc_and_tgc_prd.eps}
%\end{center}
%\caption{Screen shot of a monitoring application displaying the MDT
%hit occupancy for all chambers. Each chamber is represented by a
%small box. The color of the box is related to the average number of
%raw hits per tube. The boxes are arranged in
%an $\eta$-$\phi$ grid: a column represents an $\eta$ slice,
%perpendicular to the beam axis; a row represents one of the sixteen $\phi$
%sectors. Within each sector chambers of the Inner, Middle, Outer
%ring are displayed separately.} \label{fig:mdt-gnamon}
%\end{figure}

\begin{table*}[!thb]
\begin{center}
\begin{tabular}{|l|r|c|}
%\hline
%Total number of MDT channels                     & 339072         & Fraction \\
%\hline
\hline
Number of channels analyzed with sufficient event samples & 336144         &  Fraction \\
\hline
\hline
Channels not included in the read-out                          & 936  & 0.28\%  \\
Channels with read-out or initialization problems         & 744   & 0.22\% \\
Channels with HV or gas problems                              & 2942  & 0.88\% \\
Permanently dead channels (broken wires)                 & 323    &   0.10\% \\
\hline
\hline
Total problematic channels &   4945 & 1.47\%  \\
\hline

\end{tabular}
\end{center}
\caption{List of MDT channels with problems in run 91060. }
\label{tab:mdtDQ}
\end{table*}

%The full MDT system consists of 339072 read-out channels, of which
%tml - "not included in the read-out" -> "not included in the read-out for this run"?
%$\sim$1.5K belong to the chambers not included in the read-out. Out
%of the remaining channels $\sim$1K correspond to malfunctioning
%hardware or gas problems. These problems were easily spotted already
%by the on-line monitoring. Problems in the gas system or the HV
%power supply affect one or more tube-layers of one MDT chamber. In
%20 cases problems in the readout hardware, affecting eight to 24
%tubes, were found. Beside these problems related to clusters of
%tubes (single amplifier or TDC chips), 323 individual dead read-out
%channels were found. 
A detailed list of hardware problems found in run 91060 is reported
in Table~\ref{tab:mdtDQ}. 
The cosmic ray flux was not sufficient for a
detailed analysis of single drift tubes for 15 MDT chambers
($\sim$3K channels). Thus we were able to analyze individually 336K,
out of the working 339K, drift tubes.
To summarize, about 5K channels, out of
336K, have shown some problems in run 91060, corresponding to 1.5\%.
Most of these channels have been recovered during the 2008-2009
shutdown period. Only a very small fraction of problems, at the
level of a few per mill, could not be solved, such as permanently
disconnected tubes (broken wires) or chambers with very difficult access.

%%%%%%%====== Begin Part from Nectarios ======================
%\subsubsection{MDT offline monitoring}
In addition to monitoring in the DAQ framework (on-line monitoring),
the data are also processed with the offline reconstruction program
which produces monitoring histograms. This ensures that the
reconstruction works properly and that the correct {\it conditions
data} (calibration and alignment constants) are used in the first
processing of the data. The off-line monitoring gathers and presents
information on several variables for single drift tubes, e.g.
drift time and collected charge  distributions, hit
occupancy and noise rate. These variables are obtained for
individual MDTs or grouped for regions, such as $\eta$ or $\phi$
sectors, barrel or end-cap, side A or C. Variables related to
segments or tracks are also monitored.
%tml - This is an odd choice to show for offline monitoring since it is essentially the same
%as the online plot shown.  Shouldn't we instead show something from track or segment monitoring?
% An example of histogram produced by the MDT off-line data quality
% during run 91060 is given in Figure~\ref{fg:mdt_occ}. This plot
% shows the number of hits per MDT multi-layer in the BM stations.
% The vertical axis refers to the 16 $\phi$-sectors for multi-layers 1
% (from 1.00 to 1.15) and 2 (from 2.00 to 2.15), the horizontal axis
% refers to the 12 $\eta$-sectors (from -6 to -1 for side C and from 1
% to 6 for side A). The white empty regions for $\eta$-index = 3, 4,
% 5, 6, both negative and positive, and $\phi$-index = 1.11, 1.13,
% 2.11 and 2.13 correspond to the support feet of ATLAS barrel where
% no MDT are installed. The two holes at  $\eta$-index = -4,
% $\phi$-index = 1.09 and 2.09 spot two MDT chambers with read-out
% problems during this run.
% \begin{figure}[!htbp]
% \centering
% \includegraphics[width= 14cm]{plots/run_91060_NumberOfHitsInBarrelMiddlePerMultilayer.eps}
% \caption{
% Multi-layer hit occupancy for the barrel Middle MDT stations.
% The vertical axis refers to the 16 $\phi$-sectors for multi-layers 1 (from 1.00 to 1.15) and 2
% (from 2.00 to 2.15), the horizontal axis refers to the 12 $\eta$-sectors (from -6 to -1 for side C and from 1 to 6 for side A).}
% \label{fg:mdt_occ}
% \end{figure}
%%%%%%%%%%%%% End Part from Nectarios --------

\subsection{Barrel trigger chambers: RPC}
\label{DQA-RPC}
\begin{figure}[!htb]

\resizebox{0.48\textwidth}{!}{%
 \includegraphics{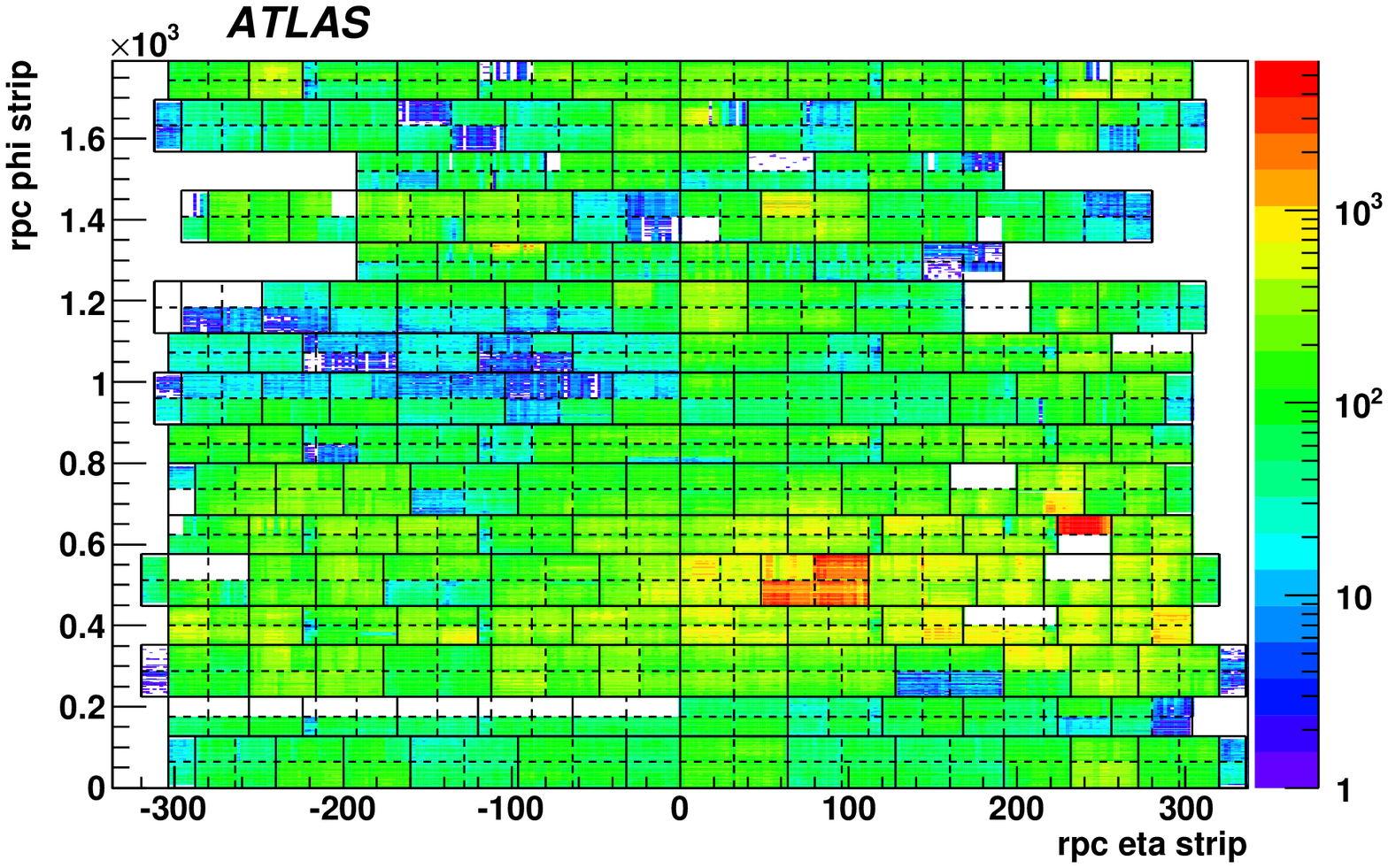}
}
\caption{RPC low-$p_T$ trigger coverage in $\eta$-$\phi$ for  Run 113860 (spring 2009).
Each $\eta$ and $\phi$ strip producing a low-$p_T$ trigger corresponds to an entry in the plot. The coverage  in spring 2009 was about  95\%.}
\label{fig1rpc}       
\end{figure}

%{\bf Editor: Michele}\\
%{\it Co-editors: Michele, ...}\\
%%%%%%% Begin part edited by Michele
\begin{sloppypar}
Commissioning of the RPC detectors progressed continuously and substantial improvements were made during the 2008-2009 shut-down. 
As an example  Figure~\ref{fig1rpc} shows a
two-dimensional distribution of RPC strips requiring a 3 out of 4
majority coincidence for the low-$p_T$ trigger demonstrating that  the trigger coverage in Spring 2009 was at the 95 \% level.
%During the last
%runs in 2009 the RPC covered more than 97\% of the $\eta$-$\phi$
%acceptance.
%As a result of all these problems the trigger and readout coverage of the RPC during run 91060 was reduced to approximately 60\%.
%In the left plot of Figure~\ref{fig1rpc}  a two-dimensional trigger distribution obtained  requiring a 3 out of 4 majority coincidence in the low-$p_{T}$ trigger
%boards for run 91060 is shown.
%During the winter shutdown a big effort has been done to fix the aforementioned problems, new firmware has solved the
%synchronization problems, optical links and fiber have been replaced as well as the gas lines.
%To solve the high temperature problem a new air conditioning flow distribution system was installed.
%This made possible to reach, for rum 113860 collected in spring 2009, the  95\% trigger coverage as shown in the right plot of Figure~\ref{fig1rpc} .
\end{sloppypar}

%\begin{figure}[!htb]
%\begin{center}
%%\includegraphics[width=7cm,]{plots/TriggerCoverage_91060.eps}
%\includegraphics[width=10cm,]{newplots/rpccoverageColor.eps}
%\end{center}
%\caption{RPC low-$p_T$ trigger coverage in $\eta$-$\phi$ for  113860 (spring 2009).
%Each $\eta$ and $\phi$ strip producing a low-$p_T$ trigger corresponds to an entry in the plot. The coverage  in spring 2009 was about  95\%.}
%\label{fig1rpc}
%\end{figure}

\begin{figure}[!htb]
\begin{center}
\includegraphics[height=5.2cm]{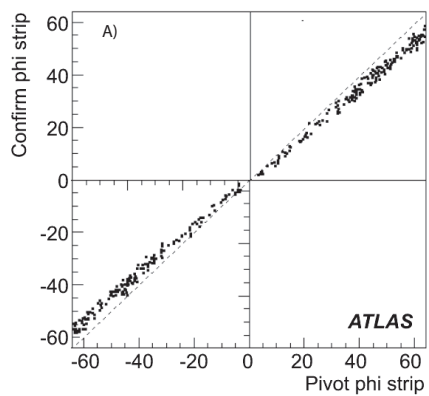}
\includegraphics[height=5.6cm]{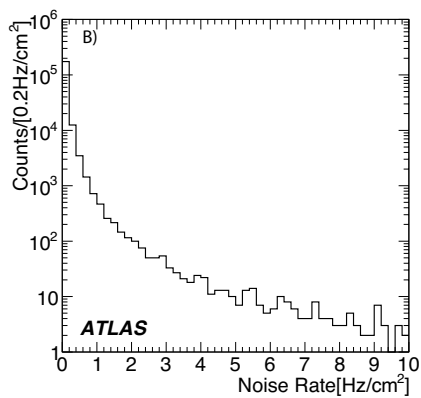}
\end{center}
\caption{ A): RPC spatial correlation between the pivot strip
number and the confirm strip number in the $\phi$ projection for
a programmed trigger road. 128 strips correspond to a RPC plane 3.8
m long. B): distribution of strip noise rates per unit area
measured with a random trigger for 310K RPC strips. The larger noise present on some strips is probably due to local weaknesses of grounding connections.}
\label{fig2rpc}
\end{figure}

Studies of the trigger performance were made using the data of run
91060 after implementation of the {\it trigger
roads}~\cite{LEVEL1-RPC}. For the low-$p_T$ trigger the four RPC
layers in the Middle station are involved, both in $\eta$ and
$\phi$ projections.
%tml - Removed lines below because pivot and confirm layers are defined already above.
%The {\it Pivot} plane groups the two external
%layers, the {\it Confirm} plane the two internal layers, and are
%spaced by $\sim$ 0.7~m.
Tracks were accepted by the trigger if any strip of the pivot plane
was in coincidence with a group of strips of the confirm plane
aligned with the IP, realizing a majority combination of 3 out of 4 RPC
layers. Figure~\ref{fig2rpc} left shows the spatial correlation
between $\phi$ strips in the pivot planes and $\phi$ strips in
the confirm planes. 
The correlation line is slightly rotated with respect to the diagonal due to the different distance  of the confirm and pivot planes  with respect to the IP.
%The band is close to the line at 45$^{\circ}$.
%Since the strip pitch of $\sim$3 cm is the same in both planes, the
%angle of the band deviates from 45$^{\circ}$ by an angle close to
%the ratio of the plane spacing to their distance from the IP.

A random trigger was used to measure the counting rate
for each read-out strip. This is a measurement of the RPC system
noise rate. About 310K strips were analyzed over a total of 350K
working strips. Figure~\ref{fig2rpc} right shows the distribution of
single channel noise rate, normalized to an area of 1 cm$^2$. For each
strip, the noise rate is calculated as the number of hits divided by
the number of random triggers and the width of the read-out gate of
200 ns, and is normalized to the area of the strip (typically 550 cm$^2$ for a BM eta strips and 900 cm$^2$ for a BO eta strips). Only a few
hundred strips showed a counting rate above 10 Hz/cm$^2$ which is
the background rate expected when the LHC will run at
high-luminosity. The average noise rate of the RPC was  stable during the different running periods.  

The fraction of dead channels, considering only the part of the
detector included in the read-out in the Fall 2008 runs, was 1.5\%,
mainly due to problems in the front-end electronics.

%%%%%%% End part edited by Michele

\subsection{End-cap trigger chambers: TGC}
\label{DQA-TGC}
%{\bf Editor: Akimasa}\\
%{\it Co-editors: Masaya, Susumu, ...}\\
%%%%%%%%%% Start Part from Akimasa
In the end-caps the muon trigger is provided by the TGC chambers
installed in three layers that surround the MDT Middle chambers. All
together they form the so-called Big Wheels (BW), one in each end-cap.
In addition, TGC chambers are also installed close to the EI
chambers in the Small Wheels (SW), but these are only used to
measure the muon $\phi$ coordinate. 
%The TGC of the SW are not included in the actual trigger 
%configuration but In case of of high cavern background they 
%could be added for trigger confirmation. 
In Fall 2008 all the BW TGC sectors were read-out. Given the
installation schedule for the ATLAS detectors, the Inner TGC station were
the last chambers installed and they were not fully operational
during 2008 runs. For this reason they are not discussed in the
following.

\begin{figure}[!htb]
  \begin{center}
\includegraphics[width=7cm]{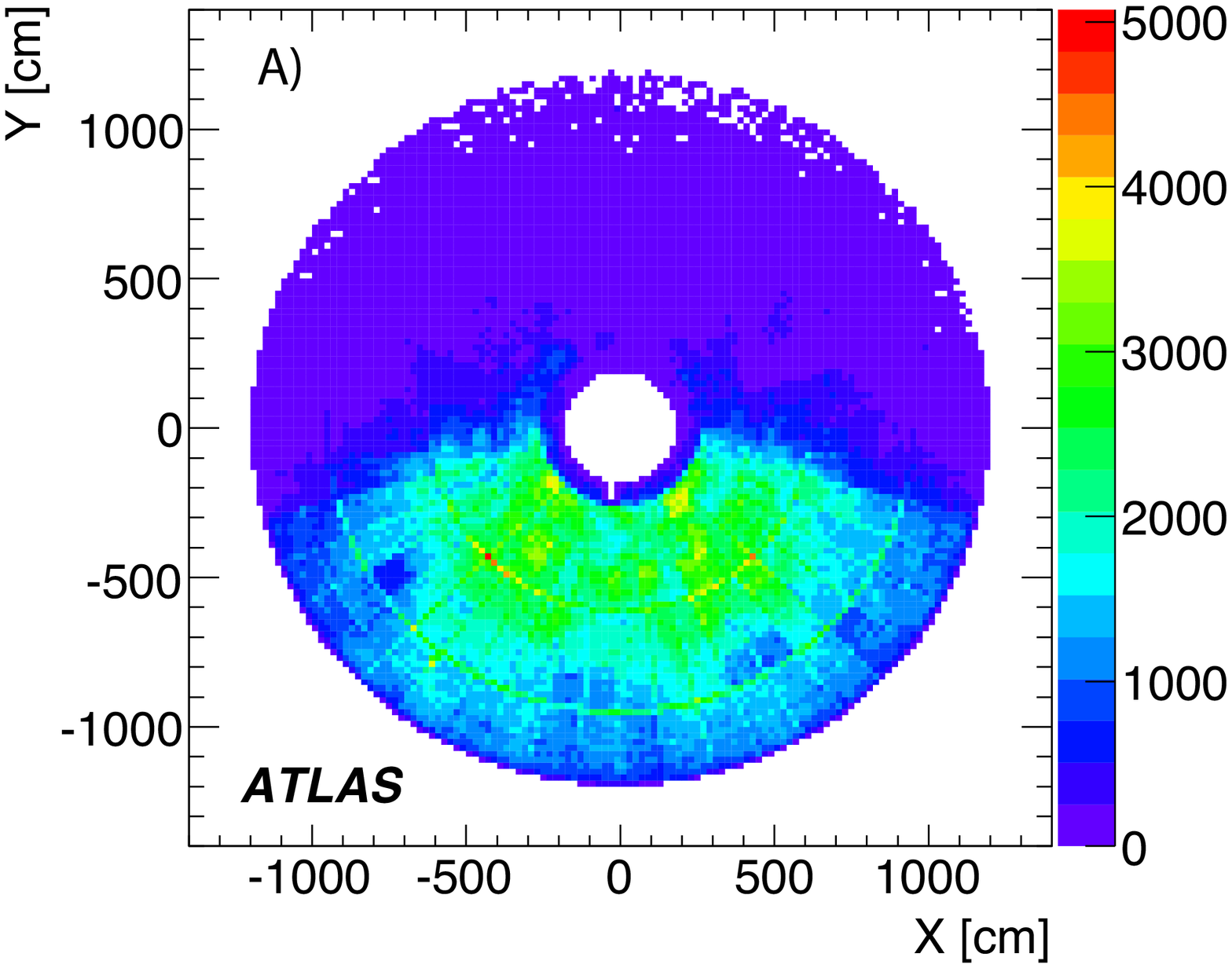}
\includegraphics[width=7cm]{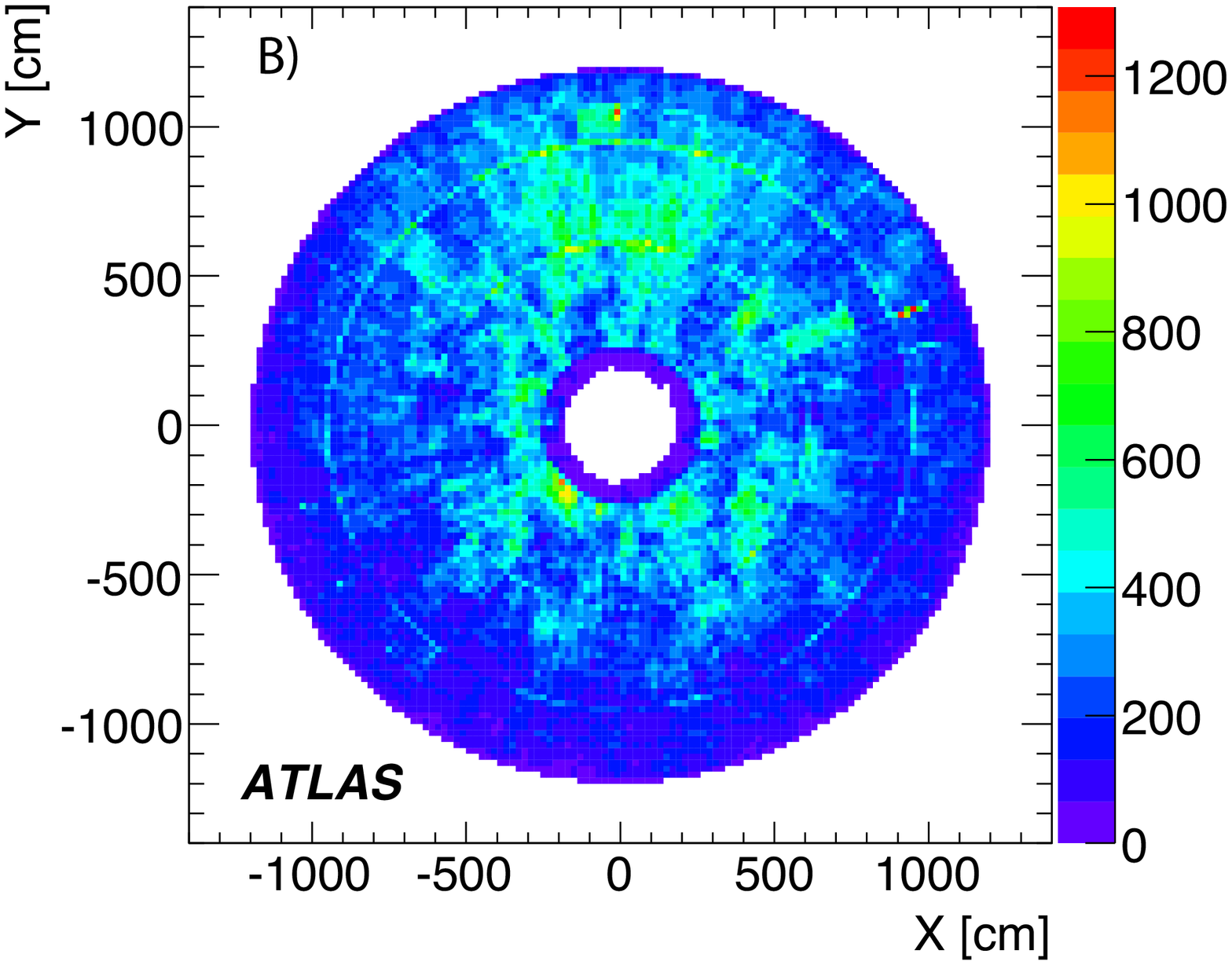}
  \end{center}
  \caption{Map of coincidences of wire and strip hits in the $x$-$y$ plane. A): the five bottom sectors
  (sectors 8--12, $195^{\circ} < \phi < 345^{\circ}$) used for timing the ID tracking detectors in run 91060.
  B): with all sectors participating in the trigger during run 91803.}
  \label{fig:TGCCoverage}
\end{figure}

Two types of trigger configuration were adopted in Fall 2008. One
was optimized to study the end-cap muon detectors with cosmic rays.
In this configuration all TGC BW sectors were used in the
trigger. The other setting was optimized to provide the trigger for
the ID tracking detectors and was used for timing the ID. In order
to mimic muons coming from the IP, only the five bottom sectors were
used to trigger. The typical detector coverage in these two trigger
configurations is shown in Figures~\ref{fig:TGCCoverage} by plotting
the coincidence positions in the $x$-$y$ plane for wire and strip
hits for run 91060~(left) and run 91803~(right). Only about 0.8\% of chambers were not operational 
due to HV or gas problems. Since for the trigger a majority logic is required these inactive chambers do not produce any dead regions in the trigger acceptance.\par

The HV and front-end threshold setting, the gate widths for wires
and strips, and the trigger sectors are listed in
Table~\ref{tab:TGCDet} for these two runs.

%Table with active sector logics for run 91060 and maybe a more recent one\\
\begin{table*}[!hbt]
\begin{center}
\begin{tabular}{|l|c|c|c|c|}
\hline
Run & Trigger sector & HV    & Threshold & Gate widths for wire / strip\\
\hline
\hline
91060 &  8 to 12   & 2800 V & 100 mV     & 35 / 45 ns \\
91803 & 1 to 12    & 2650 V &  \ 80 mV     & 35 / 45 ns \\
\hline
\end{tabular}
\end{center}
\caption{TGC sectors participating in the trigger, high voltage setting, threshold and gate width.}
\label{tab:TGCDet}
\end{table*}

\begin{figure}[!htb]
  \begin{center}
 \includegraphics[width=7cm]{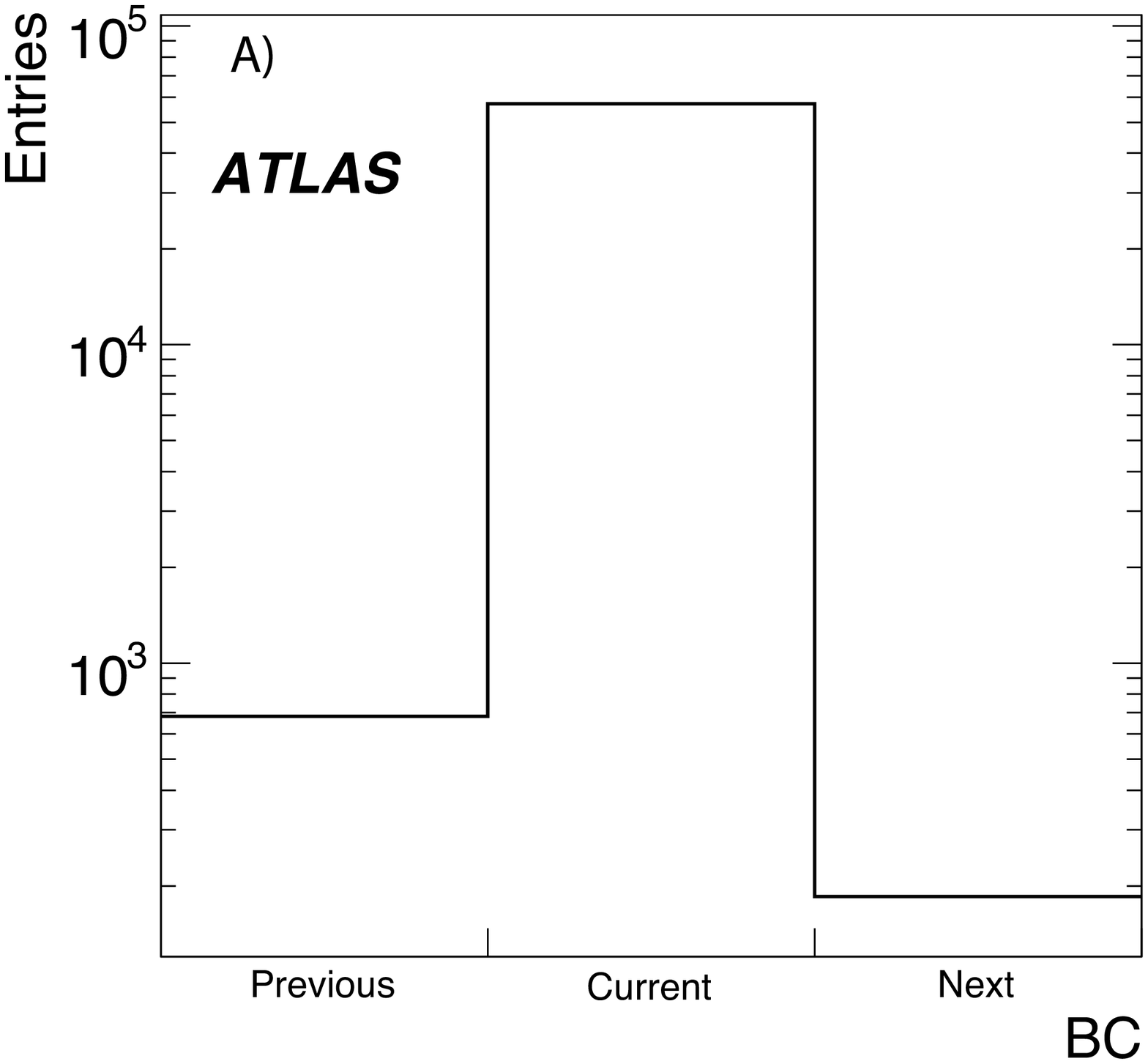}
 \includegraphics[width=7cm]{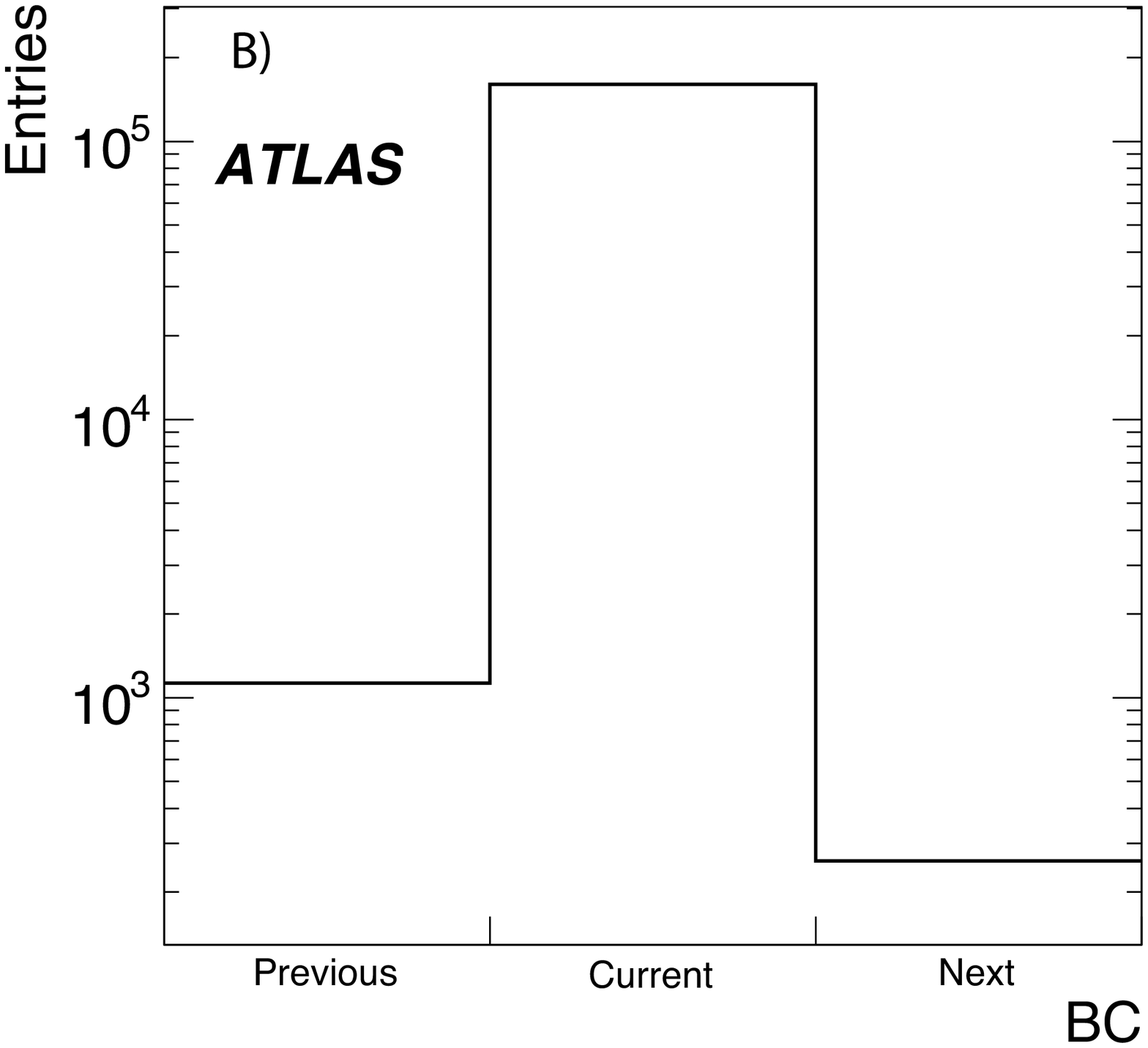}
 \end{center}
  \caption{A): TGC front-end and B): sector logic buffers for BC identification.
  Three BC crossing, previous, current and next are readout.
}
  \label{fig:TGCTiming}
\end{figure}

%Trigger and RO timing\\
For each trigger issued by the CTP, the TGC Read Out Driver (ROD)
sends to the DAQ system the data corresponding to three
Bunch Crossings (Previous, Current and Next BC) contained in two
separate buffers. Of the two buffers, one is located in the
front-end board where the wires and strips providing the low-$p_{T}$
coincidence are separately rec\-orded. In the second buffer, located in the Sector Logic Board in the service
counting room, the coincidence of the wire and strip signals is done. Each buffer has a programmable identifier that has to
be adjusted in order to read out the correct (Current) BC data.
Figure~\ref{fig:TGCTiming} shows the readout timing for the
front-end and the sector logic buffers for level-1 triggers issued
by the TGC. About 98.6\% of data in the front-end buffer, and 99.8\%
of data in the sector logic buffer are read out with the correct
timing. The small population in the previous or next BC is due to
cosmic ray showers.

\section{MDT chamber calibration}
\label{Calibration}
%{\bf Editor: Felix}\\
%{\it Co-editors: Ed, Oliver, Dan, Mauro, Domizia, ....}
%%%%%%%%%% Text from Felix here %%%%%%%%%%%%
\subsection{Calibration method}
%{\itshape References are missing.}
The MDTs require a calibration procedure~\cite{CALIB} to precisely convert the 
measured drift time into a drift distances from the anode wire (drift radius)
that is subsequently used in pattern recognition and track fitting.
The calibration of the MDT chambers is performed in three steps. In
the first step the time offset with respect to the trigger signal, $t_{0}$, for each tube or group of
tubes is determined; in the second step the drift-time to space
relation, $r(t)$ function, is computed; in the third step the
spatial resolution is determined.\par
 The calibration constants are
loaded in the Conditions Data Base (known as `COOL')~\cite{COOLdb}
and then retrieved, according to an Interval Of Validity (IOV) mechanism, to be used in the offline reconstruction.  
The IOV determines for which group of runs
the calibration constants are valid. 
The gas mixture composition varied during the data taking period since the water injection part of the gas
 system was under commissioning  resulting in a not constant admixture of water vapour, 
as can be seen in Figure~\ref{drift_time_vs_time}. 
Nonetheless  the calibration procedure based on the IOV mechanism was 
able to provide good calibration constants for all the running period. 
\begin{figure}[!htb]
\begin{center}
      \includegraphics[width=\columnwidth]{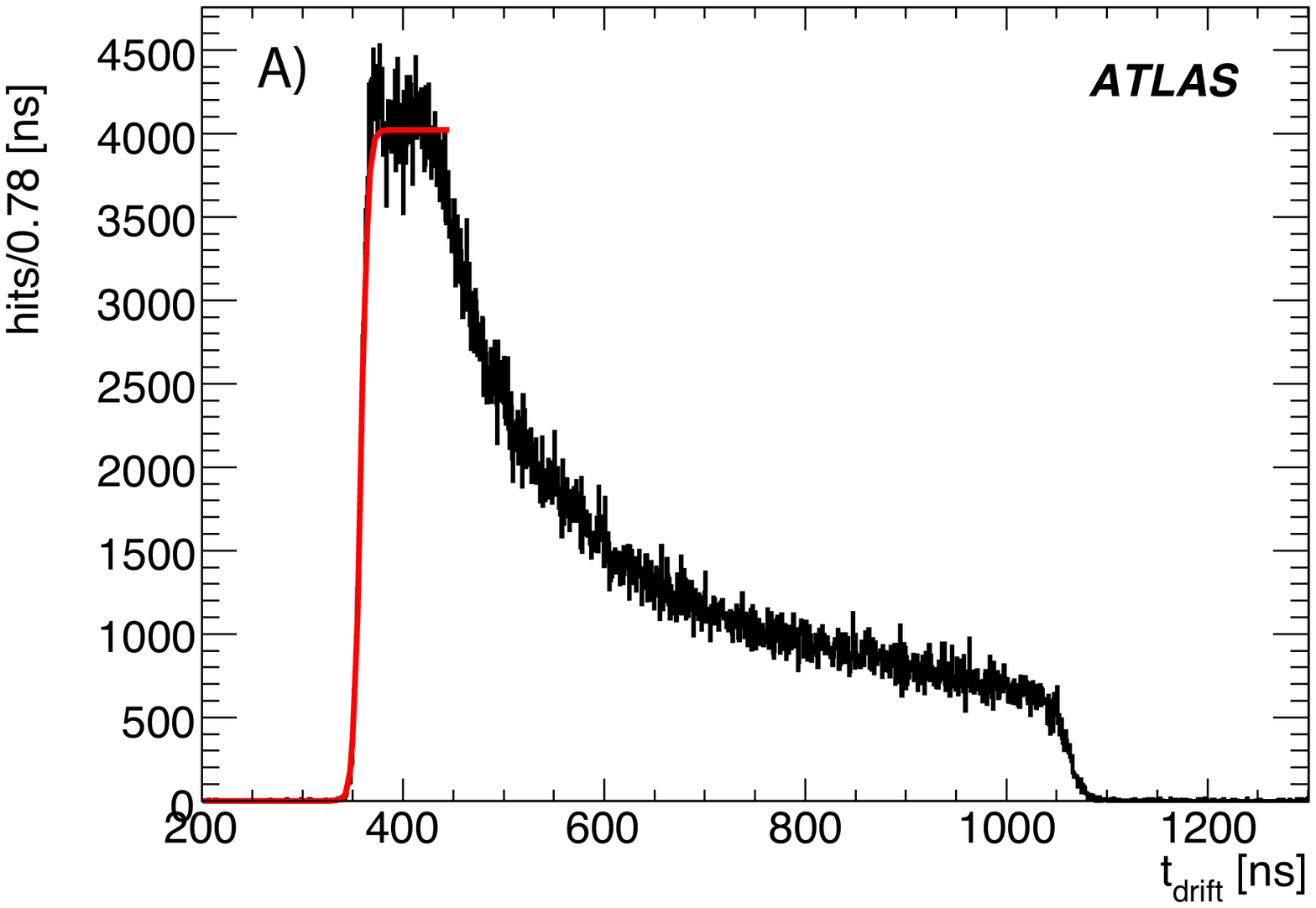}
   \includegraphics[width=\columnwidth]{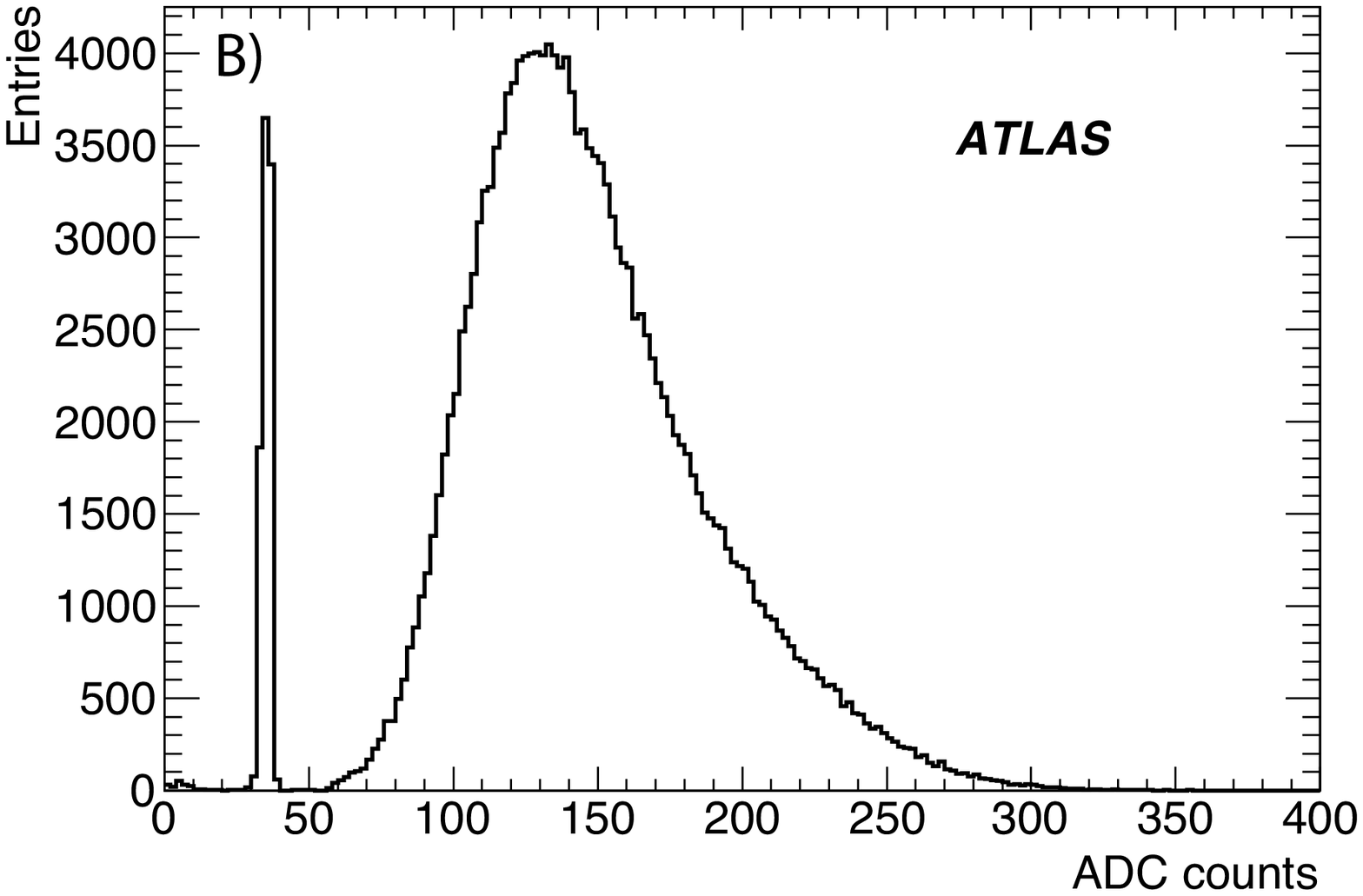}
  \caption{A): typical drift time spectrum in cosmic ray events for an MDT chamber.
   The position of the inflection point of the leading edge of the spectrum, $t_{0}$, is determined by fitting a Fermi
   function (shown in red) to the beginning of the spectrum. B): typical spectrum of ADC for all tubes in a chamber.
  Hits below 50 ADC counts are identified as electronic noise.}
 \label{calib_t0_fit}
\end{center}
\end{figure}

\begin{figure}[!htb]
\begin{center}
       \includegraphics[width=7cm]{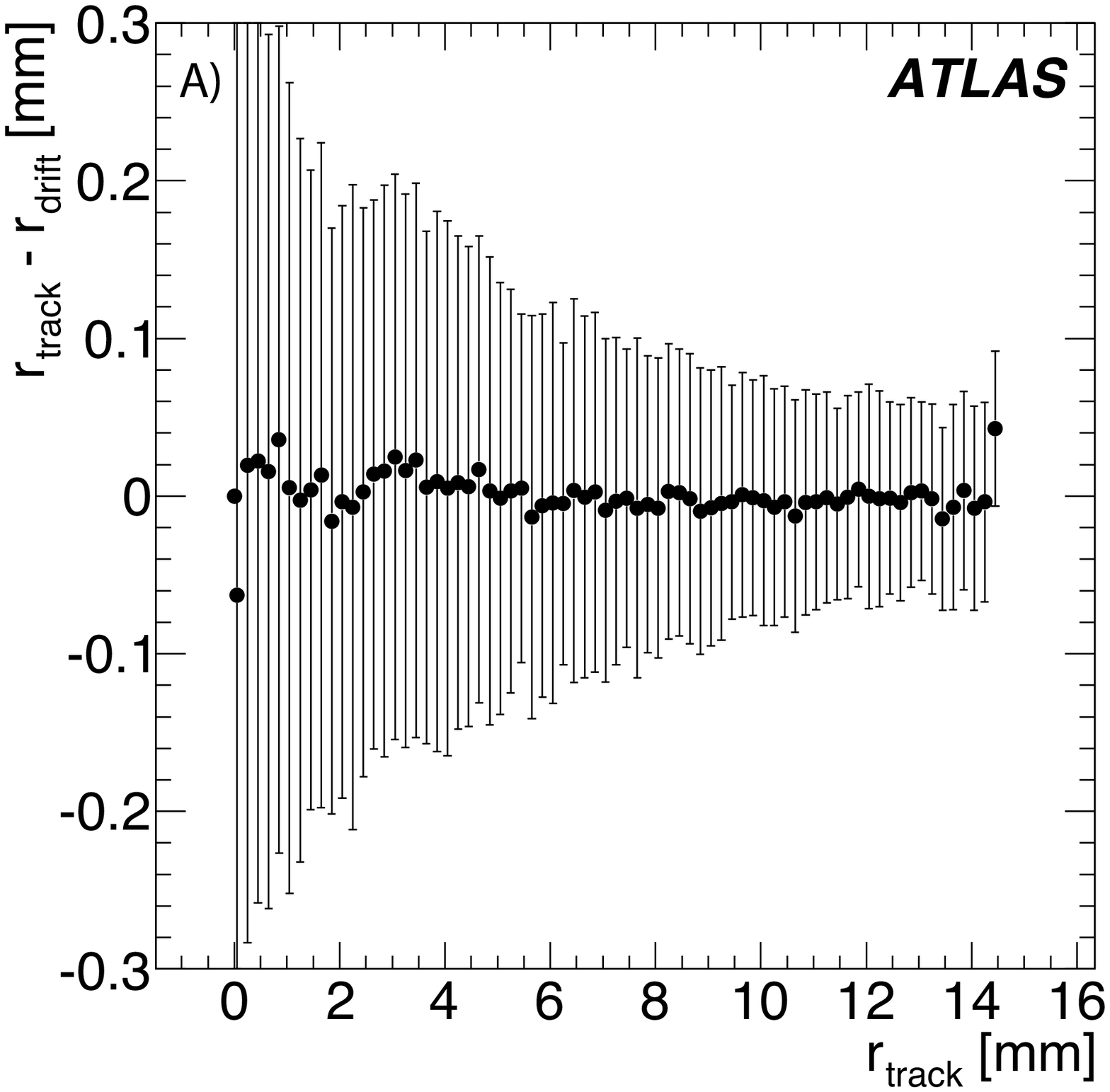}
    \includegraphics[width=7cm]{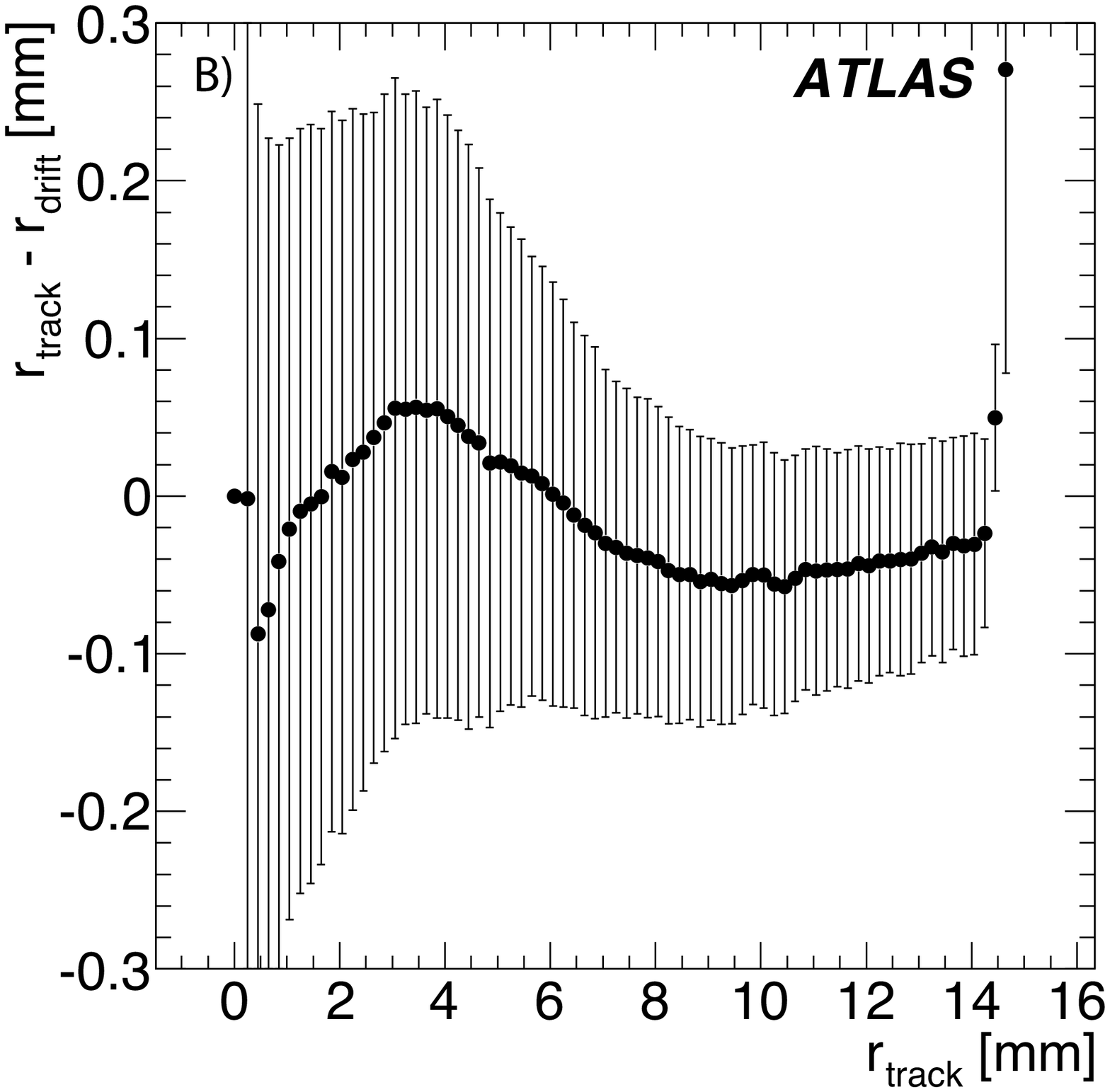}
    \caption{A): Residuals as a function of the track segment distance from the wire after the $r(t)$
    auto-calibration and RPC-time corrections. 
The points correspond to the mean value of the distribution of residuals and the error bars to its
RMS value. B): Residuals as a function of the track
    segment distance from the wire after
    the $r(t)$ auto-calibration using the G$t_{0}$--refit method. The points correspond to the mean value of the
    distribution of residuals and the error bars to its RMS value. Residual systematics at the level of 50$\mum$ are present using this correction.}
\label{calib_residuals}
\end{center}
\end{figure}

%The $t_{0}$ value represents a time offset that is needed to convert the measured time into the real electron drift
%times associated to the muon drift distance from the MDT wire.
The $t_{0}$ offset depends on many fixed delays like cable lengths,
front-end electronics response, Level-1 trigger latency, time of
flight from the IP and has to be determined for each drift tube.
The offset is obtained by fitting a Fermi function to the leading edge
of the drift time distribution as shown in
Figure~\ref{calib_t0_fit} left. The precision expected in LHC collision
data is better than 1~ns with a dataset of about 10K muons crossing the drift tube.
This uncertainty does not significantly degrade the position 
resolution of the MDT tubes which corresponds to a time span of about 5~ns.
In Figure~\ref{calib_t0_fit} right also  the typical spectrum of ADC for all tubes in a chamber is reported.
Charge information in each tube is obtained using a Wilkinson ADC~\cite{ADC}.
%tml - In most places I changed "r(t) relations" to "r(t) functions".  "r-t relation" is OK, but
%"r(t)" is a function.
 As the MDT chambers are operated at different
temperatures depending on their positions in the MS, the $r(t)$
functions
 differ depending on location and are determined separately. In
addition, variations of the toroidal magnetic field along the drift
tubes produce different Lorentz angles, thus different $r(t)$
functions. An initial rough estimate of the $r(t)$ function is
obtained with an accuracy of 0.5~mm by integrating the drift-time
distribution. This is correct under the approximation of a uniform
$dn/dr$ distribution, where $n$ is the number of hits at a drift radius $r$

%\begin{eqnarray}
%    \frac{dn}{dt} = \frac{dn}{dr}\frac{dr}{dt} = \frac{N_{hits}}{r_{max}} \frac{dr}{dt} \qquad
%    \Rightarrow \qquad r(t) = \frac{r_{max}}{N_{hits}} \int\limits_{0}^{t}
%                            \frac{dn}{dt'}\,dt'.
%    \nonumber
%\end{eqnarray}
\begin{eqnarray}
    \frac{dn}{dt} = \frac{dn}{dr}\frac{dr}{dt} = \frac{N_{hits}}{r_{max}} \frac{dr}{dt}
    \Rightarrow r(t) = \frac{r_{max}}{N_{hits}} \int\limits_{0}^{t}
                            \frac{dn}{dt'}\,dt'.
    \nonumber
\end{eqnarray}
$N_{hits}$ is the total number of hits in the time spectrum and $r_{max}$ is the maximum drift radius (14.4 mm).
In cosmic rays this 
approximation is only good at the level of a few hundred$\mum$ mainly because of the production of $\delta$-ray electrons along the track. An $r(t)$
relation with a higher accuracy, of about 20$\mum$, is obtained from
this initial estimate by applying corrections, $\delta r(t)$, which
minimize the residuals of track segment fits with an iterative
procedure. This minimization procedure, called {\it
auto-\-calibration}, takes into account the dependence of the
parameters of the fitted segments on the applied corrections $\delta
r(t)$ and is mainly based on the geometrical constraints from the
precise knowledge of the wire positions.
Figure~\ref{calib_residuals} shows a typical residual distribution
of a chamber, as a function of the distance of the track segment
from the anode wire, after the auto-calibration.

%tml - This discussion seems largely redundant for the t0 refit.
%This should appear only once and then be referred to as necessary.
In cosmic ray events additional sources of time jitter, beyond the
intrinsic resolution, spoil the MDT measurement. The first
cause of time jitter is due to cosmic ray muons crossing the tubes
with an arbitrary phase with respect to the front-end electronics
clock~\cite{MDTelectronics}. This implies a time jitter
corresponding to a 25 ns uniform distribution. The second cause is related to the spread of the trigger time for triggers generated in  different parts of the detector  (up to about 100 ns due to the initial stage of the trigger timing).
%different trigger towers. 
%as described in Section~\ref{Trigger}. From the MDT drift-time spectra only an average values of $t_{0}$ per tube can thus be obtained.
%In cosmic ray events several additional sources of time jitters are spoiling the MDT drift measurement perfromance: the FE electronics clock is not synchronous with the muon crossing time (additional 25ns time jitter), the time of flight between trigger
%chambers and precision chambers can vary depending on cosmic trajectories, the barrel trigger chamber timing was still not fully commissioned during the analyzed runs.
Two different methods have been alternatively used to reduce the impact of these
effects: the {\it RPC-time correction} and the MDT {\it
G$t_{0}$--refit}. The achieved performance with both methods are
discussed in Section~\ref{MDTperf}. In the following a brief
description of the former method is given.

The RPC-time correction uses  the
trigger time measured by the RPC chambers on an event by event basis. This time correction was
applied only to the MDT chambers of the BM stations since these chambers
are close to the two RPC stations used to issue the trigger and so  no corrections due to time of flight and, more importantly, no corrections due to the spread in  timing of the trigger signals issued by different parts of the detector are needed.
%Extension to the MDT chambers in the BI and BO stations would require a more complex
%algorithm able to make ad-hoc corrections depending on which RPC
%trigger tower issued the trigger for the particular event. 
This method cannot be applied to the end-cap region since the TGC do
not provide a measurement of the trigger time but rather they select
the appropriate BC.

%The results of the RPC-time correction method are presented in Figure~\ref{calib_residuals}. They are obtained by subtracting  from %the trigger time, the time measured by RPC front-end
%electronics. 
\begin{sloppypar}
With this correction the time jitter due to the two
effects mentioned above is reduced from $\sim$100~ns to few ns (see
Section~\ref{MDTperf}).
The distribution of the residuals obtained after calibration using  the RPC-time correction method is presented in Figure~\ref{calib_residuals} left. The precision of the auto-calibration  is better than   $\sim$20$\mum$ using this correction. 
\end{sloppypar}

% G$t_{0}$--refit, consists introduces
%in every track segment fit a free parameter corresponding to a
%global time offset, common to all hits used to fit the segment.
%To compensate for the fluctuation of the trigger time, the segments are fitted with an additional free parameter, which is a drift time offset applied to all hits.
%This method is independent of the track topology and of the detector that
%issued the trigger. 

The G$t_{0}$--refit has also been used to 
 improve the single tube
resolution, as discussed in Section~\ref{MDTperf}.  Also the precision of the auto-calibration is much improved with respect to the uncorrected situation. As shown in  
Figure~\ref{calib_residuals} a precision of  $\sim$50$\mum$ is obtained for  the residuals of the segment fit after auto-calibration with small residual systematics on the auto-calibration.
%\begin{figure}[!hbt]
%\begin{center}
 %  \includegraphics[width=10cm]{plots/residual_t0refit.eps}
  %  \caption{\label{res_with_t0_refit} Residuals as a function of the track
   % segment distance from the wire after
    %the $r(t)$ auto-calibration using the G$t_{0}$--refit method.
   % The points correspond to the mean value of the
  %  distribution of residuals and the error bars to its RMS value.}
%\end{center}
%\end{figure}

\subsection{End-cap chambers calibration with monitoring chamber}
For the end-cap MDTsystem, due to the limited number of cosmic ray events,
a different method to determine the $r(t)$ relation was used. A
small MDT chamber installed on the surface of the ATLAS underground
hall was set up \cite{gas_monitoring_chamber} to monitor
%tml - I think this may be the first use of "multilayer", and it's undefined.
continuously the MDT gas composition. One multi-layer is connected to the supply line of the gas recycling system while the other is connected to the return line.
This chamber benefits from a very large cosmic ray
rate and can therefore determine the $r(t)$ function with high
precision in short time intervals. Cosmic ray muons are triggered by
scintillator counters mounted on the monitoring chamber. The trigger time is
measured and subtracted event-by-event from the tube drift times, in
this way the jitter related to the asynchronous front-end clock is
automatically removed.

\begin{figure*}[!htb]
\begin{center}
  \includegraphics[width=\textwidth]{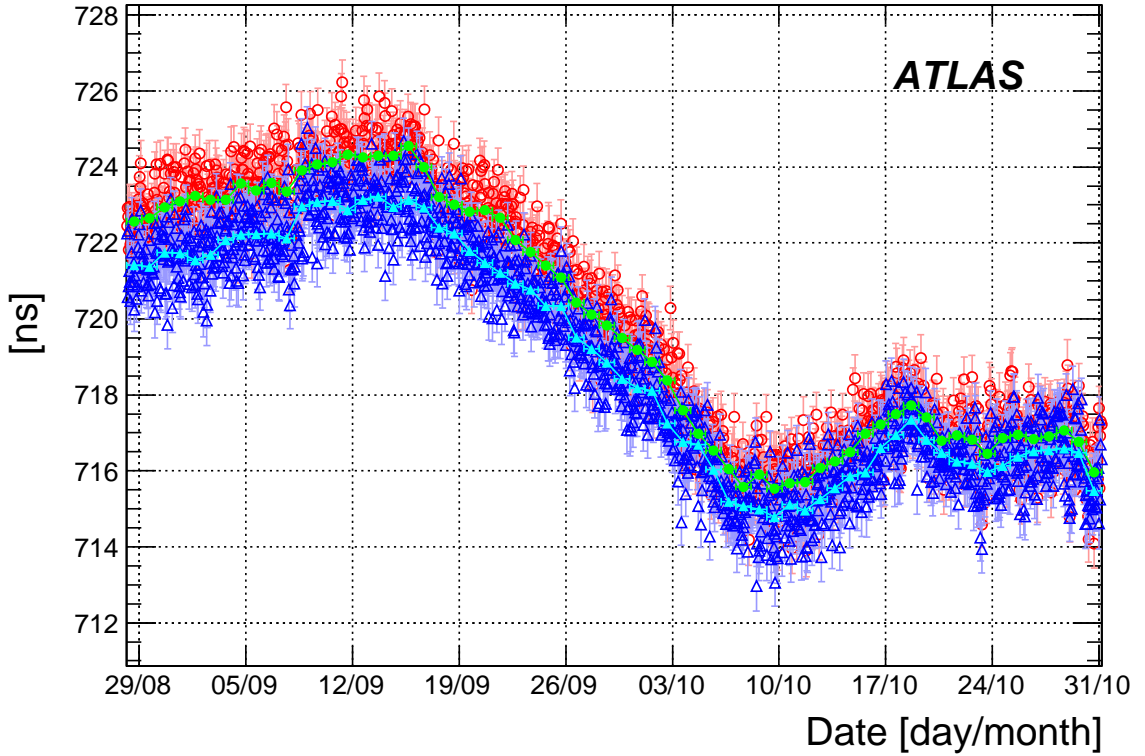}
    \caption{\label{drift_time_vs_time} Maximum drift time measured by the gas monitor chamber versus  time during September-October 2008. The red points refers to the return line and the blue points to the supply line (green and light blue points are the time average of the supply and return line measurements).  The large variation seen between middle of September and 10$^{th}$ of October is due to the change of the quantity of water vapour added to the standard mixture.}
\end{center}
\end{figure*}

Data from the monitoring chamber are used to derive a $r(t)$
function every 6 hours to monitor the gas drift properties.
Figure~\ref{drift_time_vs_time} shows the variation of the maximum
drift time (the drift time of muons crossing the drift tube close to
its edge) over the period September-October 2008. Two $r(t)$
functions were used to cover the Fall 2008 run period, 
%the first for run numbers from 87760 to 90270, the second for the following runs.
for each period the $r(t)$ function for each chamber of the MS was
corrected to account for the temperature difference using the data
measured by the sensors mounted on any particular chamber.  The temperature
corrections to the $r(t)$ function were derived from the Garfield-MagBoltz
simulation program \cite{Veenhof:1993hz}, \cite{Biagi:1999},\cite{garfield}.
The output of the simulation
was validated by several measurements with a muon
beam~\cite{h8b}.
% and in the laboratory~\cite{h8a,h8b}.
In the end-cap region, the
temperature varies by about 4$^{\circ}$C  from top to bottom of the MS, resulting in a variation of the
maximum drift time of about 10~ns, nn the other hand the temperature of the cavern was remarkably stable in time.

\begin{figure}[!htb]
\begin{center}
 \includegraphics[width=\columnwidth]{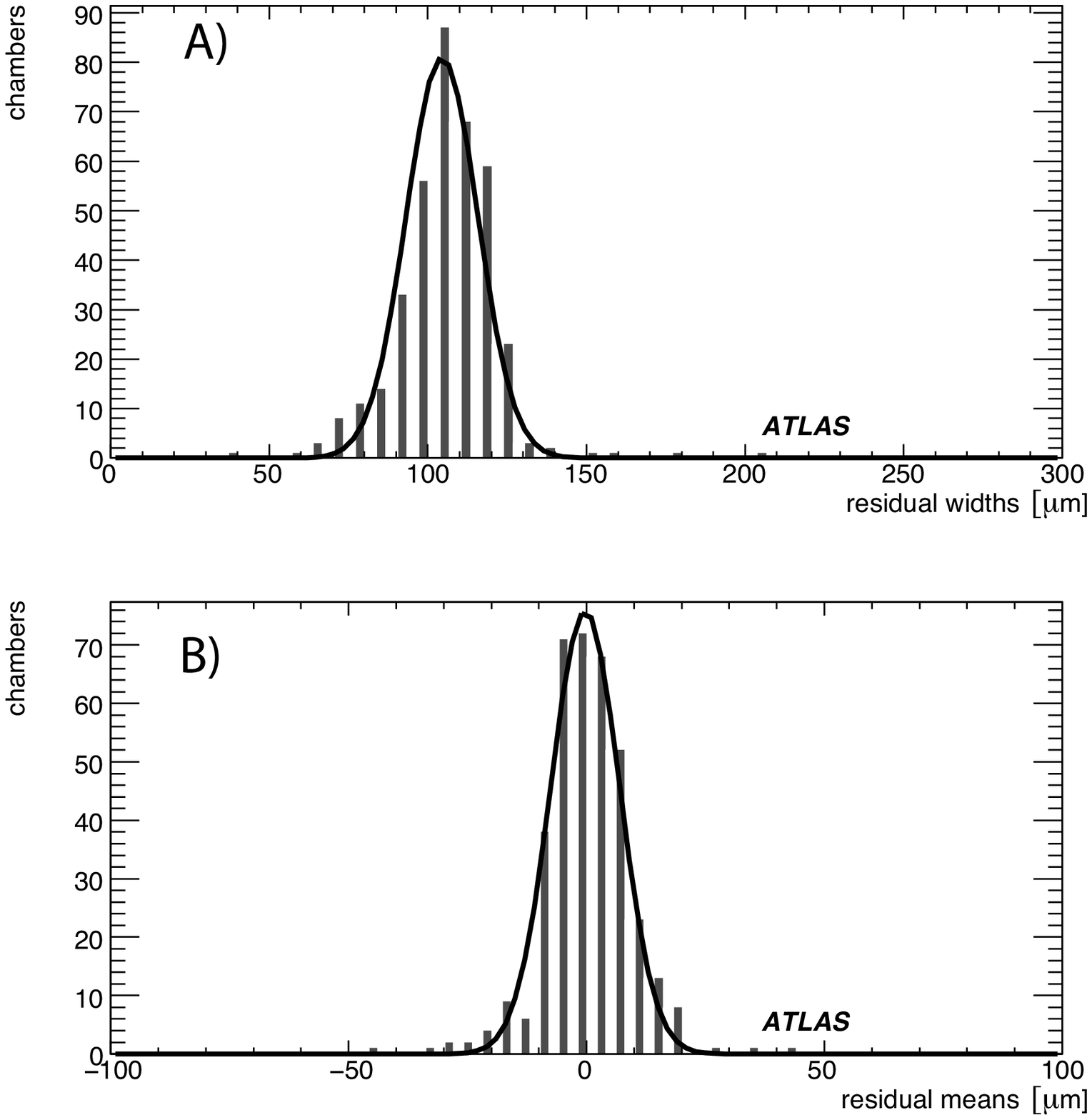}
    \caption{\label{mon_chamber_res} A) Distribution of the RMS  and  B) of the mean 
    values of the residuals from the fit
    to track segments in 373 end-cap chambers using the $r(t)$ function derived from
    the gas monitoring
    chamber. The black lines represent Gaussian fits to the distributions.}
\end{center}
\end{figure}

Figure~\ref{mon_chamber_res} shows the distribution
of the mean and RMS value of the residuals from the fit to track segments in
all end-cap chambers (run 91060). A Gaussian fit is superimposed. The $r(t)$
function derived from the gas monitor chamber with temperature corrections provides an acceptable calibration for all the MDT chambers of the end-cap: the
average standard deviation of the residuals is about 100$\mum$.

 q
%%%%%%%%%% End text from Felix

\section{Detector performance: efficiency and resolution}
\label{Performance}

%\subsection{MDT}
\label{MDTperf}
\subsection{MDT}
% {\small {\bf remove this sentence -} In this section the cosmic ray collected in the ATLAS cavern are used to measure the MDT performance.
% Several aspect related to cosmic rays have been taken into account since they have a large impact on the measured performance.
% In particular the asyncronous behaviuor of cosmic ray muons with respect to the LHC clock required the development of dedicated techniques as explained in the following subsections.}
%{\bf Editor: Mauro - 2 pages}\\
%{\it Co-editors: Felix, Dan, Peter, Egge, ....}\\
%%%%%%%%%%%%%%%%%%%%%%%%%%%%%%%%% Begin New part from Mauro %%%%%%%%%%%%%%
%\subsubsection{MDT drift time distribution and G$t_{0}$--refit}
\subsubsection{MDT drift time distribution}
The behaviour of the drift time  distributions of individual tubes is
an important  quality criterion for the MDT performance. The minimum and maximum
drift times, $t_{0}$ and t$_{max}$, respectively, correspond to
particles passing very close to the wire and close to the tube
walls, and their stability indicates the stability of the calibration.
 The number of hits recorded in a small time window before the rising edge of the drift time distribution 
$t_{0}$ can be used to estimate the rate of noise due to hits not
correlated with the trigger.
%and of cavern background (that will be present in LHC collisions).
A precise knowledge of $t_{0}$ for each tube is essential for high
quality segment and track reconstruction. As explained in
Section~\ref{Calibration}, for cosmic rays some additional time
jitter is present and must be accounted for.
%The value of this quantities represents a time offset that is needed to convert the measured time into the real electron drift times associated to the muon drift distance from the MDT wire.
% In LHC collision mode it only depends on fixed delays (i.e., cable lengths, trigger latency, ...) and the time of flight of muons from the ATLAS IP.
%The first source of time jitter is the fact that cosmic muons are crossing the tubes with an arbitrary phase with respect to the Front-End electronics clock (corresponding to the LHC distributed clock).
%In LHC collisions the FE electronics is synchronous with the LHC Bunch Crossing (BC): the phase
%between the clock and the recorded hit is at fixed delay that is absorbed in the $t_{0}$ definition.
%For the cosmic events this implies an additional time jitter corresponding to a 25ns uniform distribution.
%An additional source of time jitter during the cosmic rays runs analyzed in this paper is relayed to the time spread between the different barrel trigger towers as described in Section~\ref{Trigger}.
%From the MDT drift-time spectra only an average values of $t_{0}$ per tube can thus be obtained as explained in Section~\ref{Calibration} .

%tml- More t0-refit redundancy here
%In order to improve the quality of track reconstruction, a modified
%segment fitting algorithm has been implemented accounting for a
%global time offset as an additional parameter to the segment
%fit~\cite{TNSGTFit}, referred to as G$t_{0}$--refit. 
In order to improve the quality of track reconstruction the G$t_{0}$--refit time correction has been used.
The performance
of the G$t_{0}$--refit algorithm has been investigated in the past,
both using simulated data and using data taken with a BIL (Barrel
Inner Large) chamber in a cosmic ray test stand under controlled
trigger conditions~\cite{TNSGTFit}. The achieved G$t_{0}$ resolution
ranged between 2 and 4 ns depending on the chamber geometry (8 layer
chambers have better resolution than 6 layer chambers) and hit
topology. In particular  the G$t_{0}$--refit  algorithm cannot work if all the hits are on the same side of the wires, typically for tracks at 30$^{\circ}$ with respect to the chamber plane. 
%(e.g., it is not possible to determined the g$t_{0}$ if all hits are on one side of the wire).
The selection of good quality segments requires a minimum of five MDT hits  and segments with all hits on the
same side of the wires are removed.

%In the analysis of present data a fraction of {\bf XXXX} segments did not fulfill the above mentioned requirements.
%The G$t_{0}$--refit has an impact on the precision of the segment parameters: in comparison to a segment with a precisely known $t_{0}$ (i.e., error below 1 ns)
%the resolution on the position and angle is about a factor 2 worse.
%The time measured by the RPC strips that have triggered the cosmic muon can be used to correct for the cosmic muon time jitter.

In addition to the G$t_{0}$--refit also the RPC-time correction
method was used for  the MDT chambers in the middle barrel station (BM) which are located closely to the RPC trigger chambers. The time measured by these RPC can be used to correct
for a global time offset.
%as describer in Section~\ref{Calibration}.
An example of the effectiveness of the method is given in
Figure~\ref{calib_t0_fit} where the drift time distribution for a
BML chamber is shown after RPC-time corrections. The steepness of
%tml - Not clear how "steepness" is defined - is it 10% to 90%.  If so,
% "rise time" would be a better term.  Also "of the distribution" is not
%well defined.
the rising edge, measured as one of the parameters of the Fermi distribution, is improved, passing from 22 ns 
without correction, to 3 ns with RPC time corrections, a value in agreement with results from
muon beam tests~\cite{h8b}. The precision of the RPC-time correction
is about 2 ns as explained in Section~\ref{RPCperf}. This also
includes the contribution of the signal propagation time in the RPC
strips.

\begin{figure}[!hbtf]
\begin{center}
\includegraphics[width=\columnwidth]{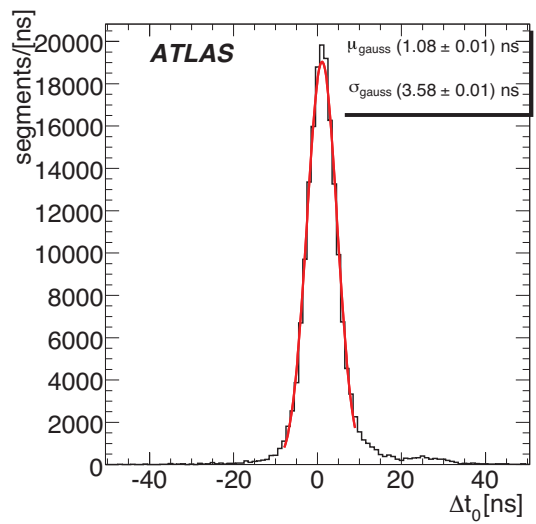}
\end{center}
\caption{Difference between the g$t_{0}$ obtained with the G$t_{0}$--refit method
and with RPC-time correction. The width of the distribution is a convolution of the uncertainties of
the RPC-time correction and the G$t_{0}$--refit method.}
%$t_{0}$ free parameter distribution obtained from the $t_{0}$-refit method on segments where RPC-timing corrections were applied.
\label{deltat0}
\end{figure}

%Segments can thus be reconstructed in two ways: using the RPC-time
%correction or ignoring RPC timing and using the G$t_{0}$--refit
%method. 
The distribution of the difference between the fitted
G$t_{0}$ and the RPC timing correction per segment is shown in
Figure~\ref{deltat0} for a BML chamber. The standard deviation of
about 4 ns is consistent with an uncertainty of 2 ns from the RPC-time 
correction added to an uncertainty of 3 ns introduced by the
G$t_{0}$--refit method. Tails up to 30 ns are present in the
distribution due to bad hit topologies and background hits.

\subsubsection{Drift tube spatial resolution}
%==============================
% {\small {\bf This sentence is replaced} The single MDT tube spatial resolution as a function of the drift distance can be studied on chambers
% for which the segment reconstruction with the RPC timing corrections is applied. As an example, a BML chamber (BML2A03, position 2 in eta, side A, 3rd sector in phi) has been chosen.
% The method is based on an iterative procedure. At first iteration an approximate input resolution function is assumed.
% Segments with a minimum of 6 hits are considered.
% These segments are fitted again after removing one hit at the time and the width of the distribution of the residuals for the excluded hit
% is plotted  as a function of the drift distance from the wire.
% %The widths of  the residual  distribution  ($\sigma_{resid}(r)$)  is measured as a function of the drift distance.
% The errors on straight line fit (depending on the assumed tube resolution) are then propagated to the selected hit ($\sigma_{SL}(r)$).
% The resolution is then computed by quadratically subtracting from the standard deviation of the residuals the
% extrapolation error due to the fit:
% \begin{equation}
%  \sigma_{resol}(r) = \sqrt{\sigma_{resid}(r)^2 - \sigma_{SL}(r)^2}
%  \nonumber
% \end{equation}
% The procedure is iterated using the new resolution function until the input and output resolutions agree
% within statistical errors; a small number of iterations (two to four) is usually needed. {\bf with the following}} \vskip .4cm
%==============================
The MDT single tube  resolution, as a function of drift distance, was studied using different time corrections. 
The extraction of the resolution function is based on an iterative method. At the first iteration an
approximate input resolution function is assumed. Only segments with a
minimum of six hits are considered. These segments are fitted again
after removing one hit at the time. Subsequently,  the width of the distribution
of the residuals for the excluded hit, $j$, is computed as a
function of the drift distance from the wire, $\sigma_{fit,j}(r)$.
The errors of the straight line fit (depending on the assumed tube
resolution) are then propagated to the excluded hit. The resolution
$\sigma_j(r)$ is then computed by quadratically subtracting from the
standard deviation of the residuals the fit extrapolation error,
$\sigma_{extr,j}(r)$:
\begin{equation}
  \sigma_j(r) = \sqrt{\sigma^2_{fit,j}(r) - \sigma^2_{extr,j}(r)}
  \nonumber
\end{equation}
The procedure is iterated using the new resolution function until
the input and output resolutions agree within statistical
uncertainties; a small number of iterations (two to four) is usually
needed.

\begin{figure*}[!htb]
\begin{center}
\includegraphics[width=\textwidth]{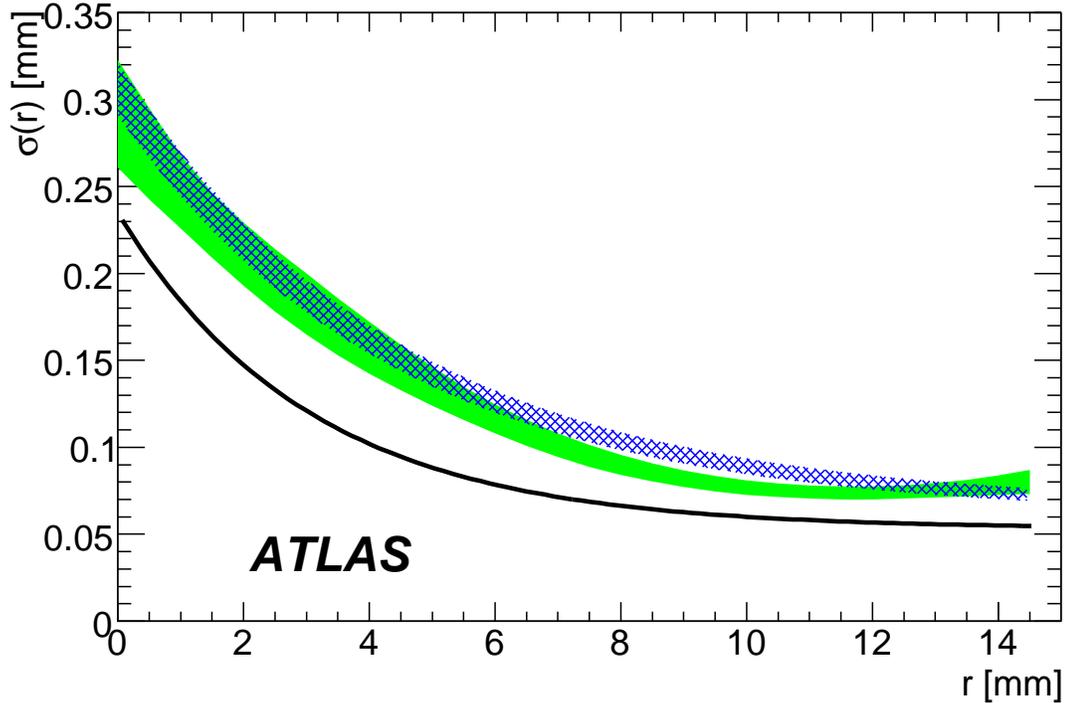}
\end{center}
\caption{Drift tube resolution as a function of the radius.
The green shadowed (RPC correction method) and the blue hatched (G$t_{0}$--refit method) bands represent the resolution
function measured with cosmic rays with the two different methods described in
the text. The solid line represents the resolution measured with a high momentum muon beam~\cite{h8b}.}
\label{resolBML2A03}
\end{figure*}

In Figure~\ref{resolBML2A03} the tube resolution obtained for a BML
chamber is shown as the green band. The width of the band accounts
for the systematic uncertainty of the method. Also shown (solid
line) is the resolution function obtained for an MDT chamber at a
high energy muon test beam~\cite{h8b} with well controlled trigger
timing. This can be considered as reference for the single-tube
resolution. The resolution function measured with cosmic rays is
consistent with a time degradation of the reference resolution of
about 3 ns.  This is in reasonable agreement with the  2 ns time
resolution quoted for the RPC-time correction in addition to a
small contribution from  multiple scattering and individual tube
differences in $t_{0}$.

%{\bf THE FOLLOWING PARAGRAPH AND THE RED LINE IN THE FIGURE SHOULD BE DISCUSSED !}
%==============================
% \vskip .4cm {\small {\bf This sentence is replaced} An alternative method to determine single-hit space resolution, based on segments reconstructed with the G$t_{0}$--refit method and without RPC timing corrections,
% has been applied to the same chamber.
% This method also relies on an iterative procedure: the initial resolution $\sigma_{resol}(r)$ is parametrized by a smooth function.
% Using this parametrization the pull of a selected MDT tube is determined by fitting a straight line to all the hits with the exception of the hit under study (selected hit).
% The errors on straight line fit are then propagated to the selected hit ($\sigma_{SL}(r)$).
% The total error on the residual is than calculated as:
% $ \sigma_{resid}(r) = \sqrt{\sigma_{resol}(r)^2+ \sigma_{SL}(r)^2}$
% The pull is then defined as the residual of the MDT hit divided by the total error ($\sigma_{resid}$). This procedure is iterated until the pull distribution has a standard deviation of 1.
% The obtained resolution is also reported as a dashed line in Figure~\ref{resolBML2A03}.
% The uncertainty on this method is estimated to be of about 5\% and it is consistent with an additional time uncertainty of about 3 ns introduced
% by the G$t_{0}$--refit method. {\bf with the following}} \vskip .4cm
% ==============================

\begin{sloppypar}
The single hit spatial resolution was determined also by applying the
 G$t_{0}$--refit method to track segments reconstructed
in the same chamber. The procedure was similar to that presented
above with the convergence of the method driven by the estimate of
the residual pulls. The resolution function is shown as the blue
hatched band in Figure~\ref{resolBML2A03}. 
%The uncertainty of this
%method is about 5\% and is consistent with an additional time
%uncertainty of 3 ns introduced by the G$t_{0}$--refit method.
The measured resolution is consistent with the test beam measured resolution provided that an additional time uncertainty of  about 2-3 ns is taken into account. 
%also in this case small contributions due to multiple scattering and individual tube differences in $t_{0}$ should be taken into account.
\end{sloppypar}

\subsubsection{Drift tube noise}
\begin{figure*}[!hbt]
\begin{center}
\includegraphics[height=0.8\columnwidth]{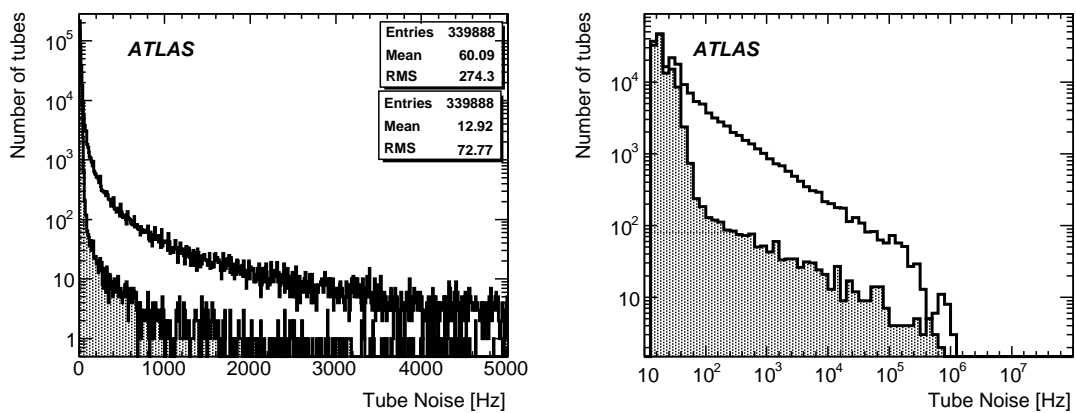}
   \caption{Distribution of the drift tube noise rate with (shadowed histogram, bottom statistical box) and without (empty histogram, top statistical box) the ADC cut described in the text. 
In the right plot the logarithmic scale allows observation of the very few noisy tubes. The tail of the distribution is due to very few noisy tubes that are suffering from pick up of high frequencies through the HV cables or interferences due to the  digital clock present in the front end electronics.} 
\label{MDTnoise}
\end{center}
\end{figure*}

%tml - I think the comma I inserted after "time interval" is correct,
%but I'm confused by the sentence.  The total number of hits includes
%real muons as well - should this be the total number of hits *prior to
% t0*?  In that case I guess the "time range considered" is only the time
% prior to t0?  It's confusing.
%The noise rate is obtained by dividing the total number of hits by
%the corresponding time interval, that is the total number of
%triggers multiplied by the time range considered. In addition to the
%drift time the MDT electronics are also able to measure the charge
%tml - Does "Wilkinson ADC" need a reference?
%associated with each hit based on a Wilkinson ADC. Typical signals
%tml - "nominal running conditions" sounds like conditions of the accelerator.
% Can we say "Nominal high-voltage setting"?
%produced by muons crossing the drift tubes, at nominal running
%conditions, have charges well above the ADC noise pedestal, which
%corresponds to about 50 counts. In the reconstruction algorithms
%only hits with ADC charges above this value are considered. The
%measurement of the single-tube noise with and without this ADC
%charge cut are reported in Figure~\ref{MDTnoise} for all the MDT
%drift tubes readout during the cosmic ray data taking.
%tml - removed this - it's in the figure caption.
% in linear (left)and logarithmic (right) scales.
%The average noise rate is well below
%70 Hz even before the ADC cut selection. This corresponds to an
%average tube occupancy of less then $2 \times \ 10^{-4}$.
The level of noise can be measured in each drift tube by looking at the 
drift time distribution in a given interval before $t_{0}$ where only hits 
uncorrelated with the trigger are present.
The noise rate is obtained by dividing the number of hits normalized to the 
number of triggers by the chosen time interval.
The charge of drift tube signals, at nominal running conditions, is well above the 
ADC pedestal corresponding to about 50 counts, see Figure~\ref{calib_t0_fit}. In the reconstruction algorithms only 
hits with charge above this value are considered.
The distribution of noise rate with and without the ADC charge cut is shown in 
Figure~\ref{MDTnoise} for all MDT drift tubes.
The average noise rate is only 60 Hz without the ADC cut  and 13 Hz with ADC cut, the former figure corresponds 
to an average tube occupancy of less than $10^{-4}$.

\subsubsection{Drift tube efficiency}
\begin{sloppypar}
The single tube efficiency was studied by reconstructing segments in
a chamber using all tubes except the one under observation i.e.
excluding one MDT layer at the time in segment reconstruction. Two
different types of inefficiencies can be defined: {\em i)} absence
of a hit in the tube; {\em ii)} a hit is present but is not
associated to the segment because its residual is larger than the
association cut. The inefficiency of type {\em i)}, referred to as
{\it hardware inefficiency}, is 
%expected to be 
very small, mostly occurring at large drift distances, near the tube wall, 
where the short track length results in fewer primary electrons or due to the track passing through the dead material between adjacent tubes. The
inefficiency of type {\em ii)}, referred to as {\it tracking
inefficiency}, is dominated by $\delta$-electrons, produced by the
muon itself, which can mask the muon hit if the $\delta$-electron
%tml - Can we say "smaller drift distance" instead of "impact parameter"?
%Impact parameter has multiple, other, meanings.
has a smaller drift time than the muon. Tube noise can be
an additional source of this type of inefficiency.
\end{sloppypar}

\begin{figure*}[!hbt]
\begin{center}
\includegraphics[width=\textwidth]{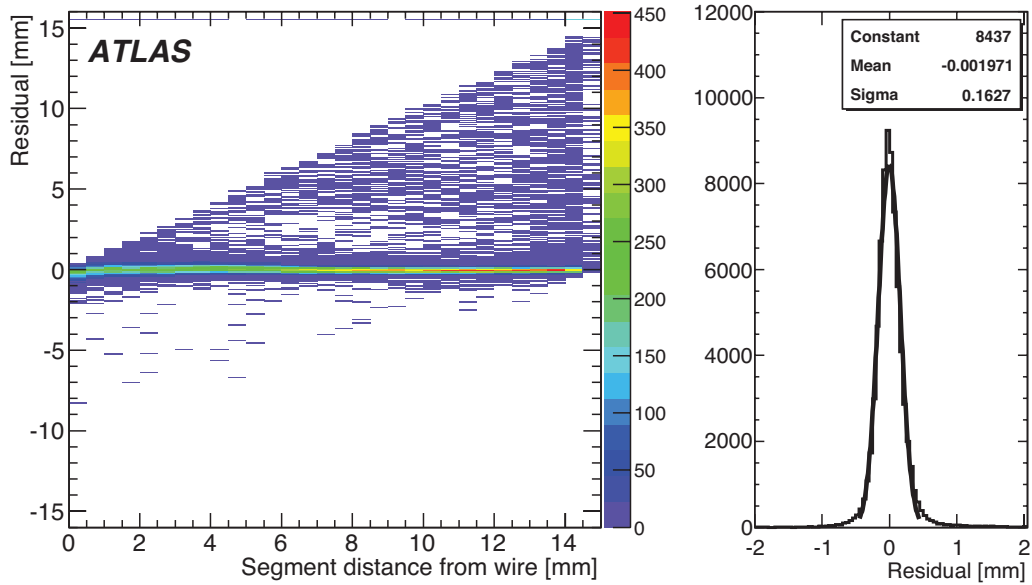}
  \caption{Distribution of the hit residuals for tubes excluded in the segment
  fit, as a function of the distance of the track from the wire.
 Small residuals are associated with efficient hits.
 The triangular region is populated by early hits produced by $\delta$-electrons.
 Missing hits, as explained in the text, are assigned a residual value of 15.5 mm.}
\label{MDTeffiRes}
\end{center}
\end{figure*}

Figure~\ref{MDTeffiRes} shows the distribution of the signed
residuals for hits in the tube of one barrel
chamber
%, BML2A03,
as a function of the distance of the segment from the wire. A large
population at small values of the residual, compatible with the spatial resolution,
is visible. Large positive residuals are associated with early hits
mainly due to $\delta$-electrons. If a hit is not found in the tube
traversed by the muon (thus a residual cannot be computed) a value
of 15.5 mm is assigned, larger than the tube radius of 15 mm. The population
of missing hits is visible at the top of Figure~\ref{MDTeffiRes} and it peaks 
close to the tube wall.

%tml - This plot has a color scale which is not shown.  It needs to be
%otherwise it's not very meaningful.

\begin{figure}[!hbt]
\begin{center}
\includegraphics[width=\columnwidth]{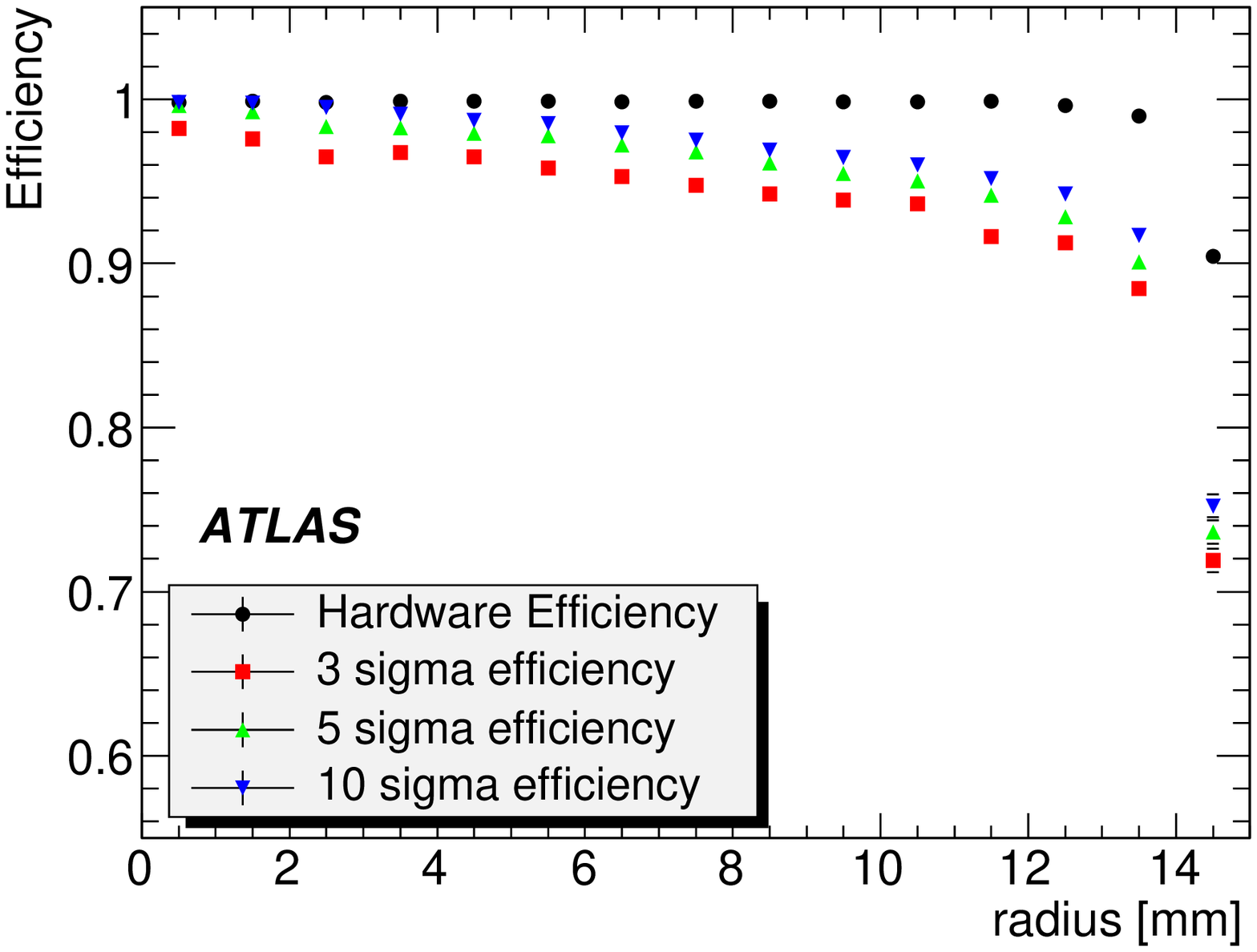}
  \caption{Tube efficiency as a function of the drift distance averaged over all tubes of a
  BML chamber.  Shown are the hardware efficiency and the tracking efficiency for hit residuals
  smaller than 3, 5, and 10 times the standard deviation of the distribution.}
\label{MDTeffiVsRadius}
\end{center}
\end{figure}

The tracking efficiency is defined as the fraction of hits with a
distance from the segment smaller than $n$ times its error, this error
being a convolution of the tube resolution and the track
extrapolation uncertainty. Figure~\ref{MDTeffiVsRadius} shows the
hardware efficiency and the tracking efficiency as a function of the
drift radius for $n$ = 3, 5, and 10.
%tml - I don't think "tracking efficiency" is the right term here.  It
% should be something like "hit-association efficiency".
The tracking efficiency decreases with increasing radius, mainly due
to the contribution of $\delta$-electrons. The average tube hardware
efficiency is 99.8\%; the tracking efficiency is 97.2\%, 96.3\% and
94.6\% for $n$ equals to 10, 5 and 3 respectively.

\begin{figure*}[!hbt]
\begin{center}
   \includegraphics[height=6.0cm]{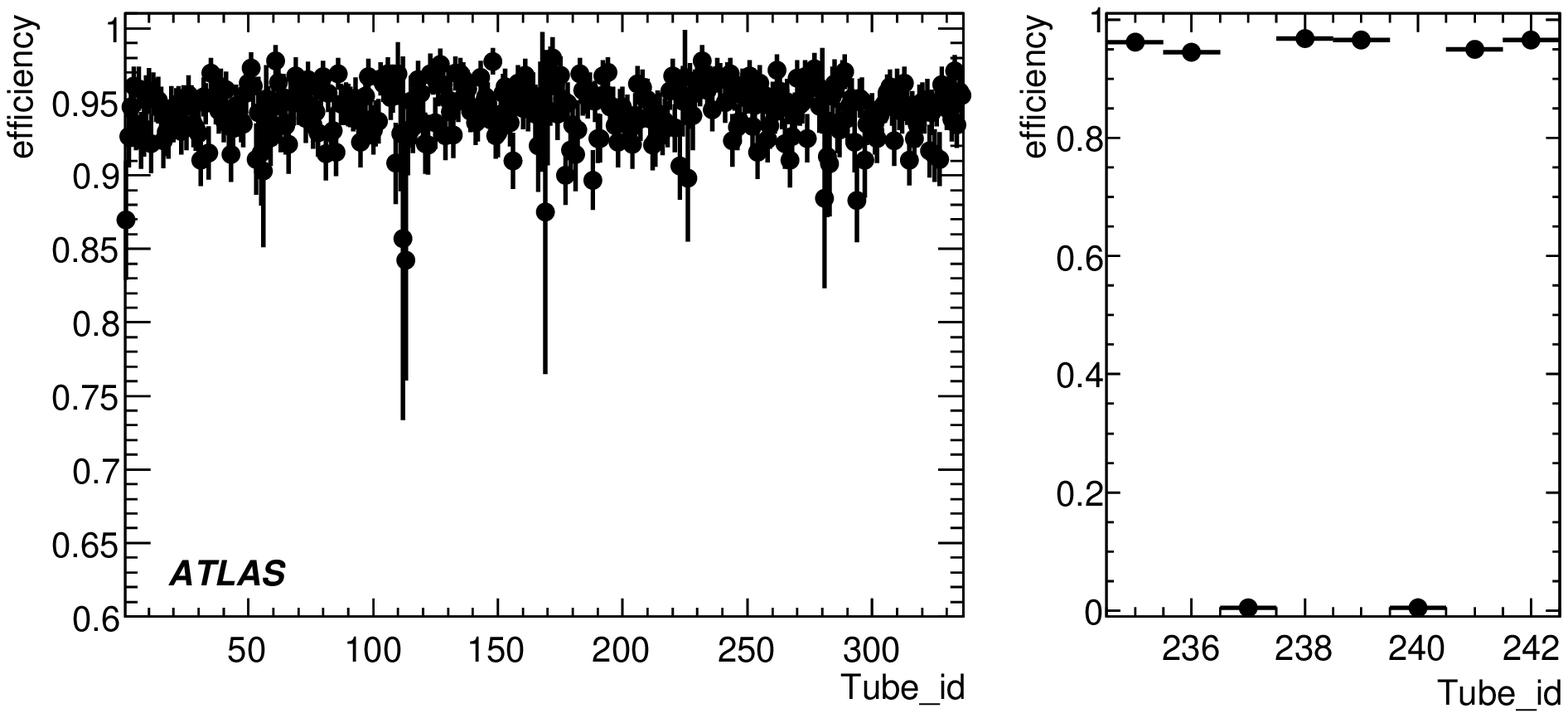}
  \caption{
Single tube tracking efficiencies with a 5$\sigma$ association cut, as explained in the text,
for a BML chamber.
The plot on the right shows an expanded view in the region where two tubes with the wire
disconnected were found.}
\label{MDTeffi1}
\end{center}
\end{figure*}

Figure~\ref{MDTeffi1} shows the average value of the tracking
efficiency for each tube of a BML chamber for $n$ = 5. The average
value is about 96\%. An efficiency consistent with zero was obtained
for two tubes as can be seen in the expanded view on the right plot.
These were recognized as tubes with disconnected wires and were not
considered in the average value.

\begin{figure}[!hbtf]
\begin{center}
   \includegraphics[height=\columnwidth]{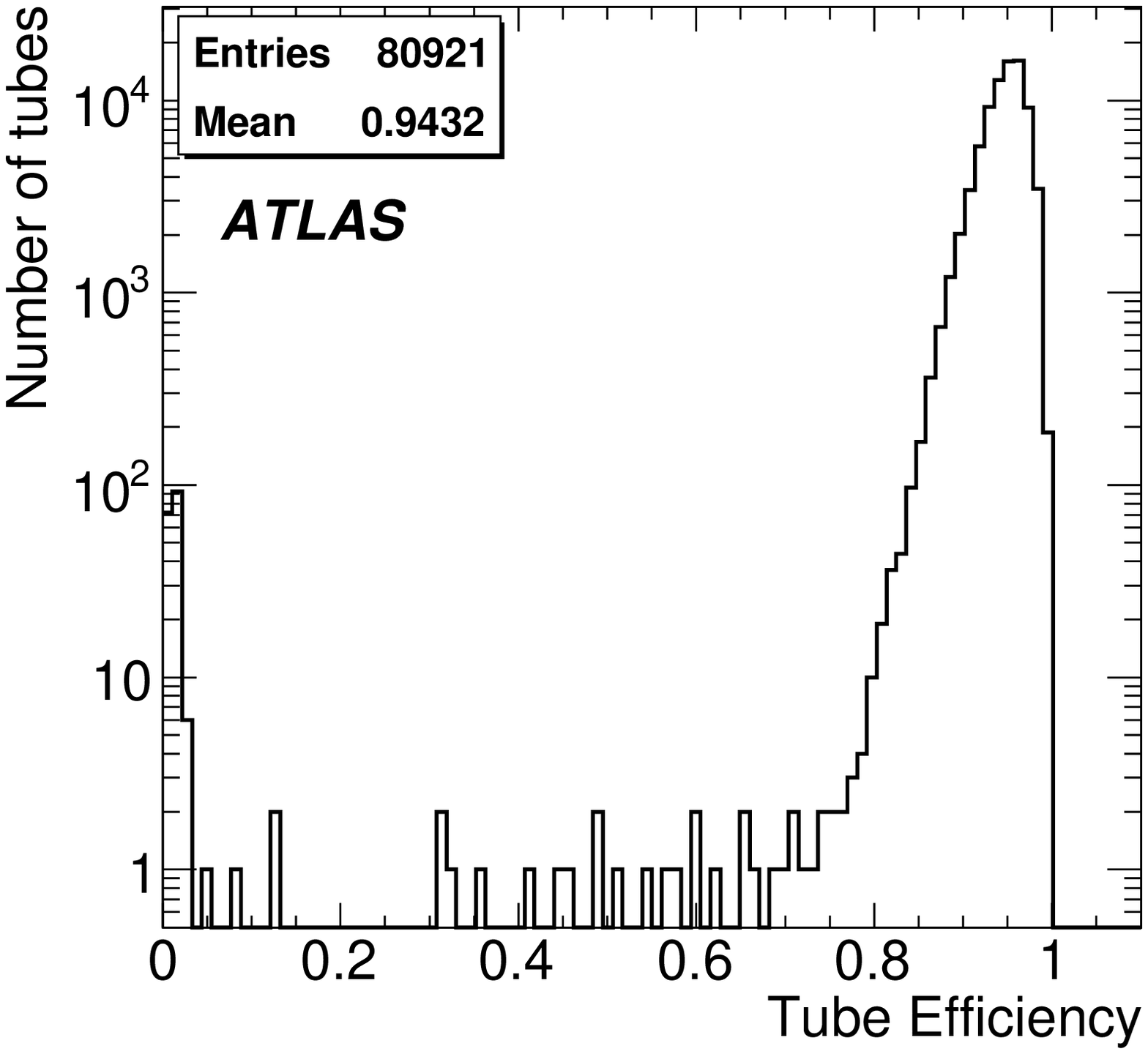}
  \caption{
  Distribution of the tracking efficiency, with a 5$\sigma$ hit association cut,
  for $\sim$81K drift tubes in the barrel MDT.  About 0.2\% of tubes were not
  working and have efficiency compatible with zero.}
  \label{MDTeffi2}
\end{center}
\end{figure}

%tml - Not sure I understand this.  The discussion above is about
%single-tube efficiencies, not entire chambers as this implies.
The results of a study on all the barrel chambers with enough   cosmic ray
illumination to allow the determination of the single tube
efficiency  is presented in Figure~\ref{MDTeffi2}. The distribution of the tracking efficiency for a 5$\sigma$ hit association cut is shown for about 81K drift tubes.  
%Beside detailed studies of the single chamber efficiency, a
%global study was done on the barrel chambers with enough cosmic ray
%illumination to allow the determination of the single tube
%efficiency.
In addition to about 0.2\% of dead channels,
less than 1\%  of tubes have tracking efficiency below 90\%,
mainly due to calibration constants determined with insufficient
precision.

\subsection{RPC}
\label{RPCperf}
\begin{figure}[!htb]
  \centering
  \includegraphics[width=\columnwidth]{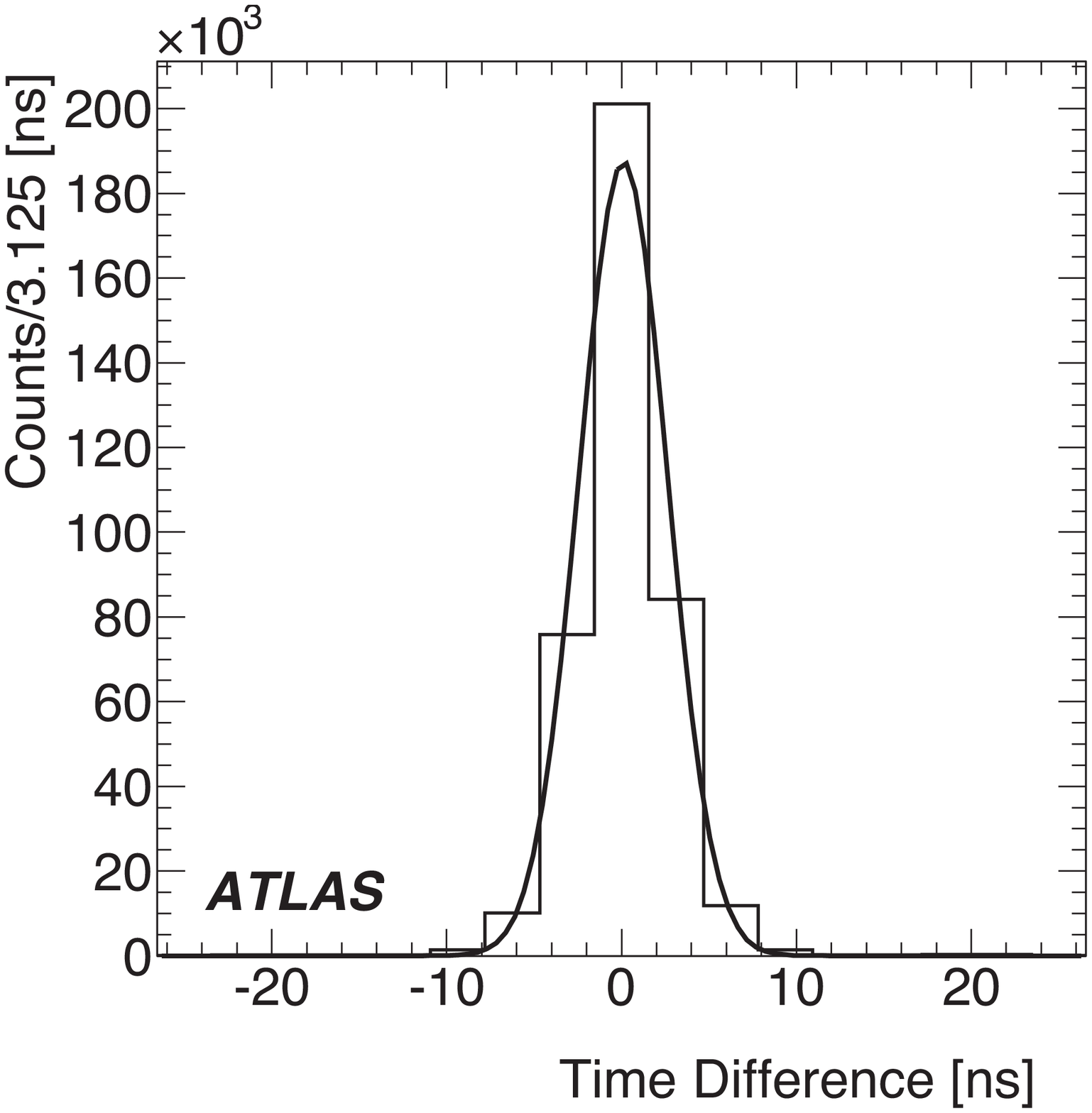}
  \caption{
  Distribution of the time difference between the two RPC layers of a pivot
  plane in the $\phi$ projection. 
  The binning of the histogram corresponds to the strip read-out time 
  encoding of 1/8 of BC.}
  \label{fig:rpcperf:timeres}
\end{figure}

%{\bf Editor: Andrea}\\
%{\it Co-editors: Giordano, Michele, ...}\\
In addition to providing the barrel muon trigger, 
the RPC system is also used to identify the BC of the interaction that produced the muon. This
requires a time resolution  much better than the bunch
crossing period of 25 ns. For this, the time of the strips that form
the trigger coincidence is encoded in the front-end with a 3-bit
interpolator providing an accuracy of 3.125 ns~\cite{LEVEL1-RPC}.
The distribution of the time difference between the two layers of a
pivot plane in the $\phi$ projection was used to determine the
RPC time resolution. With this method there is no need to correct
for the muon time of flight and the signal propagation along the
read-out strips. The RMS width of the distribution shown in
Figure~\ref{fig:rpcperf:timeres} is 2.5 ns. From this a time
resolution of  1.8 ns
%${2.5\over  \sqrt{2}}=1.8$ ns 
is derived for the two RPC layers forming the
coincidence. For this measurement only strips associated to a
reconstructed muon track and belonging to events with one and only
one RPC trigger were considered.

\begin{figure*}[!htb]
\begin{minipage}{36pc}
\begin{minipage}{17pc}
\includegraphics[angle=0,width=7cm]{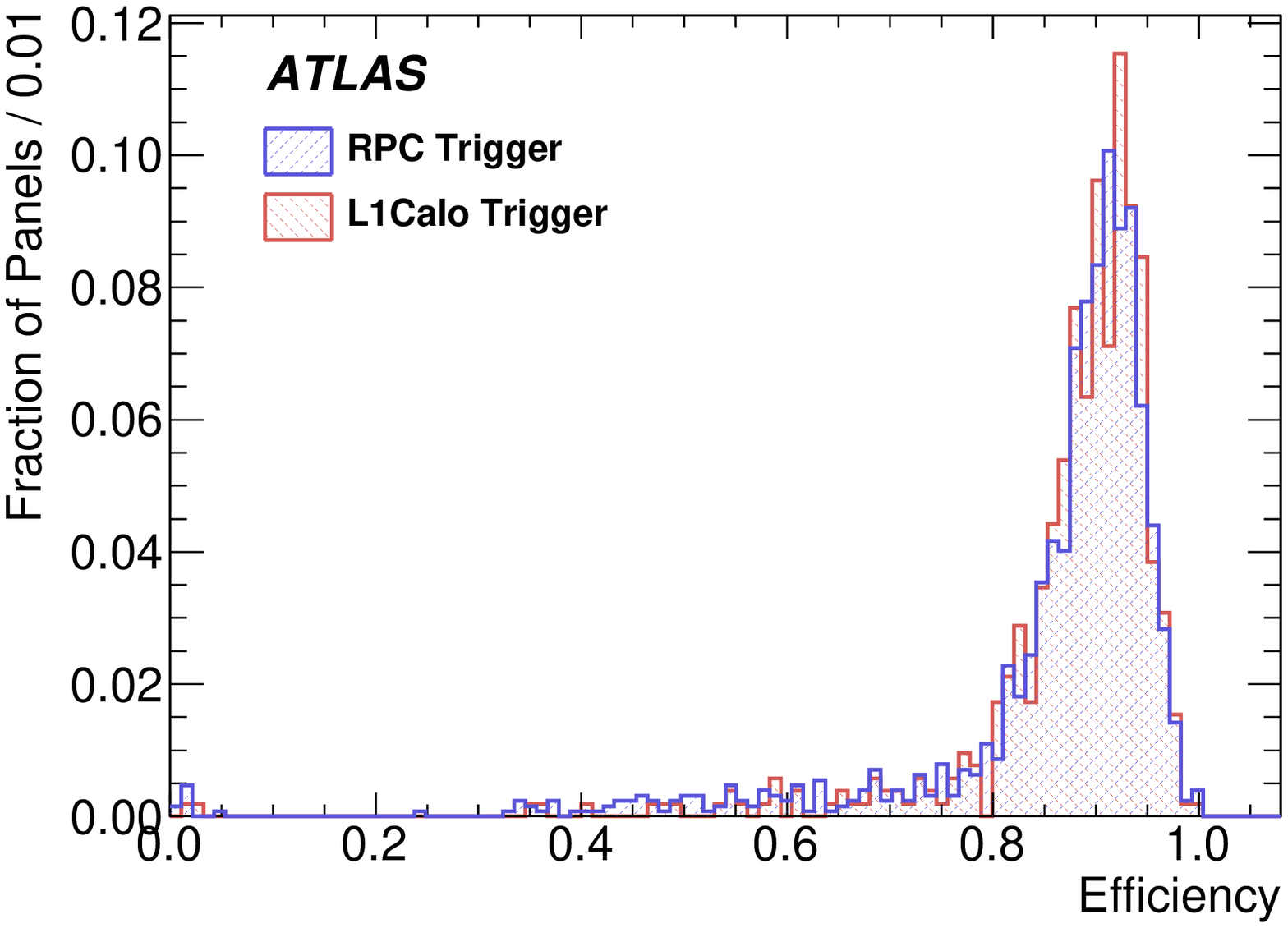}
\end{minipage}\hspace{2pc}%
\begin{minipage}{17pc}
\includegraphics[angle=0,width=7cm]{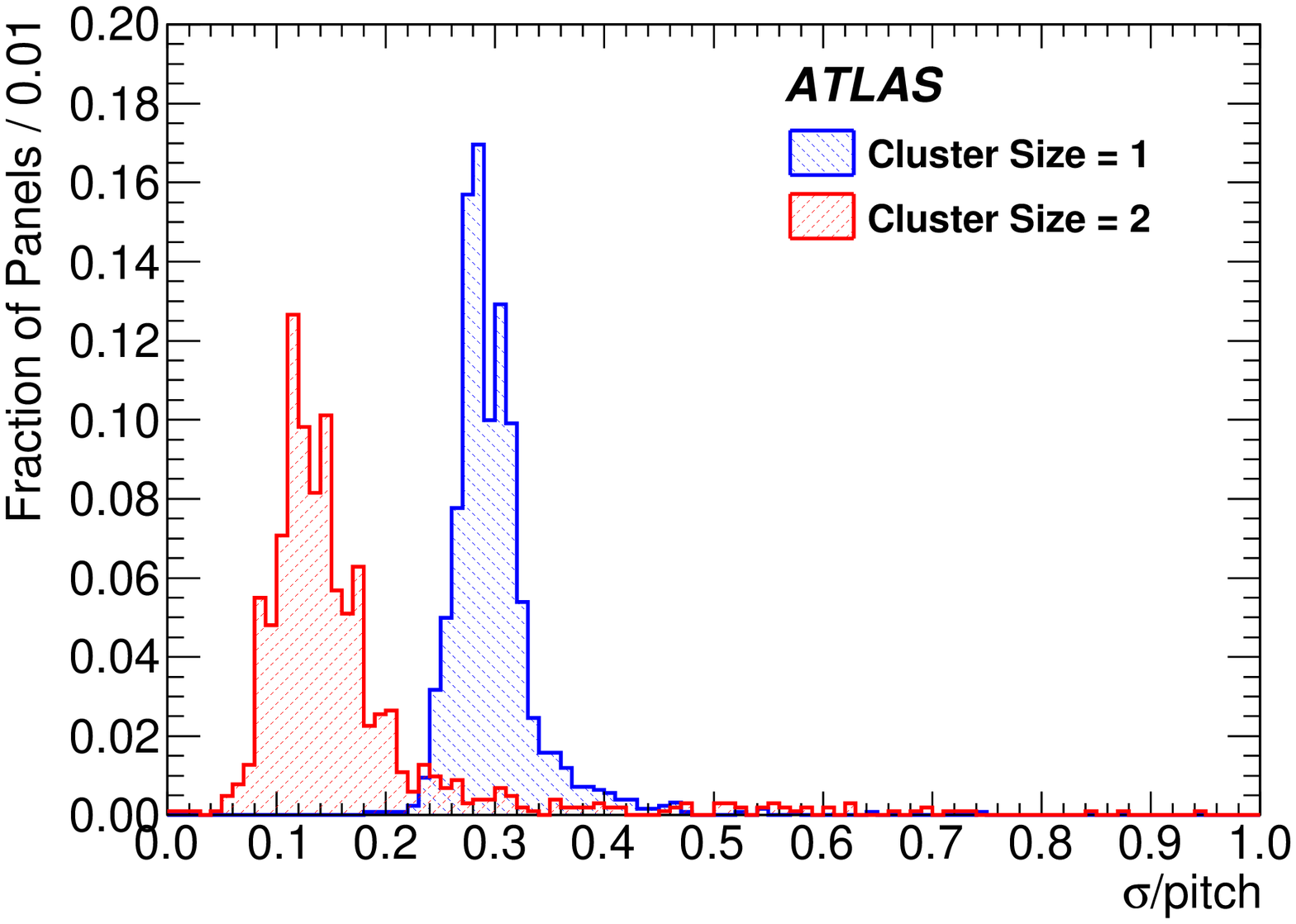}
\end{minipage}
\end{minipage}

\begin{minipage}{36pc}
\begin{minipage}{17pc}
\caption{\label{fig:rpcperf:calotrig} Distribution of the average
efficiency for RPC of the Middle stations for run 91060. The two
distributions refer to two different triggers: RPC trigger (full
line, 91.33\% peak efficiency) and calorimeter trigger (dashed line, 92.0\% peak efficiency). Both distributions are
normalized to unit area. The measured efficiency is lower than expected mainly because the read-out timing was still not optimal.}
\end{minipage}\hspace{2pc}%
\begin{minipage}{17pc}
\caption{\label{fig:rpcperf:sigmares} Distribution of the spatial
resolution provided by the $\eta$ strips for RPC of the Middle
stations. The spatial resolution is divided by the strip pitch. The
distributions are normalized to unit area. }
\end{minipage}
\end{minipage}
\end{figure*}

Two other important RPC quantities related to the detector
performance are the efficiency and the spatial resolution.
%This last quantity is very sensitive to the hit {\it cluster size} that is defined as the number of adjacent hits produced by the crossing muon.
%Read-out efficiency and spatial resolution have been measured in cosmic ray events as expalined in the following.
In order to determine the RPC efficiency two main issues have to be
taken into account. The first one is due to the fact that  the RPCs are actually providing the
muon trigger thus resulting in a {\it trigger bias} on the efficiency calculation.  The second one  is caused by the fact that the RPC  hits are also used in the track reconstruction; in particular, they measure the
coordinate in the non-bending, $\phi$ projection.
%In order to determine the RPC efficiency two main issues have to be
%taken into account: {\em i)} The RPCs are actually providing the
%muon trigger ({\it trigger bias}) and {\em ii)} Their hits are also
%used in the track reconstruction; in particular, they measure the
%coordinate in the non-bending, $\phi$ projection. 
The second
effect has negligible contribution if the efficiency is measured for
the $\eta$ strips, since in this projection the track reconstruction
is driven by the MDT. For the efficiency measurement, MDT tracks were extrapolated to the
% tml - "defined as efficient" sounds like efficiency is a binary thing.
% I think what's mean is that the layer was considered to have a hit from
% the track, but I am not sure.
RPC plane and the layer was counted as efficient if at least one
$\eta$ hit was found with a distance of less than 7 cm from the
extrapolation.  The effect of the {\it trigger bias} has been removed from the efficiency measurement of an RPC plane  by selecting all the events in which the other three planes (in the case of a Middle Station) were producing hits, since the trigger requirement is a 3 over 4 planes majority. 
The distribution of the efficiency, averaged over
each layer, for the RPC chambers in the Middle stations is shown in
Figure~\ref{fig:rpcperf:calotrig}, the distribution is peaked at an efficiency of 91.3\%. 
%This analysis refers to run 91060
%in Fall 2008; the tail at low efficiency values is due to known
%problems, which were mostly fixed for subsequent runs.
To check the remaining impact of the trigger bias on the efficiency measurement, the same analysis was
repeated with a sample of cosmic rays selected with a calorimeter
trigger ({\it Level-1Calo} trigger) independently of the RPC trigger
response. The result for the efficiency is superimposed in
Figure~\ref{fig:rpcperf:calotrig}: a good agreement between the two
distributions is observed.
%Several dedicated runs were taken to determine the working point of the chambers from the point of view of the High Voltage (HV) setting. Figure \ref{fig:rpcperf:effDiff} summarizes
%such measurements showing, for each panel, the variation in efficiency observed between two different HV values.
%The plot shows that at 9.6kV, almost all panels have already reached the efficiency plateau.\\
%As far the the reconstruction bias is concerned, it cab be removed by excluding from pattern recognition and track fit the layer for which efficiency has to be measured.
%This was not yet done in the analysis presented here, but is expected to give a negligible contribution in the efficiency determination of the eta (bending) readout panels.
%On the other hand, removing the trigger bias is less trivial. Knowing the trigger configuration (in particular its majority) one can extract from the data an unbiased sample for a given layer. For example if the trigger is requiring a coincidence of at least 3 %RPC layers out of the 4 in the low-$p_{T}$ station one can have an unbiased efficiency measurement for layer 1, by using only events where layers 2, 3 and 4 had a hit. This correction has been applied in the analysis, and its effectiveness is clearly %visible in Figure \ref{fig:rpcperf:calotrig} shows the efficiency distribution already discussed in Figure \ref{fig:rpcperf:effBM}, compared to the same distribution obtained with a different trigger source.\\

The spatial resolution is related to the clusters size, that is the number of strips associated to a muon track. A muon crossing the detector near the center of a readout strip, will in general produce a cluster of size one, while clusters of size two are only observed when muons hit a narrow region at the boundary between two strips. The actual sizes of the regions corresponding to clusters of size one and two depends on the detector operating parameters, but it is in general true that the latter is smaller than the former. This implies that the spatial resolution must be smaller when measured on a subset of data with only clusters of size two. The spatial resolutions of $\eta$ strips was determined selecting muon tracks reconstructed in the MDT as explained above. For each RPC read out plane, the distribution of the distance from the extrapolated track was obtained separately for clusters of size one and two and then was fitted with a Gaussian. The RMS widths of the fit were divided by the strip pitch (ranging from 27 to 32 mm depending on the chamber type)  to allow for comparison between different RPC and are shown in Figure~\ref{fig:rpcperf:sigmares}.  This technique has been used only for the $\eta$ panels since the MDT are measuring in the Z-Y plane. 
On average, clusters of size two give a spatial resolution about half as for clusters of size one, which is below 10 mm as expected.

\subsection{TGC}
\label{TGCperf}

The basic structure of the TGC chambers and their assembly in the MS
end-cap wheels is presented elsewhere~\cite{ATLASdet}. Inactive
regions due to the gas-gap frame and the wire supports account for a
loss of active area varying from 3\% to 6\% depending on the chamber
type. In order to optimize the trigger efficiency these inactive
regions are staggered with respect to the trajectory of high
momentum muons produced at the IP. In the active area the TGC wires
are expected to have an efficiency of more than 98$\%$. For the
cosmic ray run 91060 the trigger logic required a coincidence of 3
out of 4 layers in the doublet chambers (referred to as TGC2 and
TGC3 as in Figure~\ref{fig:muon-spect}).
%, as described in Section~\ref{Trigger} and~\ref{DQA-TGC}.
%tml - This statement about non-pointing and asynchronous has been made
%at least once before and probably isn't needed here.  Also the trigger bias
%issue was mentioned in the RPC section and doesn't need to be repeated.
%However, no one else mentioned showers.  If they're an issue for the TGC efficiency
%then they are for the others too.
In evaluating the detector efficiency one has to take into account
the trigger bias and the fact that cosmic rays are non-pointing to the IP,
asynchronous, and do not only consist of single muons but also of extended showers. 
%To evaluate the wire efficiency with the cosmic muons, the following differences with respect to the collision events should be handled;
%the random arrival timing of the cosmic muons with respect to the LHC clock,
%the trigger bias since events triggered by the TGC are used to determine the TGC wire efficiency,
%the presence of cosmic shower events with high multiplicity,
%the non pointing trajectories of the cosmic muons.
%The TGC wire efficiency has been evaluated with muon tracks selected by applying the following criteria.

To evaluate the efficiency of a layer in the doublet chambers, it is
required that  there is one and only one hit in each of the
other three layers and  that these three hits are
associated to the current BC. This is intended to remove high
multiplicity events (showers) and out-of-time tracks.
%This requirement limit possible ineffiency due to random phase of cosmic muons with respect to LHC clock.
%The requirement of only one wire hit per layer also removes the high multiplicity events due to the cosmic showers.
As a result of this selection, the 3 out of 4 trigger condition is
satisfied independently of the presence  of a hit in the layer
under evaluation. The efficiency of this layer is thus determined in
an unbiased way.

 A similar procedure is used for the triplet chambers (TGC1). When evaluating,  the efficiency of a layer, it is required {\em i)} that
the other two layers satisfy the 2 out of 3 trigger coincidence and
{\em ii)} that the line joining the two hits (track) crosses the
layer in its active area.

\begin{figure*}[!htb]
\begin{center}
  \includegraphics[width=14.0cm]{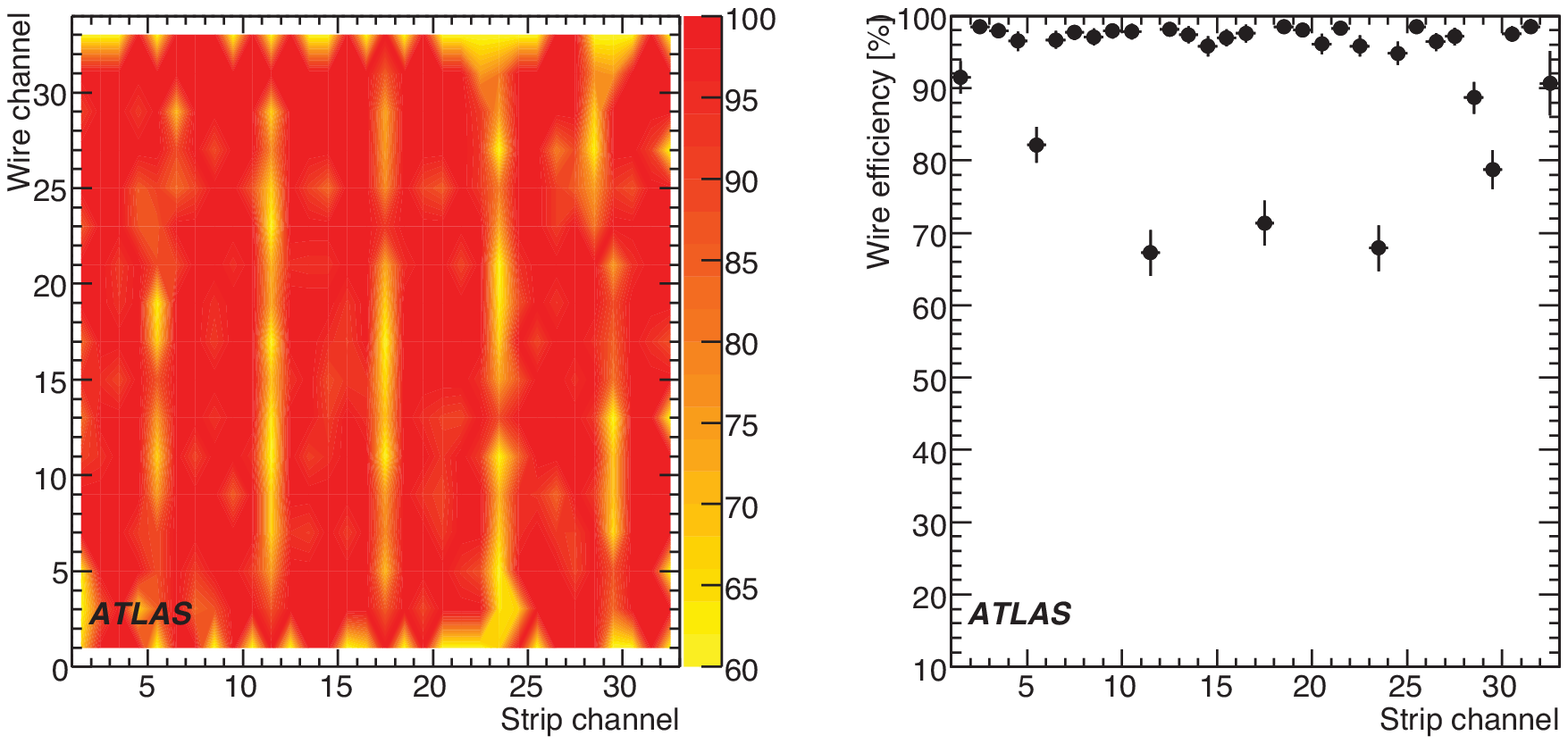}
  \caption{Left: efficiency map for a TGC chamber layer. The horizontal axis is the
  strip channel and the vertical axis is the wire channel. Right: efficiency
  projection to the strip channels. Observed efficiency drops are consistent
  with the wire support locations.}
  \label{sec53:fig1}
\end{center}
\end{figure*}

%tml - Here too the use of "efficient" I think actually means that the chamber is
%considered to have a hit from the track.  This needs to be fixed.
In both cases, the layer under test is considered  efficient if there is at least one hit
associated to any of the previous, current or next BC.
Figure~\ref{sec53:fig1} on the left shows an efficiency map in the
wire-strip ($\eta$-$\phi$) plane, and on the right its $\eta$
projection, i.e. the strip efficiency. Some inactive regions are
clearly visible: the bands in Figure~\ref{sec53:fig1}-Left indicate
the location of the wire supports.

\begin{figure*}[!thtb]
\begin{center}
%  \centering
 % \includegraphics[width=14.0cm]{plots/TGC_Fig1.eps}
% \includegraphics[width=14cm]{newplots/TGC_Fig3.eps}
  \includegraphics[width=14cm]{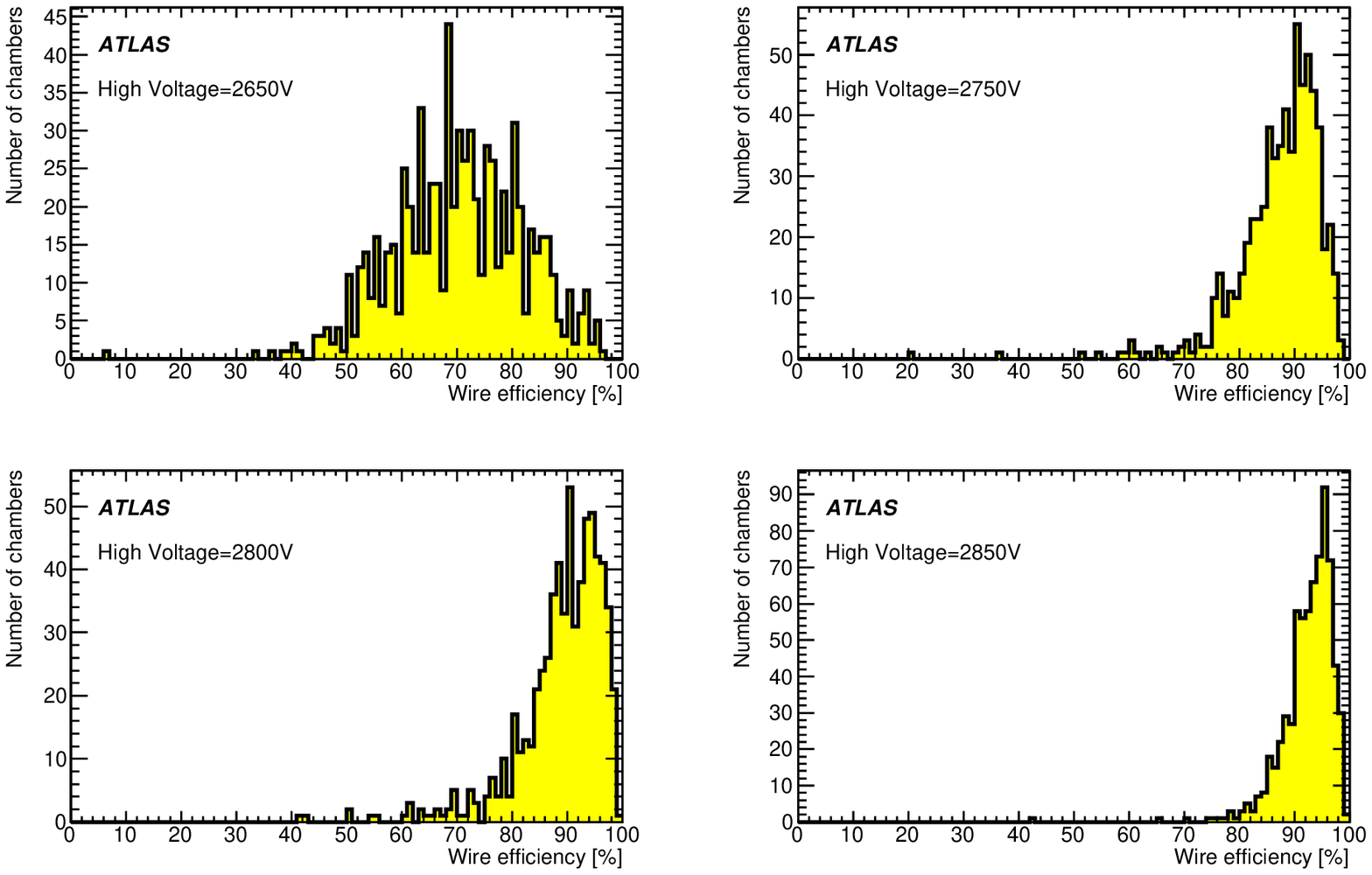}
  \caption{Distribution of the TGC wire efficiency of individual layers for
  different  high voltage values. The distribution for 2800V, the
  nominal voltage in 2008, was obtained with run 91060.}
  \label{sec53:fig3}
  \end{center}
\end{figure*}

\begin{sloppypar}
The overall efficiency, including the inactive regions, is evaluated
for a fraction of TGC layers (about 40\% of TGC doublet layers) by
requiring that a muon track crosses the layer under test at least 10
cm away from its edge. The muon track is defined using MDT hits
combined with TGC hits in the layers that are not under evaluation.
Figure~\ref{sec53:fig3} shows the distribution of the wire
efficiency for different values of high voltage setting: 2650, 2750,
2800 and 2850 V. The average value of the efficiency, at the nominal
voltage of 2800 V, is 92\% consistent with the local efficiency
measured as explained above and the contribution from
inactive-regions.
\end{sloppypar}

%%%%%%%%%%% End NEW Susumu part

%%%%%%%%%%%%%%%%%%%%%%%%%%%%
% inserted here by C.Amelung
%%%%%%%%%%%%%%%%%%%%%%%%%%%%

\section{MDT optical alignment}
\label{Alignment}
%{\bf Editor: Christoph - 0.5 pages}\\
%{\it Co-editors: Christoph, Scott, Pierre-Francois, Igor, ....}\\

%The ATLAS muon spectrometer employs an air-core toroidal magnetic field, which has the advantage of causing only minimal multiple
%scattering due to the small amount of material present between chambers.
% {\small {\bf delete this sentence} The design of the ATLAS Muon spectrometer based on air core toroids has a consequence the relatively low magnetic field strength that can be reached:
% the bending of a $1\ TeV$ muon track in the magnetic field is such that the track sagitta varies
% between $0.5\mm$ at pseudorapidity $\eta=0$ and $1\mm$ at $\eta=2$.
% Consequently, in order to measure the momentum of a $1\ TeV$ muon to 10\% at all angles, the error on the sagitta measurement must be less
% than $50\mu$ in the bending direction of the magnetic field, transverse to the MDT tubes.} \vskip .4cm

The design transverse momentum resolution at 1 TeV  of the MS is about 10\%, this translates into a  sagitta resolution of  50 $\mum$. The intrinsic resolution of MDT chambers contributes a 40$\mum$
uncertainty to the track sagitta, hence other systematic
uncertainties (alignment and calibration) should be kept at the level of 30$\mum$ or smaller.
%Muon tracks are detected in
%three about equally spaced layers of chambers.  
%The intrinsic resolution of the MDTs results in a sagitta error of about $50\mum$, and the
%additional error from the chamber alignment should be smaller than that value.
Since long-term mechanical stability in a large structure such as
the MS cannot be guaranteed at this level, a continuously running
alignment monitoring system~\cite{MSALIGNMENT}  has been
installed. This system is based on optical and temperature sensors
and detects slow chamber displacements, occurring at a timescale of
hours or more. The information from the alignment system is used in
the offline track reconstruction to correct for the chamber
misalignment. No mechanical adjustments were made to the chambers
after the initial positioning.
The system consists of a variety of optical sensors, all sharing the
same design principle: a source of light is imaged through a lens onto an
electronic image sensor acting as a screen. In addition to optical position measurements, it is also necessary to determine the thermal expansion of the chambers. In total, there are about 12000 optical
sensors and a similar number of temperature sensors in the system.
%There is a large variety of optical sensors in the alignment system, all sharing the same principle: a source of light is imaged through a
%lens onto an electronic image sensor acting as a screen. The source of light is either a back-illuminated coded chessboard pattern (RASNIK
%mask), or one or several pairs of point-like light sources (BCAM and SaCam systems). In addition to optical position measurements it is
%also vital to determine the thermal expansion of chambers, by measuring their temperature. In total, there are about 12000 optical
%sensors and a similar number of temperature sensors in the system. They are mounted on chambers and on auxiliary reference
%objects, forming a complex network, the layout of which was validated and optimized by Monte-Carlo simulations. During ATLAS running, sensor
%data are acquired in a combination of sequential and parallel operations. Images are analyzed on-line, and only the analysis result
%(i.e.\ the spot position for BCAM and SaCam images, and the decoded position of the mask for RASNIKs) plus some diagnostic information is
%stored in a database, yielding a data volume of about 0.1\,KB per image. Some fraction of the raw images, having a size of about 100\,KB
%each, is retained for debugging purposes. One readout cycle, yielding one image from each sensor, takes about 30--60 minutes. The alignment
%data acquisition is integrated with the ATLAS detector control system.
%
%tml - I thought the following was not true for the barrel - hence the initial alignment
% with straight tracks
Optical and temperature sensors were calibrated before the
installation such that they can be used to make an {\it absolute}
measurement of the chamber positions in space, rather than only
following their movements with time relative to some initial
positions.

\subsection{End-cap chamber alignment}
\label{ECalign}
\begin{sloppypar}
%{\bf Editor: Christoph 1.5 pages}\\
%{\it Co-editors: Scott, ....}\\
The end-cap chambers and their alignment system~\cite{ENDCAPALIGNMENT} 
were installed and commissioned during 2005--2008, and  continuous alignment
data-taking with the complete system started in summer 2008.
%Commissioning the alignment system was a time-consuming and
%labor-intensive process. One of the challenges was to place the
%chambers to within $5\mm$ of their nominal positions to cope with the alignment 
%optics aperture.
%Another challenge was to locate the entire big wheel system to within 
%$25\mm$ of the design position.
After commissioning, more than 99\% of all alignment sensors were operational, and only a
small number failed during the  data-taking in 2008. The effect of the
missing sensors on the final alignment quality is negligible.
\end{sloppypar}
%tml - I don't think the Gaussian fits help here, since they are not very
% good fits to the central part of the distributions.  FWHM/2.35, from the
% distributions would be more useful.  And the chi2/NDF is confusing unless you
% read the text to understand what it is.
\begin{figure}[!htb]
{\centering
%\ \hfill\includegraphics[height=7.cm]{plots/plot_fig1_7_1a.eps} \hfill\includegraphics[height=7.cm]{plots/plot_fig1_7_1b.eps}\hfill\
%\ \hfill\includegraphics[height=7.cm]{newplots/pulls-data.eps} \hfill\includegraphics[height=7.cm]{newplots/pulls-simul.eps}\hfill\
\ \hfill\includegraphics[height=7.cm]{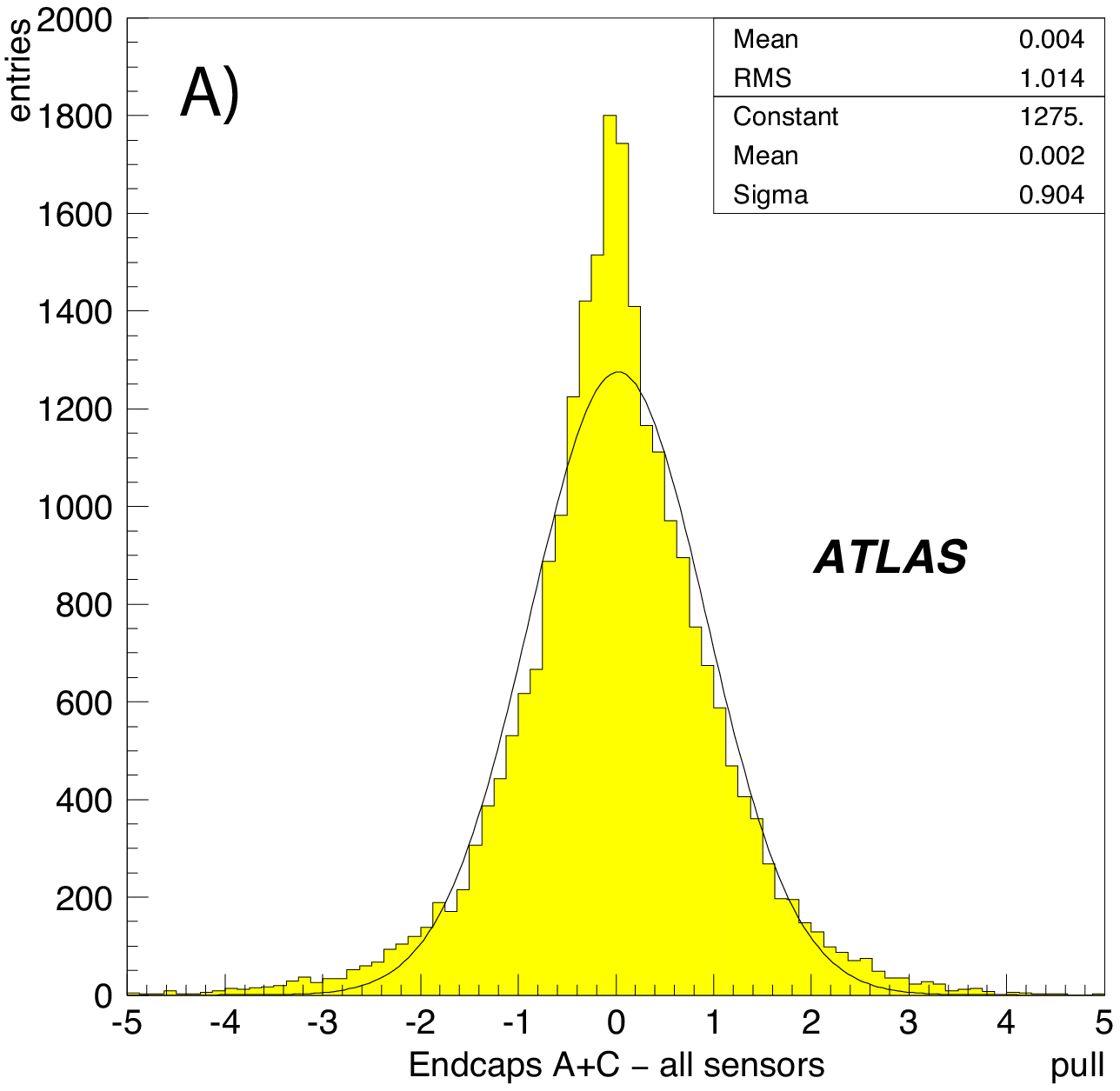} \hfill\includegraphics[height=7.cm]{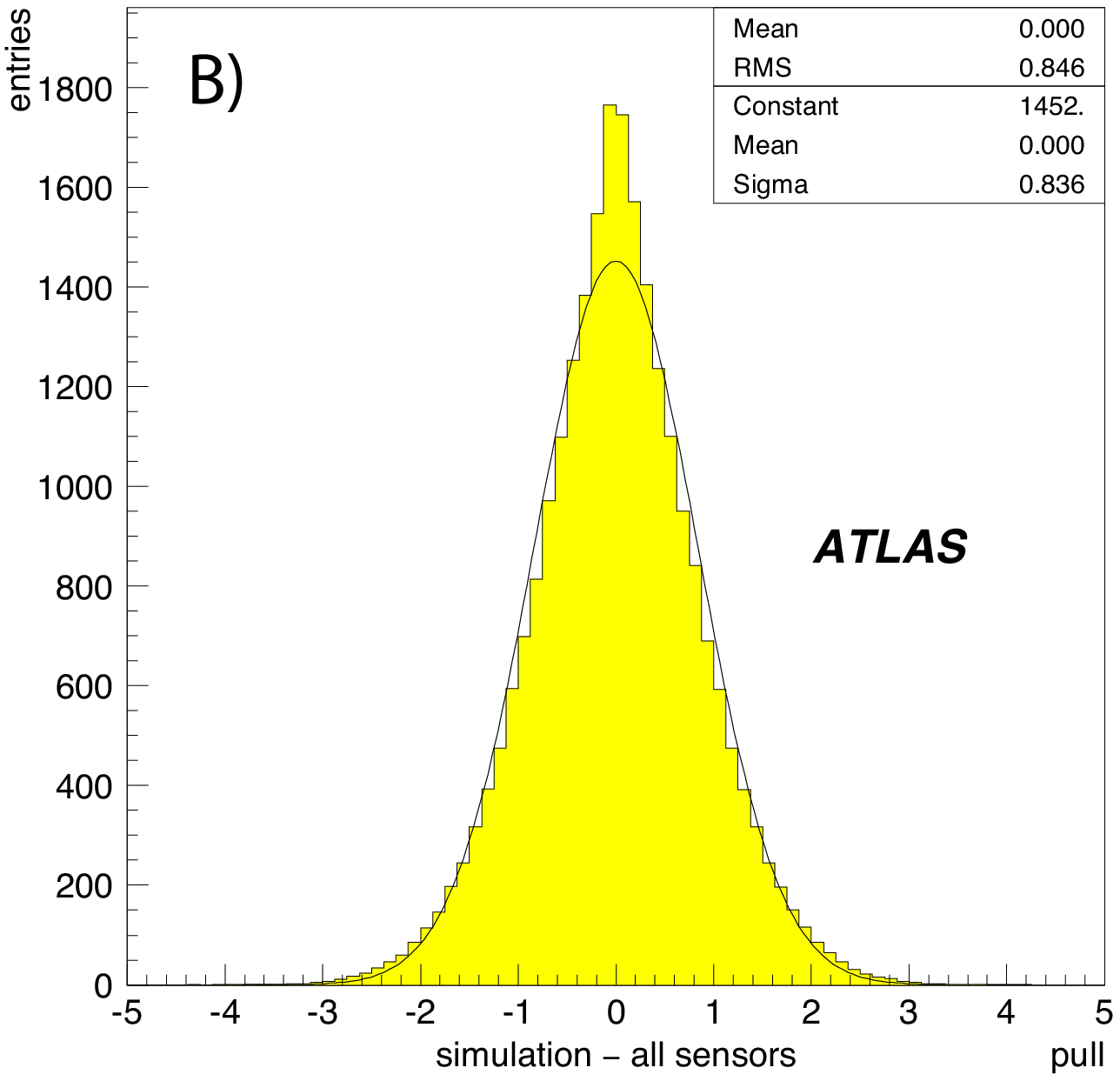}\hfill\
\caption{\label{sec71:fig1}The observed (A, from data) and expected
(B, from simulation) pull distributions for the end-caps, assuming
 design resolution for all sensor types. Correlations and weakly
constrained degrees of freedom cause the expected pull distribution to have
RMS width below unity. The observed $\chi^2/\mathrm{ndf}$ from the fit on data
is 1.4, while the one from simulation is 1.0.}}
\end{figure}

The position coordinates, rotation angles, and deformation
parameters of the chambers are determined by a global $\chi^2$
minimization procedure. The total $\chi^2$, as well as the
contributions of the individual sensor measurements to the $\chi^2$
(pulls) can be used to estimate the alignment quality from the
internal consistency of the fit. If the observed sensor
resolutions agree with the design values, one
expects approximately $\chi^2/\mathrm{ndf}=1$ and a pull
distribution with zero mean and unit RMS width.
Figure~\ref{sec71:fig1} shows the observed and expected pull
distributions in the end-caps, obtained by assuming the design
resolutions for all sensor types. \par
 In a second step, the assumed
sensor resolutions are adjusted until the observed pull
distributions, broken down by sensor type, agree with the expected
distribution. This yields the observed sensor resolutions, which are
used as input to a Monte Carlo simulation of the alignment system.
The simulation predicts a sagitta accuracy due to alignment of about $45\mum$, which is close to the design performance. \par
Validating the alignment as reconstructed from the optical sensor
measurements requires an external reference. During chamber
installation, surveys of the completed end-cap wheels were done
using photogrammetry, and the chamber positions measured with the
alignment system agreed with the survey results within 0.5 mm, the
quoted accuracy of the survey. While establishing confidence in the
optical system, the full validation of the alignment can only be
done with tracks. Thus, cosmic muons recorded during magnet-off
running were used to cross-check the alignment provided by the
optical system.

For a perfect alignment, the reconstructed sagitta of straight
tracks should be zero for each EI-EM-EO measurement tower (note
that, when averaged over many towers, the mean value can be
accidentally compatible with zero despite single towers being
significantly misaligned). For cosmic muons, the observed width of
the sagitta distribution is dominated by multiple scattering. A
shifted and/or broadened distribution would indicate imperfections
of the alignment. Triplets of track segments were selected in the
EI-EM-EO chambers, requiring the three segments to be in the same
sector and assigned to the same reconstructed track. Some segment
quality cuts were applied for this analysis: {\it i)} $\chi^2$/ndf
$<$ 10 and at most one expected hit missing per chamber; {\it ii)}
the angle between the segments and the straight line joining the
segments in EI and EO was required to be smaller than 5 (50) mrad
in the precision (second) coordinate; {\it iii)} at least one
trigger hit in the second coordinate was required to be associated
to the track. A total of 1700 segment triplets passing the cuts were
selected in run 91060.
%{\bf give the exact number of segments, 1700/3 gives a non integer number of tracks. Figure~\ref{sec71:fig2} does not show the number of entries}

\begin{figure}[!htb]
  \begin{center}
\includegraphics[height=6.cm]{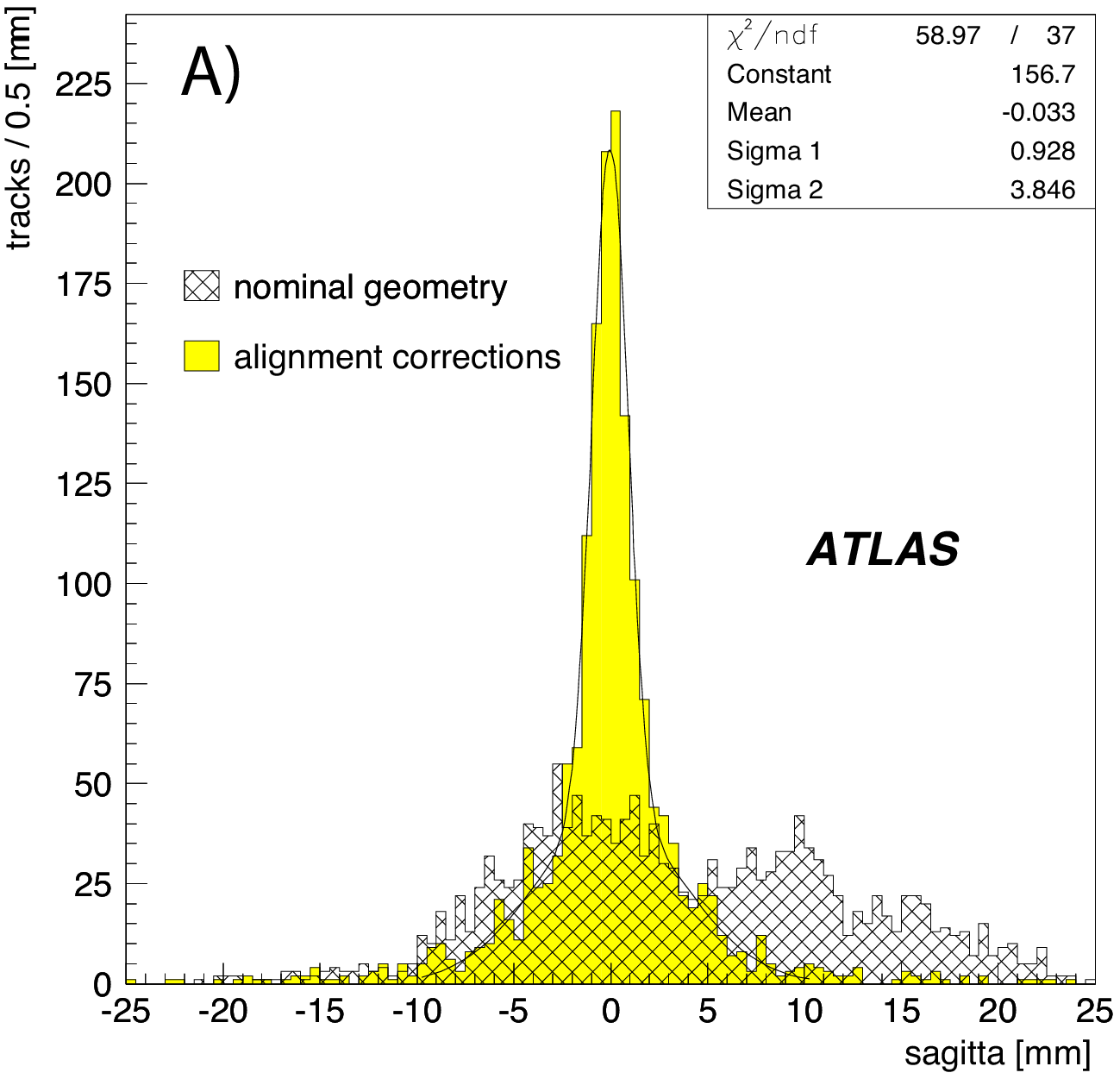}
\includegraphics[height=6.cm]{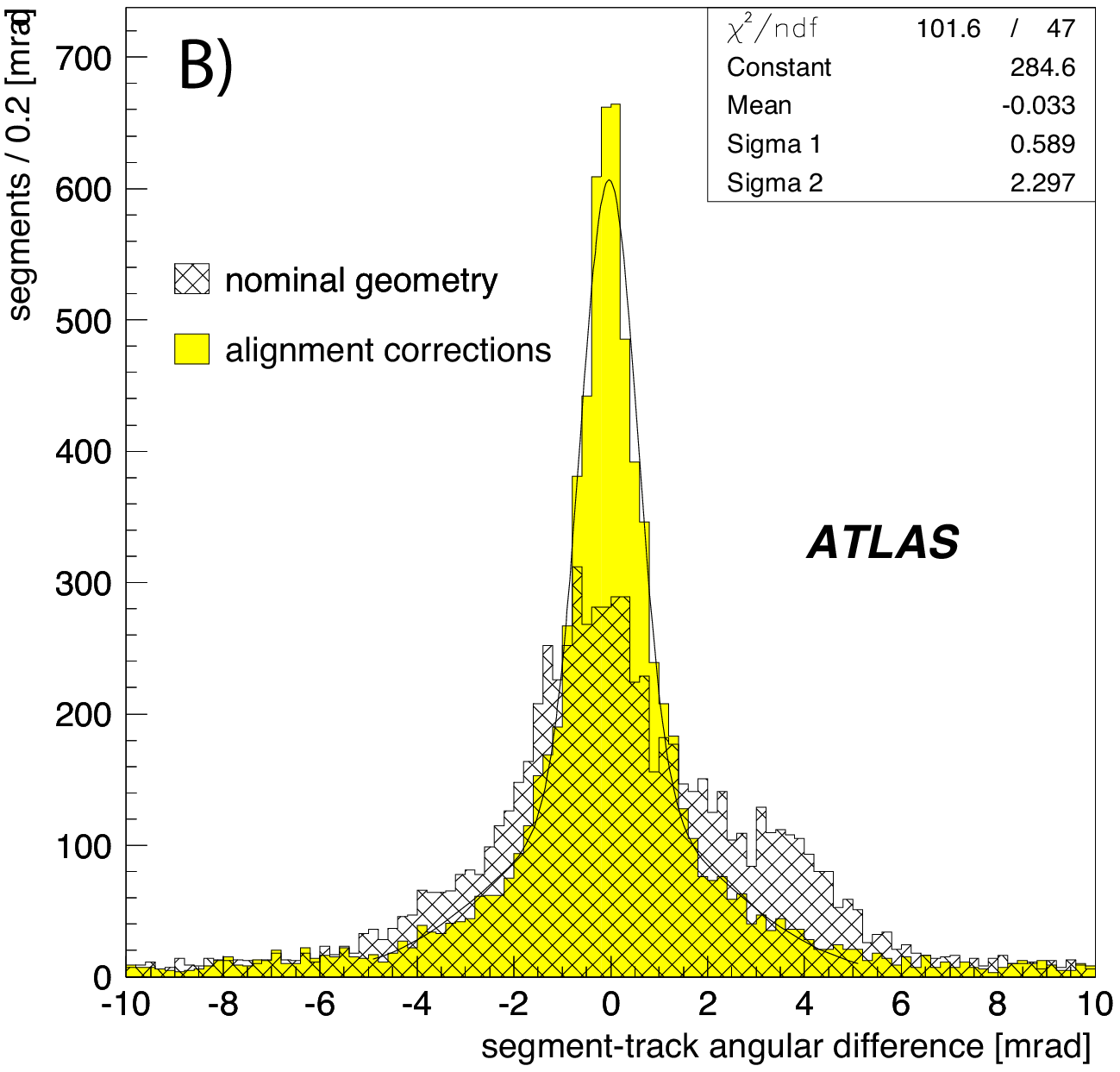}
 \end{center}
\caption{A): measured sagitta distribution for the
two end-caps. The cross-hatched histogram shows the sagitta before
alignment corrections, thus reflecting the accuracy of chamber
positioning. The filled histogram shows the
sagitta after applying alignment corrections, the curve is the fit of a double-Gaussian function, each Gaussian containing 50\% of the events.
B): measured angle
in the precision coordinate between the segments and the track
to which they are associated.}
\label{sec71:fig2}
\end{figure}

\begin{sloppypar}
Figure~\ref{sec71:fig2} left shows, for the two end-caps, the
observed sagitta distribution before and after applying alignment
corrections (i.e. the chamber positions, rotations, and deformations
as determined by the optical system, as well as a correction for the
gravitational sag of the MDT wires). Figure~\ref{sec71:fig2} right
shows the corresponding difference in angle in the precision
coordinate between each of the segments and the track (the straight
line joining the EI and EO segments). For the distribution on the
right, the cut at $5\mrad$ was omitted. The improvement in both
variables is clearly visible, the mean value of the corrected
sagitta distribution as obtained from the fit with a double-Gaussian
function is ($-33 \pm 42) \mum$
%(statistical error only){\bf no need to outline this: it is the mean value of a distribution},
and thus perfectly compatible with zero within the $45\mum$ error
estimated above from the internal consistency of the alignment fit.
The width of the corrected sagitta distribution agrees approximately
with expectations for the typical energies of triggered cosmic
muons. The width of the corrected angle distribution, on the other
hand, is about twice as large as expected. This is mainly a
consequence of the additional time jitter of MDT measurements
described in Section~\ref{Calibration} which deteriorates the
segment resolution.
\end{sloppypar}

For the two end-caps separately, the mean value of the sagitta
distribution is $(-30 \pm 61) \mum$ in side A and $(-37 \pm 57)\mum$
in side C. The sign of the sagitta is defined in such a way that
most of the conceivable systematic errors would cause deviations
from zero with the same sign in side A and  side C. The analysis
is limited by statistics even though it uses a significant fraction of the
full 2008 data sample. Breaking it down further to the level of
sectors, or even to projective towers (where the best sensitivity is
obtained) would require significantly more data.

The cross-check with straight tracks confirms that, 
with the limitations of the analysis,
the chamber positions given by the optical alignment system 
are within the estimated sagitta uncertainties, indicating that the optical 
system works properly. The design accuracy has nearly been reached in the
end-caps. It also shows that the system produces a reliable estimate
of the uncertainty of the alignment corrections.

\subsection{Barrel chamber alignment}
\label{BARRELalign}
%{\bf Editor: Igor, Florian, Pierre-Francois}\\

%\subsubsection{Alignment status}
%Also the Barrel of the MS is equipped with an optical system~\cite{barrel_optical_system} that was used to provide the 2008 cosmic
%data reconstruction with alignment corrections.

The installation and commissioning of the barrel optical system~\cite{BARRELALIGNMENT} 
began in 2005 and continued together with the installation of the chambers until 2008. 
At the time of recording cosmic ray data, the barrel optical system was fully installed and 
99.7\% of the sensors were functioning correctly. Table \ref{tab:installation_status1}
summarizes the status of the 5800 installed sensors. The complete
system is read out continuously, at a rate of one cycle every 20
minutes. The readout was functioning correctly during the complete
period of acquisition of cosmic ray data.

\begin{table}[!htb]
  \label{tab:installation_status}
  \centering
  \begin{tabular}{|lrrr|}
    %\toprule
    \hline
    Type & Total & Working & Broken \\
    %\midrule
    \hline
    Projective &  117 &  117 &   0 \\
    Axial      & 1036 & 1031 &   5 \\
    Praxial    & 2010 & 2008 &   2 \\
    Reference  &  256 &  253 &   3 \\
    CCC        &  260 &  260 &   0 \\
    BIR-BIM    &   32 &   32 &   0 \\
    Inplane    & 2110 & 2101 &   9 \\
    %\midrule
    \hline
    Total      & 5817 & 5798 &  19 \\
    \%         &      & 99.7 & 0.3 \\
    %\bottomrule
    \hline
  \end{tabular}
  \caption{Status of the barrel optical system in Fall 2008. No data were recorded during
  this period from the ``broken'' sensors. Naming and functions of the different sensors are
  detailed in reference~\cite{BARRELALIGNMENT}.}
  \label{tab:installation_status1}
\end{table}

The alignment reconstruction consists in determining the chamber
positions and orientations (referred to as ``alignment
corrections'') from the optical sensor measurements. This requires
the precise knowledge of the positions of the sensors with respect
to the MDT wires. To this purpose, the optical sensors were
calibrated before installation and their mechanical supports were glued with precise
tools onto the MDT tubes. However, the original design of the barrel
optical system suffered from a few errors that eventually degraded
the precision of the alignment corrections. Furthermore, the only
devices giving projective information in the Small sectors are the
CCC sensors which are designed to provide 1 mm accuracy. 
The alignment of the chambers of the Small sectors is, by design, 
based on tracks that cross the overlap region between the Small and 
the Large sectors. However, the statistics obtained in cosmic runs was not sufficient to perform 
a precise check of this method.

The alignment corrections discussed here cover the nine upper
sectors (1 to 9). The complete period of cosmic data taking was
divided in intervals of 6 hours, and alignment corrections were
reconstructed using the sensor measurements recorded in that
interval. This provided data for monitoring of significant movements of the
MS, e.g. when the magnetic field in the toroids was switched on. 
%%\paragraph{The alignment reconstruction principle}

The barrel alignment reconstruction is based on the minimization of
a $\chi^2$, whose inputs are, for each optical sensor $i$:
\begin{itemize}
\item the recorded response $\mathbf{r}_i$;
\item a model $\mathbf{m}_i(\mathbf{a})$, representing the predicted response of sensor
$i$ with respect to the alignment corrections $\mathbf{a}$;
\item the error $\mathbf{\sigma}_i$, the estimated uncertainty of the model $\mathbf{m}_i$ .
\end{itemize}
The critical part is the model $\mathbf{m}_i$, as it combines all
the knowledge of the precise geometry of the optical sensors and
their calibration. The free parameters in the fit are the alignment
corrections $\mathbf{a}$, and in some cases additional parameters
used to model the effect of imprecise sensor positioning or of an
incorrect calibration. For all these additional parameters
appropriate constraints are included in the fit  reflecting the best estimates of the error contributions mentioned above. Overall, 4099
parameters are fit simultaneously.
%The fit technique is based on a linear least square, optimized through the use of the sparse matrix library from ROOT \cite{root}.
The total reconstruction time for the full barrel is less than one
minute.

Given the uncertainties introduced by the additional parameters in
the fit procedure, the strategy for alignment in the barrel is
slightly different from the one  in the end-cap. Dedicated runs without
magnetic field in the toroids (but  {\it with}  field in the 
solenoid to tag high momentum tracks) will be used to get initial
alignment corrections with a precision of 30$\mum$. The optical
alignment system is then used to monitor movements due to the
switching on of the toroidal field and to temperature effects. 
The mechanical stability of the system, in periods where the magnetic field was constant, is at the level of  100 $\mum$, while movements of the magnet structures at the level of few mm were observed when the magnets were switched on and off. The optical alignment system, which continuously monitors the position of the chambers,  is able to follow these movements with the required accuracy. This so-called {\it relative  alignment mode} has already been tested with success
in the MS system-test done with a high-energy muon
beam~\cite{h8a,h8b}. After the minimization, the value obtained for
$\chi^2$/ndf is $1.9$, which shows that the sensor errors are
% tml - "slightly" ??
 underestimated.

\subsubsection{Performance of the optical alignment in the barrel}
Similarly to what is done in the end-cap, an estimate of
the contribution to the sagitta error due to the alignment system may be inferred from the
$\chi^2$, using the following formula
\begin{equation}
  (V^{-1})_{ij} = \frac{1}{2} \frac{\partial^2 \chi^2}{\partial\theta_i\partial\theta_j} \Big |_{\hat{\theta}}
  \nonumber
\end{equation}
\begin{sloppypar}
where $\theta_i$ are the fitted parameters and $V$ is the global
error matrix, of size 4099$\times$4099, of all fitted parameters. To
estimate the performance of the alignment system in terms of sagitta
measurement, straight tracks, originating in the IP and crossing
three layers of chambers, were simulated and the whole fit procedure
was applied to these tracks. The sagitta of these pseudo-tracks is a
function of some of the alignment corrections, and thus the formula
of error propagation may be used to infer the contribution of the
alignment to the error of the resulting sagitta. This technique
relies on the hypothesis that the errors of the optical sensors are
correctly estimated, and thus that the $\chi^2$ is correctly
normalized. As this is not the case ($\chi^2$/ndf = 1.9), the
results are only considered as a rough estimate of the optical
alignment performance.
%{\bf But can we exclude that the actual situation is not worse than what shown in the Figure ?}
\end{sloppypar}

\begin{figure}[!htb]
  \centering
    \includegraphics[width=\columnwidth]{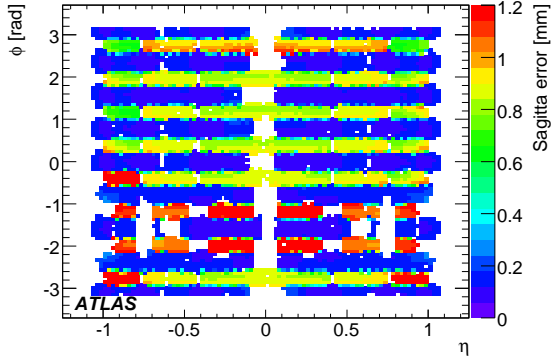}
  \caption{$\eta \times \phi$ map of the contribution to the sagitta error  due
  to  alignment, as  estimated with the method described in the text. As expected from the
  system design, the
    Small sectors  (even sector numbers) are aligned with  significantly less precision than the Large sectors  (odd sector numbers). }
  \label{fig:sag_error}
\end{figure}

The result is shown in Figure~\ref{fig:sag_error}. The Small sectors
have a significantly worse alignment than the Large sectors, as
explained above. Conservatively, one can conclude that the
performance of the optical system, in terms of sagitta precision, is
$\sim$200$\mum$ for the Large sectors, and $\sim$1~mm for the Small
sectors.

\subsubsection{Alignment with straight tracks}
Data with the toroidal field off were used to improve the alignment
precision in the barrel and to validate the alignment corrections in
relative mode. The method is to use straight muon tracks to
determine in absolute mode the {\it initial} spectrometer geometry
and, once this geometry is determined, to use the optical alignment
system to trace all chamber displacements in a relative mode.
%To align the barrel part of the muon spectrometer with straight tracks an alignment algorithm has been developed.
%~\cite{AlAlg}.
The alignment procedure with straight tracks is based on the
so-called {\sc MILLEPEDE} fitting method~\cite{MILLE}. This method
uses both alignment and track parameters inside a global fit. As a
result, all correlations between alignment and track parameters are
taken into account and the alignment algorithm is unbiased.

The track alignment algorithm has been tested with Monte Carlo
simulations and with cosmic ray data. The simulation studies show
that $10^5$ muon tracks with a momentum greater than 20~GeV and
pointing to the IP are needed to align the Large sectors with a
precision of $30\mum$. Small sectors require five times more tracks
than Large sectors, due to the multiple scattering in the toroid
coils.

Using straight cosmic muon tracks recorded in run 91060, a set of
alignment constants has been produced. A total of $10^7$ events were
used corresponding to about $3\times 10^5$ cosmic muon tracks in
each of the most illuminated barrel sectors. The statistical
uncertainty of the sagitta using this track alignment procedure was
estimated to be $30\mum$ for Large sectors.

The data of run 91060 were processed with the track reconstruction
software twice: {\it i)} using the optical alignment corrections and
{\it ii)} using the track-based alignment corrections. Both
geometries were then tested by measuring the distribution of the
track sagitta for muons crossing three chamber stations (Inner,
Middle and Outer). Only tracks passing close to the IP in the $\eta$
projection were chosen. Hits in the Inner and Outer chambers were
fit to a straight line, and the distribution of the hit residuals in
the Middle chambers was used to evaluate the sagitta. For  perfect
alignment, the mean value of the sagitta should be zero for straight
tracks and, to a good approximation, the mean value of the
distribution gives an estimate of the sagitta error.

\begin{figure*}[!thb]
  \begin{center}
      \includegraphics[width=7cm]{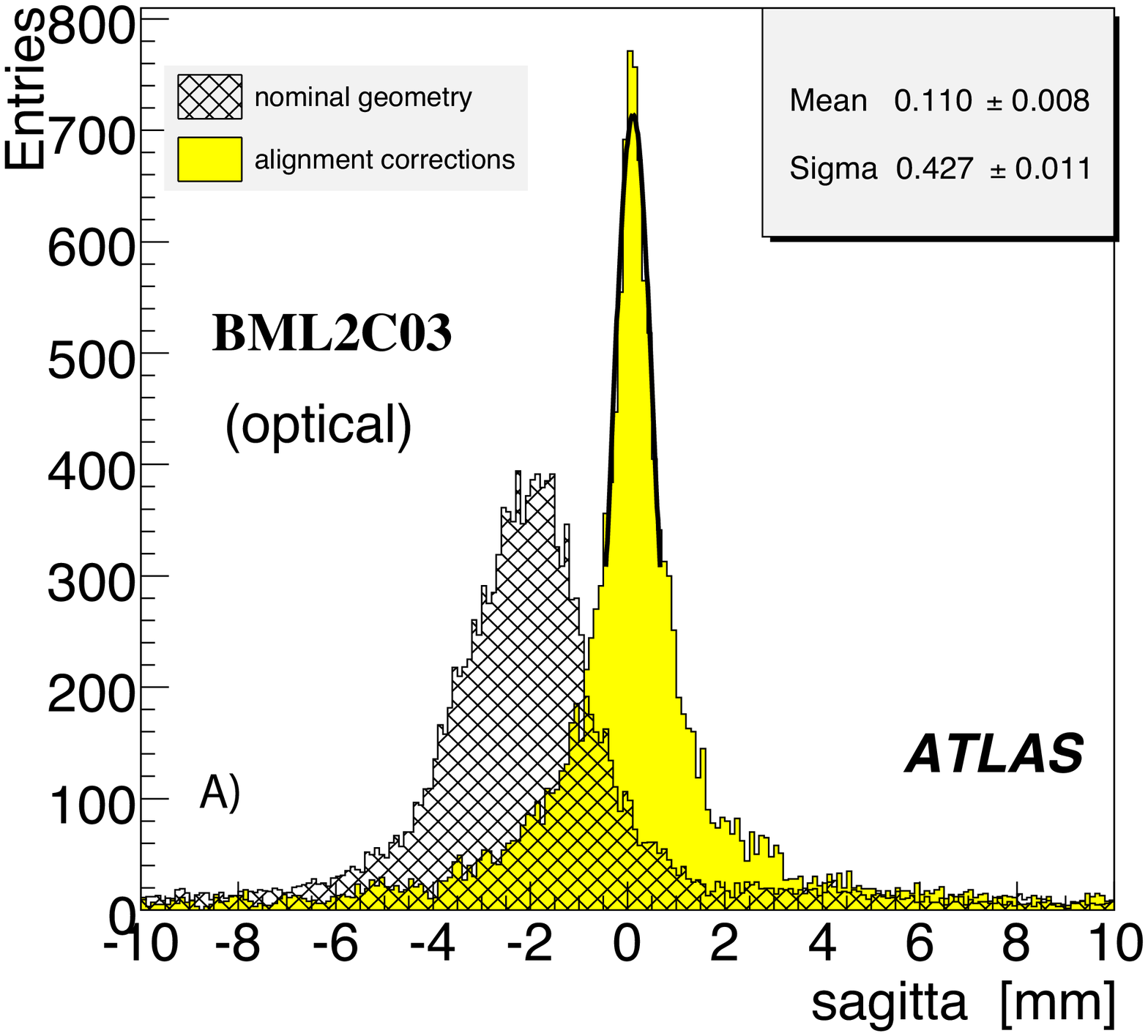}\includegraphics[width=7cm]{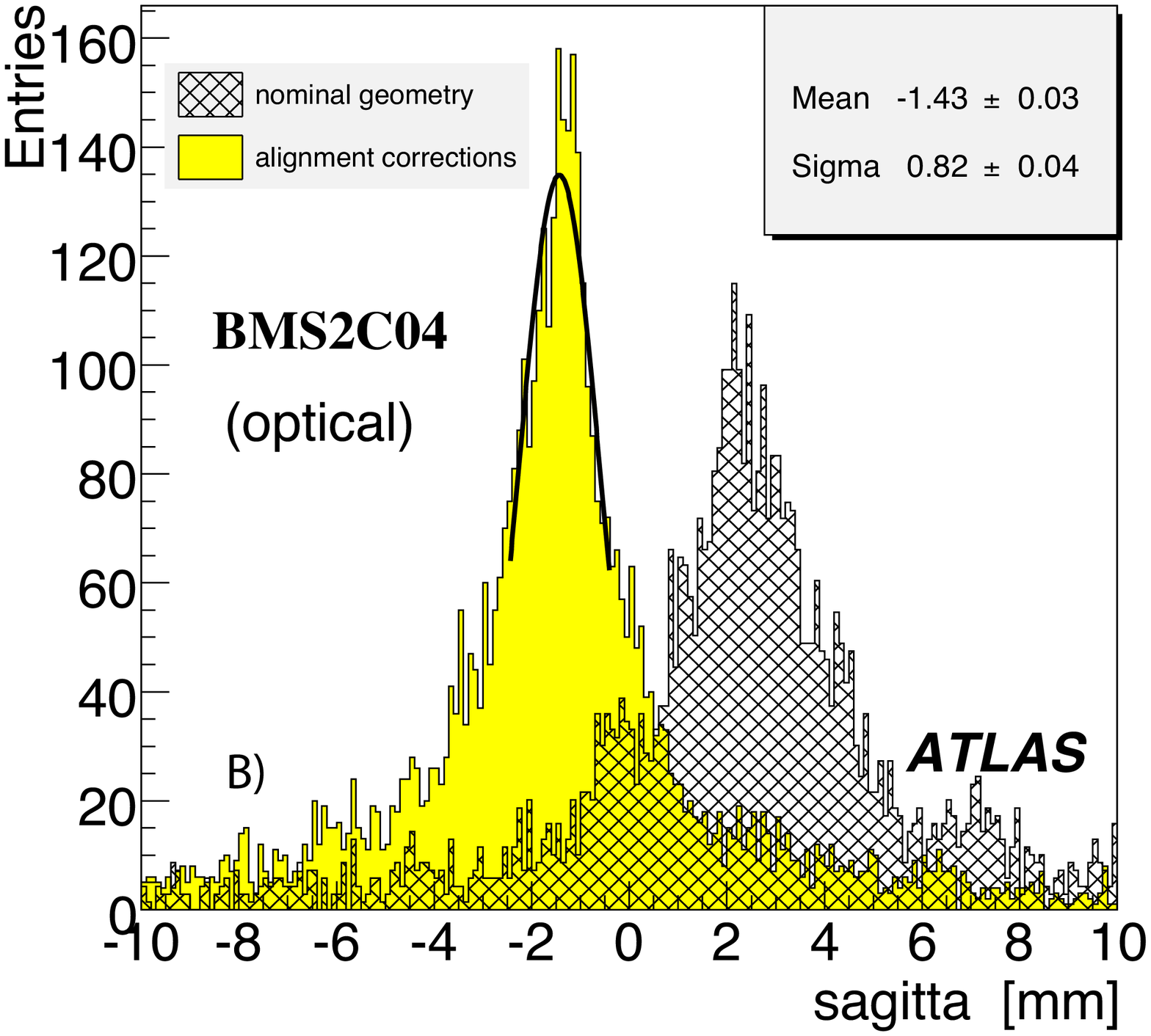}
    \includegraphics[width=7cm]{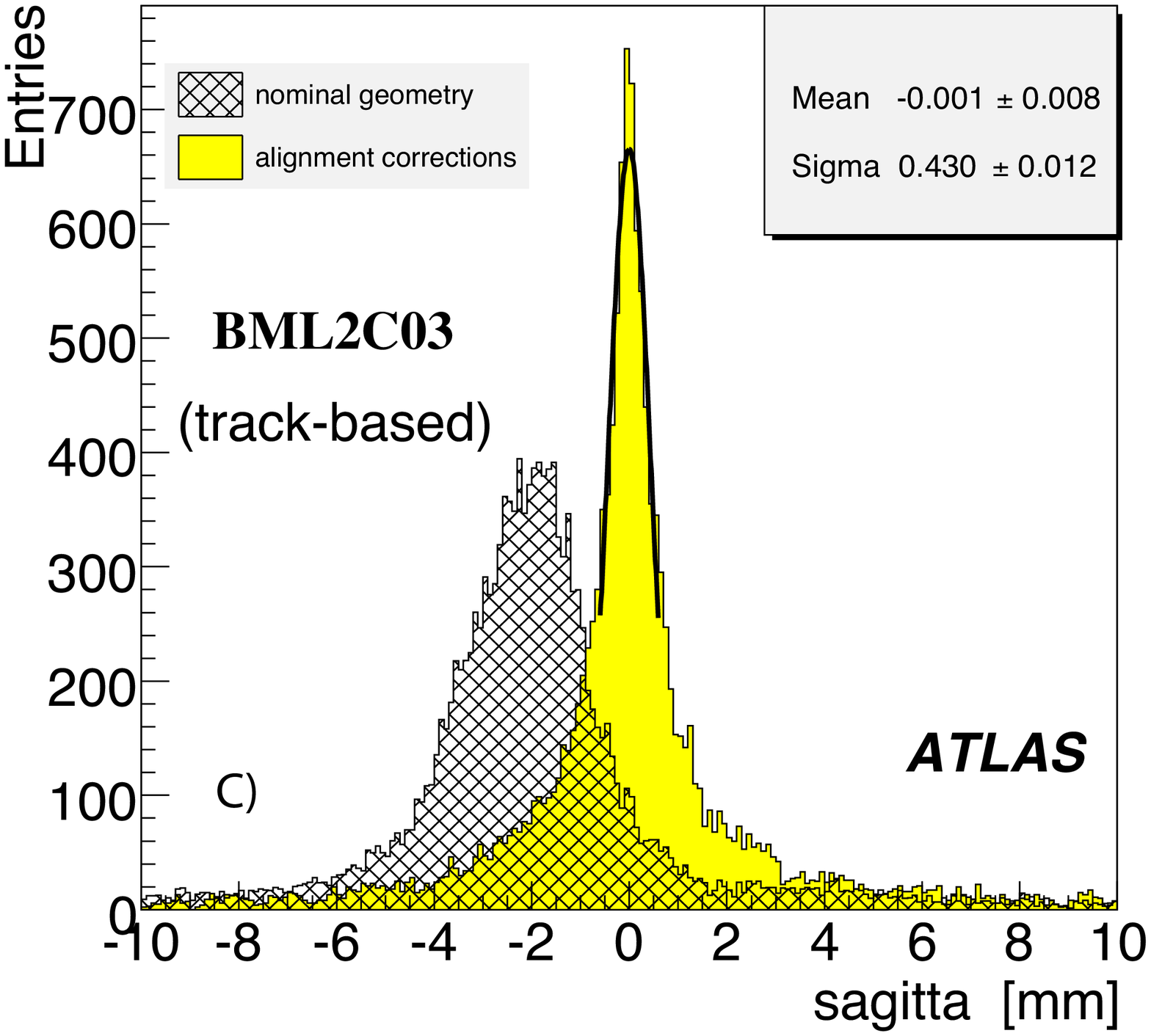}\includegraphics[width=7cm]{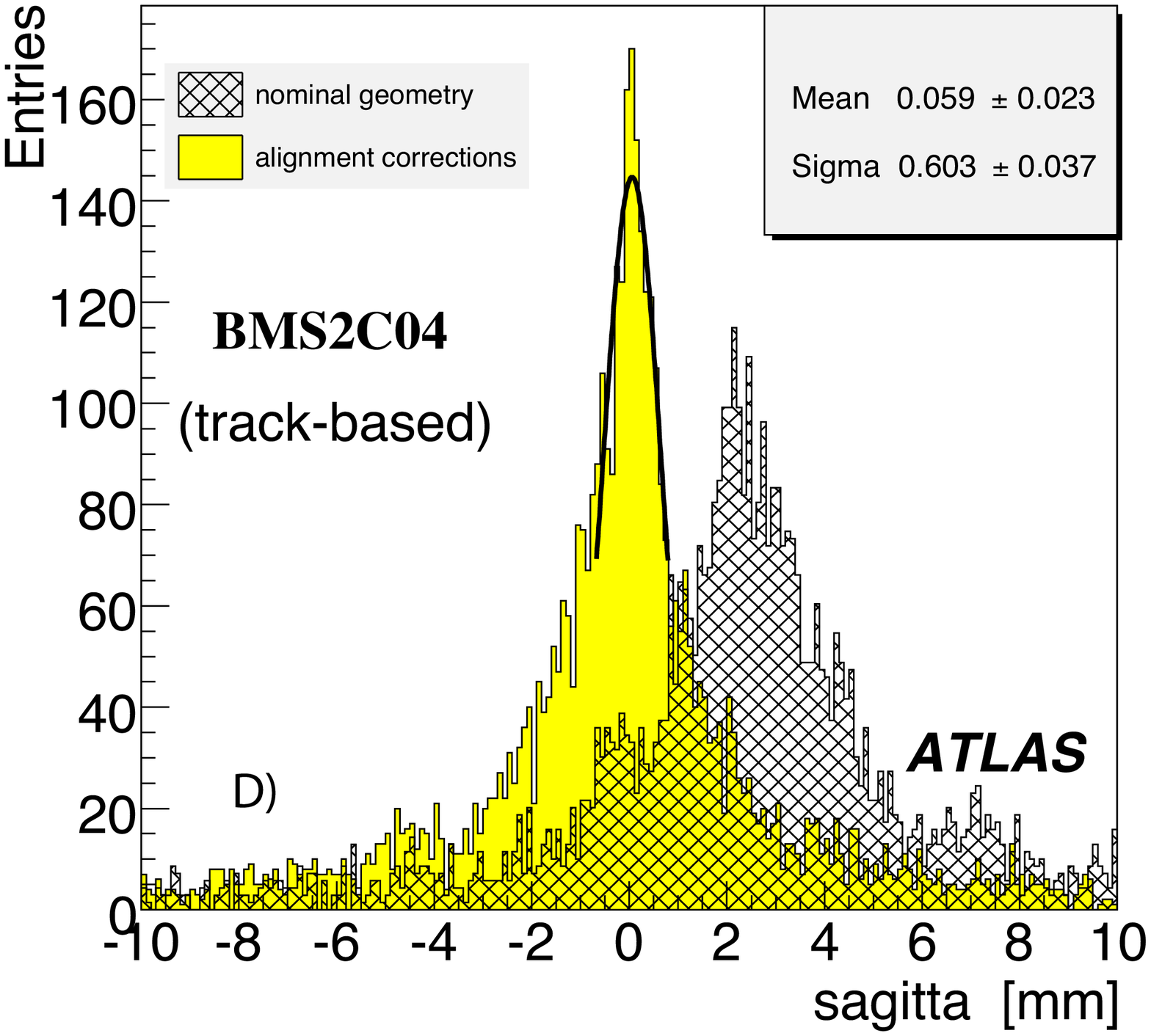}
    \caption{Distribution of the sagitta (as defined in the text) for straight tracks.
    A), B): using alignment
    corrections derived from the optical system only; C),D): using track-based alignment corrections.
    A), C): for a station in a Large barrel sector; B), D): for a station in a Small barrel sector, the optical system corrections of the small sectors have, by design, an accuracy at a level of 1 mm.
    In all panels,
    the hashed distribution is obtained using the ``nominal'' geometry. Mean and sigma in the statistical boxes refer to the distributions with alignment corrections, the peak is fitted with a gaussian in a  $\pm 1$ sigma interval}
    \label{fig:sagitta_distr}
  \end{center}
\end{figure*}

\begin{figure*}[!thb]
  \begin{center}
\includegraphics[width=7cm]{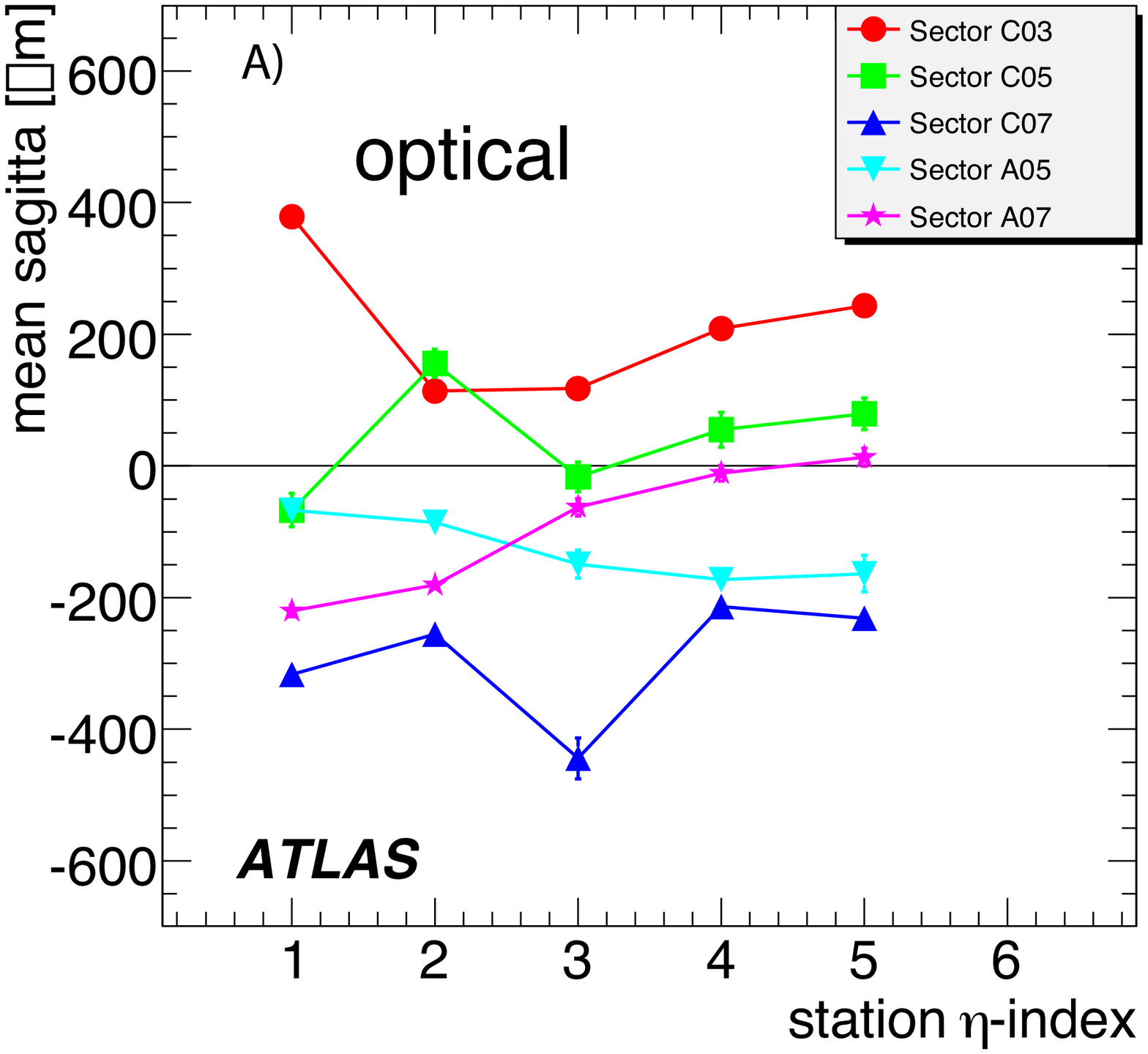}\includegraphics[width=7cm]{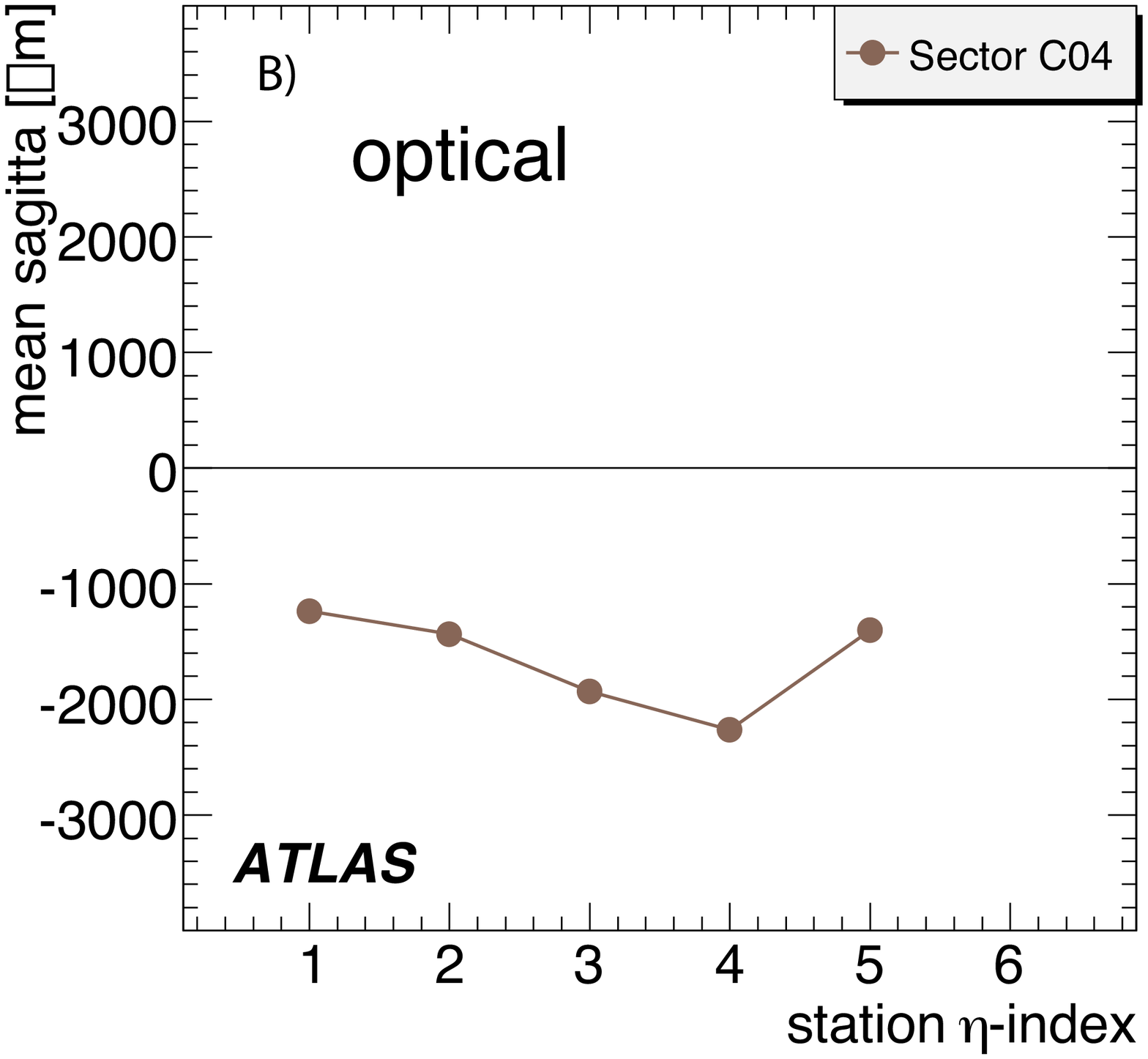}
\includegraphics[width=7cm]{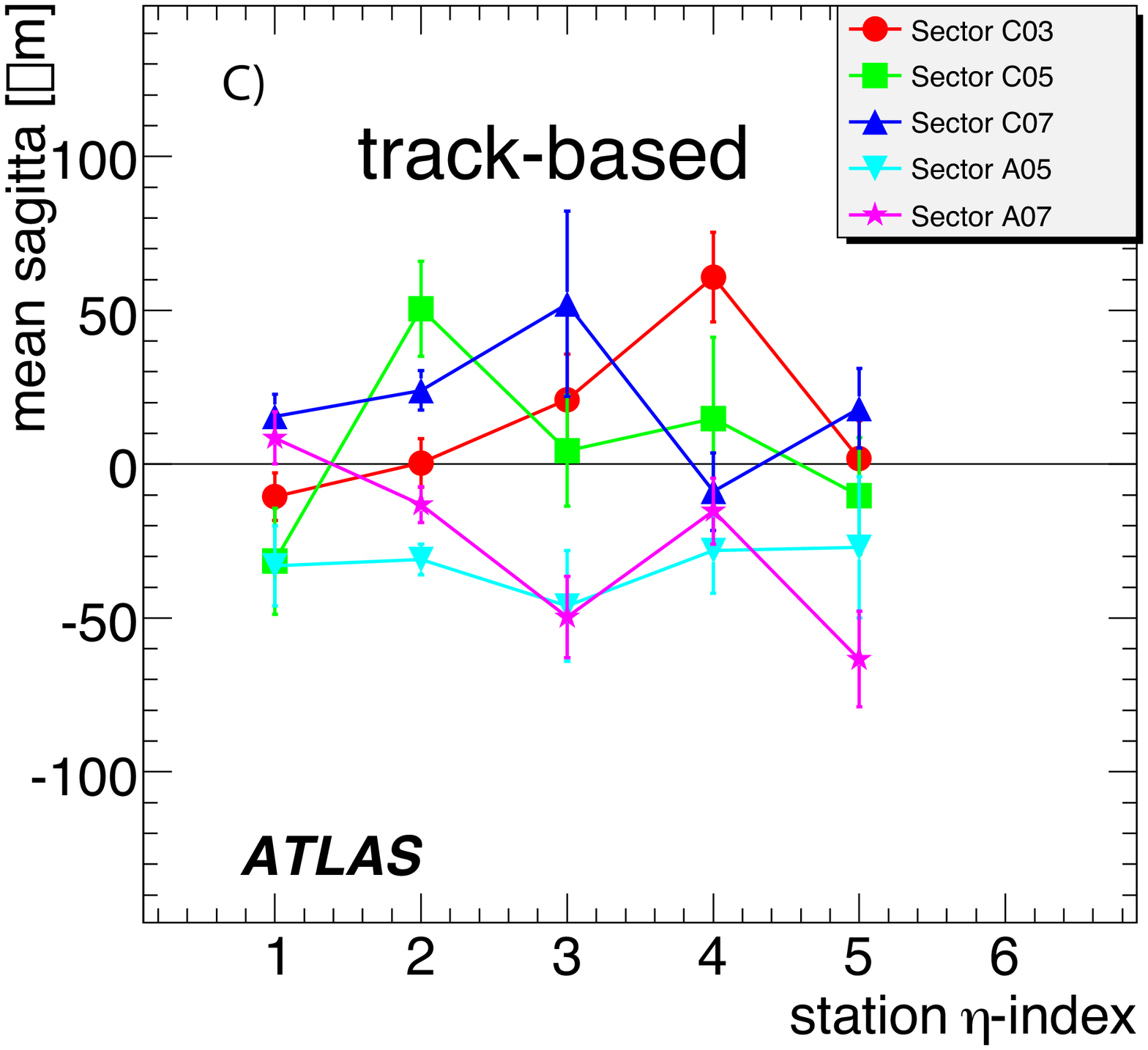}\includegraphics[width=7cm]{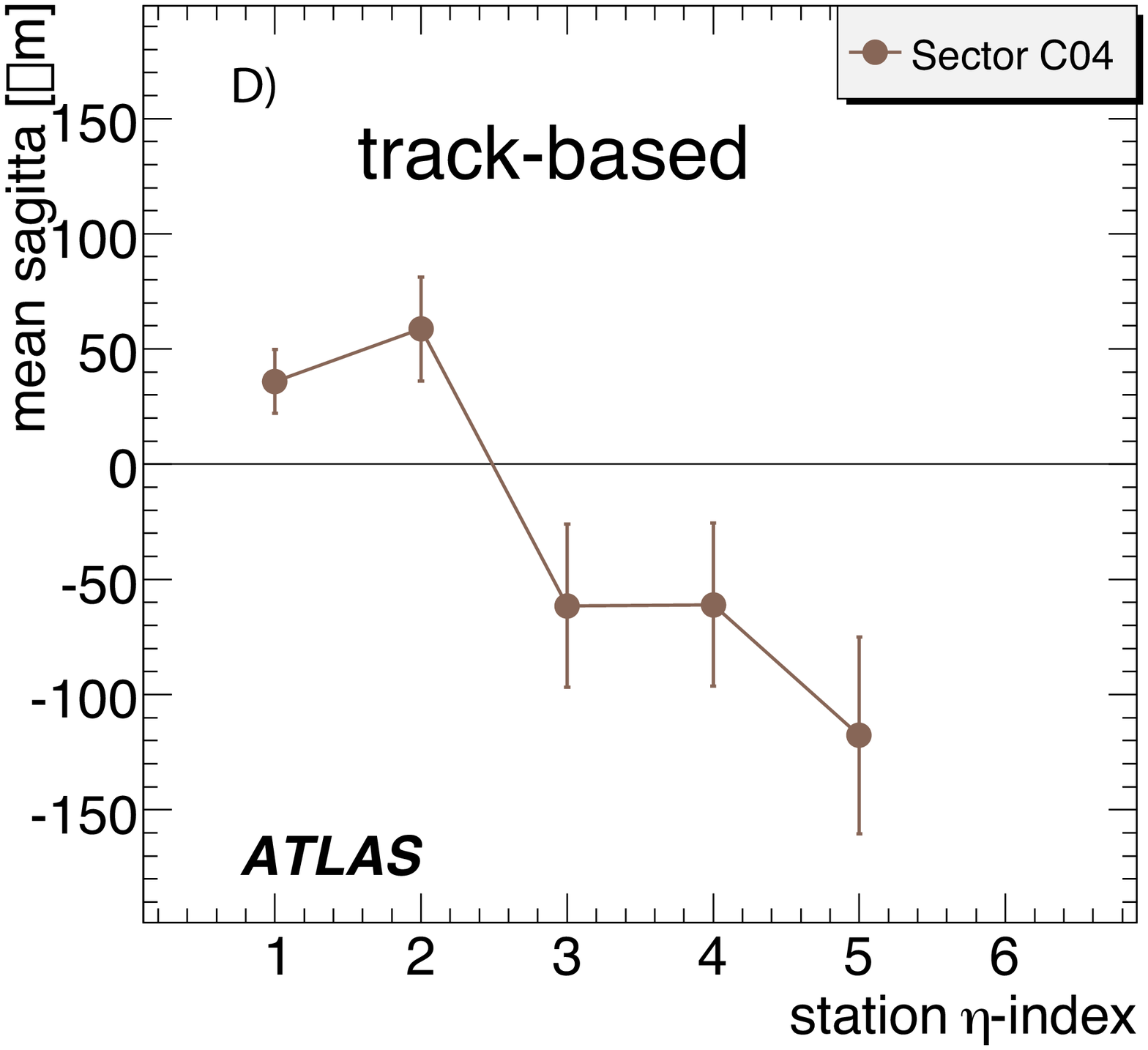}
    \caption{Mean value of the track sagitta distributions obtained A), B): with the optical alignment
     system only, C), D): and using the track-based alignment . A), C): for the upper Large
     barrel sectors.
     B), D):: for a Small barrel sector with $56^{\circ} < \phi < 79^{\circ}$. }
   \label{fig:sagitta_graph}
  \end{center}
\end{figure*}

The results are shown in Figure~\ref{fig:sagitta_distr} for the sets of
alignment corrections; on the left for a station in a Large sector,
on the right for a station in a Small sector. For reference, the
distributions using the design 
geometry are also shown. The tails of the distributions are due to
multiple scattering of muons. In the Large sector station, the two
distributions are almost identical, but the distribution with
optical alignment is centered at $\sim$100$\mum$. In the Small
sector station, the distribution with the optical alignment is
centered around 1~mm. To compare the results obtained in different
stations, Figure~\ref{fig:sagitta_graph} shows the mean values of
the sagitta distribution for the Large upper sectors (3, 5 and 7).
One Small sector is also presented, sector 4, since this was
illuminated with enough events during the same run to produce a
meaningful distribution.

\begin{sloppypar}
The results show that the optical alignment system alone provides a
precision at the level of 200$\mum$. When calibrated with sufficient statistics of high momentum straight
tracks, the optical system is able reach a
precision of 50$\mum$.
\end{sloppypar}

\begin{figure}[!htb]
  \begin{center}
\includegraphics[width=\columnwidth]{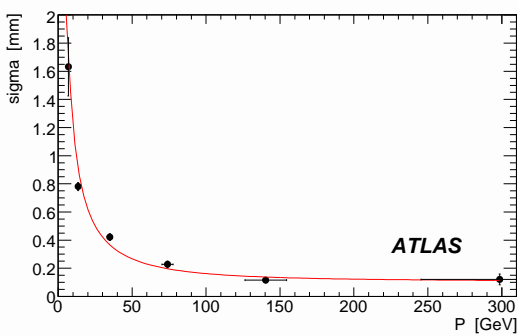}
    \caption{RMS value of the sagitta distribution in sector 5 as a function of the muon momentum measured 
    by the ID. The fit to the function described in the text is superimposed.}  
  \label{sagitta5C}
  \end{center}
\end{figure}

%%%%%%%%%%% New Part on Sagitta resolution vs Pt added by Fabio 4 December 2009 %%%%%%%%
The sagitta resolution for runs with no magnetic field in the MS 
can be studied as a function of the muon momentum
measured by the ID. The sagitta resolution as a
function of the muon momentum was parameterized as
\begin{equation}
  \sigma_s(p) = \frac{K_{0}}{p} \ \oplus \ K_{1}
  \nonumber
\end{equation}
\begin{sloppypar}
% in teRMS of two contributions: 
where the first term $K_{0}$ is due to 
multiple scattering in the material of the MS, 
%the $K_{0}$ term, 
and the second term $K_{1}$ is due to the single tube resolution and chamber--to--chamber alignment.
%the $K_{1}$ term. 
These two terms have been already measured at the
MS system test beam~\cite{h8a,h8b} and found to be $K_{0} =
9$ mm$\times$GeV and $K_{1} = 50 \mum$. A similar measurement was done with 
cosmic muons by selecting segment triplets (Inner,
Middle and Outer station) of MS projective towers. The RMS of the
sagitta of the Middle station segment with respect to the
Outer--Inner straight line extrapolation has been fit in five
momentum bins. 
%(where $p$ is measured by the ID). 
The result is shown in Figure~\ref{sagitta5C} for sector 5 (Large sector) 
%side C only It has been obtained on run 113860 and 
with RPC-time corrections
%, as described in Section~\ref{Calibration} have been 
applied in the calibration procedure. The fitted value for the two terms is $K_{0}
= (12.2 \pm 0.7)$ mm$\times$GeV and $K_{1} = (107 \pm 21)\mum$.
In the MS the multiple scattering term is expected to be
worse than the one measured at the test beam setup
%. Moreover this term is expected to be much 
and larger for Small sectors due to the
presence of the toroid coils between the Inner and Outer chambers.
%Concerning the intrinsic term the value for this specific sector is
The value of $K_1$ measured with cosmic muons in sector 5 is
only about a factor two worse than that measured at the test beam. Several
effects contribute to this, including alignment, chamber
deformations, calibration and single tube resolution. Similar
studies performed for other sectors show worse results 
due to the smaller data sample available for alignment and calibration. 
%This partially explains the larger intrinsic resolution in muon momentum discussed in
%Section~\ref{Tracks}.
\end{sloppypar}

These preliminary studies with cosmic rays indicate that the method of track-based alignment is robust and with sufficient muon data from collisions the design alignment precision will be achieved. 

%%%%%%%%%%%%% End Alignment Section %%%%%%%%%%%%%%%%%%

%tml - Why isn't this section up in Section 2 with the Moore and Muonboy
% descriptions, or those here?
\section{Pattern recognition and segment reconstruction}
\label{Segments}
%{\bf Editor: Peter - 2 pages}\\
%{\it Co-editors: Egge, Ahmimed, Eve, ...}\\
%====================================
% {\small {\bf replace this sentence} The MS hits are first associated into local segments before a full track fit is performed.
% In the Moore algorithm segments are reconstructed using patterns as a seed.
% Patterns are collection of hits selected by applying a straight line Hough transform to the
% MDT, CSC, RPC and TGC position measurements. {\bf with the following first line}} \vskip .4cm
%====================================
The pattern recognition algorithm first groups hits close in space
and time for each detector. Each pattern is characterized by a
position and a direction and contains all the associated hits.
Starting from these patterns, the segments are reconstructed with a
straight line fit. The G$t_{0}$--refit 
%described in Sections~\ref{Samples} 
is applied at this stage and,
%tml - PLEASE!  Don't tell us again about the t0 refit!
%In the G$t_{0}$--refit, the segment parameters (position and direction) are
%determined simultaneously to the segment $t_{0}$. I
if the G$t_{0}$--refit procedure does not converge, the segment parameters
are computed with the tube $t_{0}$ provided by the calibration with tube resolution increased to 2 mm. After this, a drift
radius is assigned to each tube with an uncertainty of 2 mm
(independent of the drift radius value) in order to keep high track
reconstruction efficiency even in the case where no precise
alignment constants are available. The minimum number of hits per
segment was set to 3 and no cuts were applied on the number of
missed hits.

These relaxed requirements tend to increase the number of {\it fake} segments 
while keeping a high segment efficiency. Since cosmic ray events are quite clean and have
low hit multiplicity this fake rate increase is not considered as a
problem. On the other hand, a high reconstruction efficiency allows
the use of segments to spot hardware problems in individual chambers or in calibration or decoding software. 
Most of the fake segments are rejected at the track reconstruction level.

\begin{figure}[htb]
\begin{center}
\includegraphics[width=7cm]{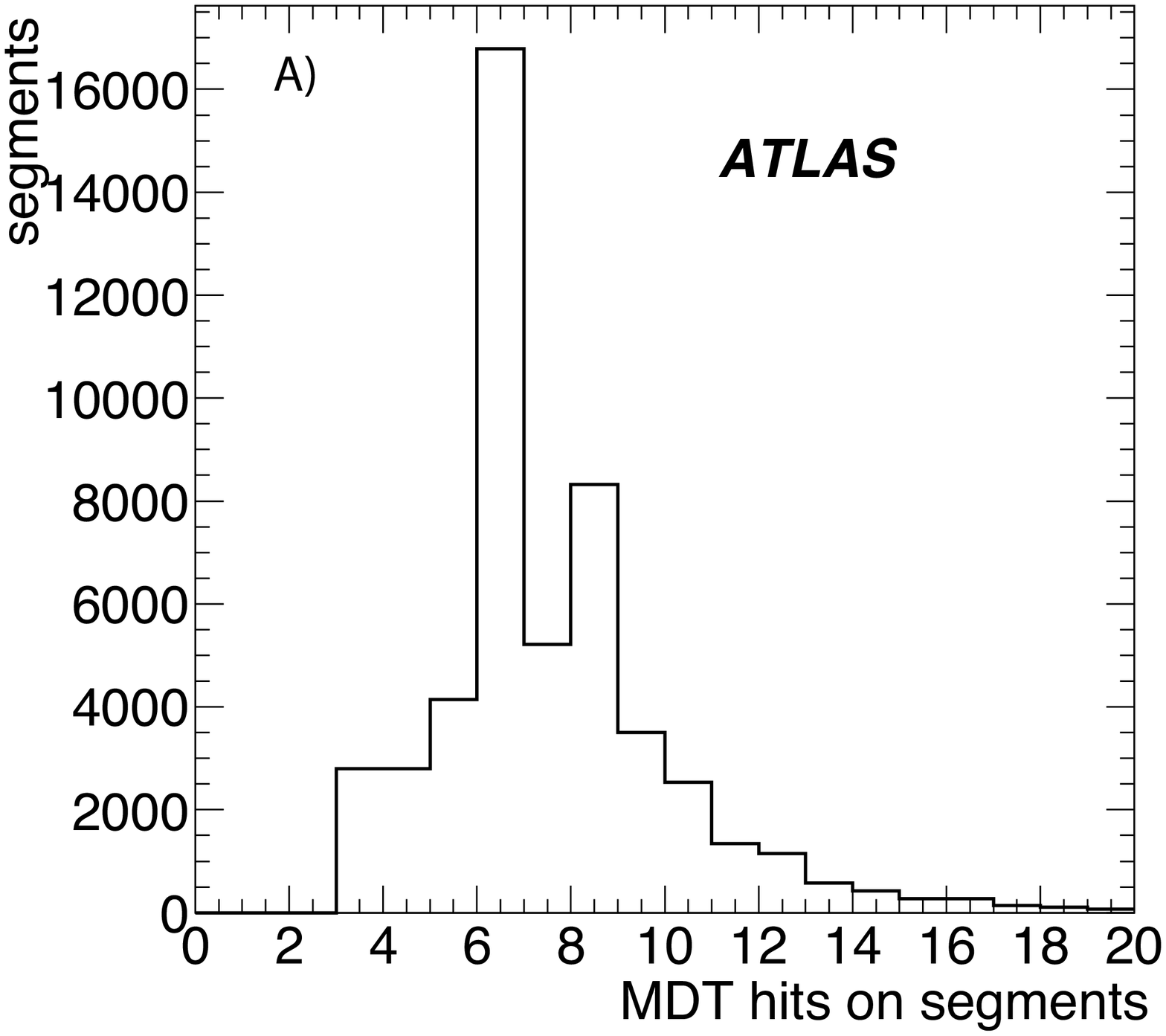}
\includegraphics[width=7cm]{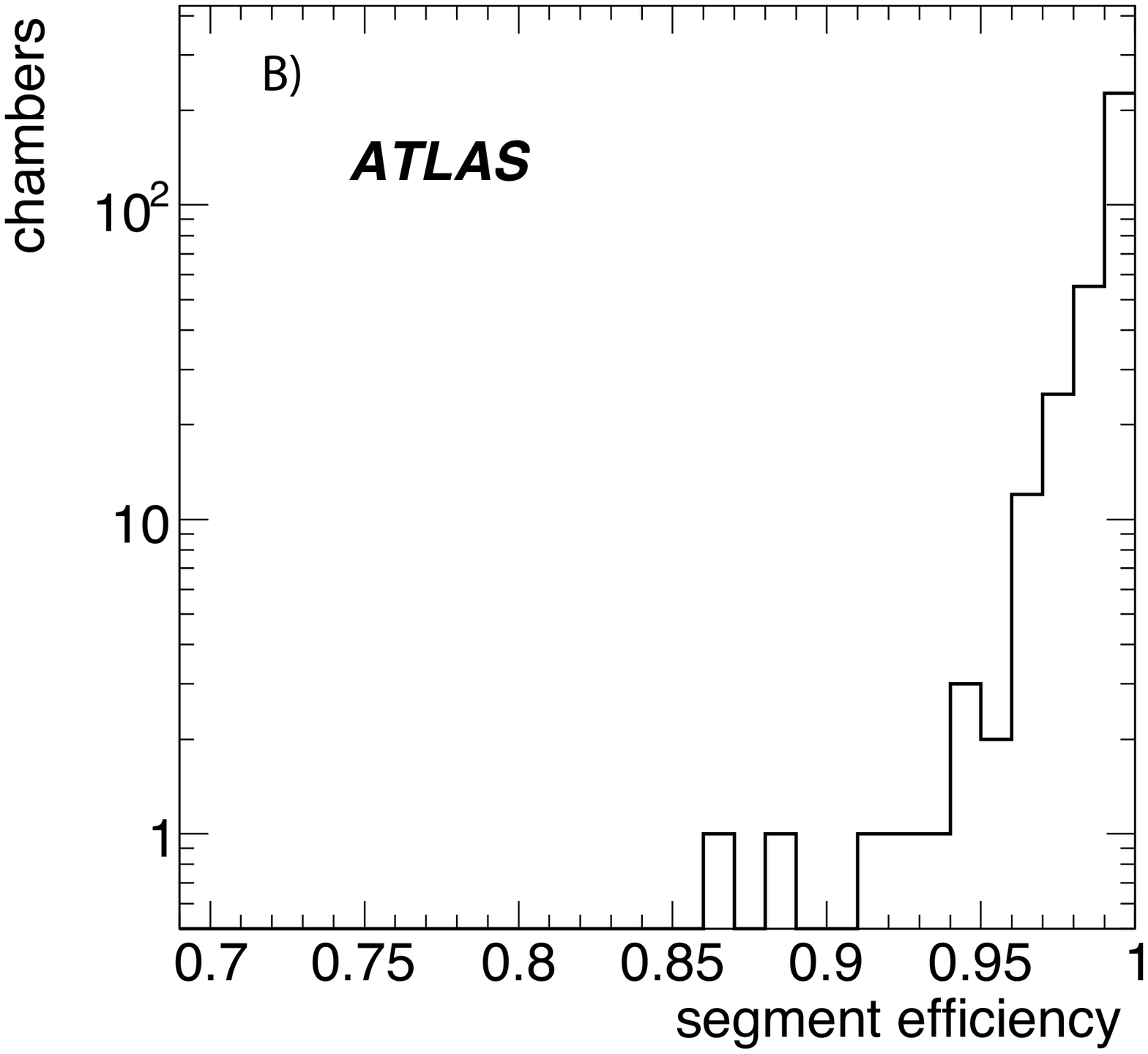}
\end{center}
\caption{A): distribution of the number of MDT hits per segment for segments
associated to a track. B): segment reconstruction efficiency for 322 MDT chambers. }
\label{mdtSegmentHits}
\end{figure}

Figure~\ref{mdtSegmentHits} shows, on the left, the number of MDT
hits per segment for segments associated to a track. In the
distribution clear peaks are observed at 6 and 8 hits
corresponding to the 6-layer (Middle and Outer) and 8-layer (Inner)
chambers.

\begin{sloppypar}
%tml - Is this procedure done in 2-D in the r-phi view? The straight
%line extrapolation would seem not to work otherwise (unless it was
%done with toroids off).
The efficiency of the segment reconstruction in run 91060 
was determined in the following way. First, cosmic shower events are
suppressed by requiring less than 20 segments in the event. 
%Then pairs of segments in two MDT layers of each station (Inner, Medium or
%Outer) are considered. Segments in each pair of stations are fitted
%to a straight line and the line is extrapolated to the third station.
Then a pair of segments in two MDT stations (Inner, Medium or
Outer)  are fitted
to a straight line and the line is extrapolated to the third station.
%(e.g., when two segments are found in Outer and Middle station, the
%track is extrapolated to the Inner station). 
In the extrapolation  multiple scattering is taken into account assuming a 2 GeV
momentum for the cosmic muon. If the extrapolated line crossed the
third station,  a reconstructed segment is searched for in that
station, but it is not required that the hits of this segment be
associated to the muon track. The segment efficiency is then
computed for each MDT chamber as the fraction of times a segment is
found. 
In order to reduce the effect of the non-instrumented regions a fiducial
cut in $\eta$ was applied for both barrel and end-cap. Chambers that
were not operational in the analyzed run were removed from the
sample. It was not possible to determine the efficiency for all
chambers due to the limited coverage of the trigger for the run used for this analysis (fall 2008) and  flux
of cosmic rays. 
For tracks crossing the overlap region between two adjacent chambers (
Small/Large sector overlap) it was not required that 
two segments be reconstructed, since this may lead to a slight overestimation of the efficiency.
%The efficiency estimate do not requires that for tracks crossing the overlap region 
%between two adjacent chambers of the same layer (small vs large sectors overlap) 
%two segments are reconstructed; this may lead to an slightly optimistic estimate of
%the efficiency.
\end{sloppypar}

\begin{sloppypar}
The distribution of the segment efficiency is shown in
Figure~\ref{mdtSegmentHits} on the right for 322 chambers in the barrel. The average value is 99.5\% and the segment
efficiency is uniform over the acceptance as shown in
Table~\ref{mdtSegmantEffi}. In the efficiency for the barrel
chambers there is a small loss of about 0.5\% due the presence of
the support structure of the ATLAS barrel. The Inner chambers have a
slightly lower segment efficiency due to the geometry of the trigger
and a larger uncertainty in the track extrapolation. 
%The extrapolation procedure is particularly critical for the Inner layer.
%tml - Could not tell what the next sentence is supposed to say.
% Should it be "In general all barrel-triggered cosmic muons cross
% the region of the MS where the MDT are instrumented since..."?
%In general all barrel triggered cosmic muons cross the instrumented
%MDT region of the MS since the trigger chambers are geometrically
%coupled to MDT. Most of the cosmic ray muons crossing Middle layer
%stations are also bound to cross an Outer one, while this is not
%true for the Inner layer. For this reason the extrapolation
%uncertainty is expected to play a role only on Inner layer segment
%efficiency. Some systematics studies on the segment efficiency have
%been performed, such as its dependence on the G$t_{0}$--refit
%procedure, on the extrapolation procedure, and on the track
%pseudorapidity. From these studies a systematic error of $0.5~ \%$
%on the segment efficiencies reported in Table~\ref{mdtSegmantEffi}
%has been assigned.
Studies on systematic effects in determining the segment efficiency, 
such as its dependence on the G$t_{0}$--refit, on the extrapolation and 
on the track angle, show that a systematic error of $\sim$0.5\% affects 
the values of efficiency listed in Table~\ref{mdtSegmantEffi}.
\end{sloppypar}

\begin{table*}[htb]
\centering
\begin{tabular}{| l | c c c | c c c |}
\hline
MDT station & BI & BM & BO  & EI & EM & EO \\
\hline
Segment efficiency & 0.987 & 0.992 & 0.996 & 0.992 & 0.998 & 0.999 \\
\hline
\end{tabular}
\caption{Average value of the segment reconstruction efficiency in
the MDT stations.} \label{mdtSegmantEffi}
\end{table*}

\begin{sloppypar}
%===========================
%tml - This alternative method seems to add nothing to the discussion.
%I suggest removing this paragraph.
%An alternative method for measuring segment reconstruction
%efficiency that is almost independent of chamber hardware problems
%is described in the following. This method can be used only on runs
%with no magnetic field in the MS area. All segment pairs with a
%polar angle difference (angle measured in the precision plane)
%smaller than 7.5 mrad are considered. The segments must be on two
%different MDT layers (Inner, Middle or Outer). The segment pairs are
%fitted to a straight track and the track is extrapolated to the
%third MDT layer. If at least three hits are found on the third layer
%with a measured charge grater than 50 ADC units and within 1 readout
%channel, about 3cm, from the track extrapolation the track is kept.
%The segment efficiency is then computed as the fraction of selected
%tracks for which a segment with at least 3 hits belonging to the
%matched ones has been reconstructed.
%in the third MDT layer.
%Since the normalization already requires the presence of three hits
%on the tested MDT layer this segment efficiency is almost
%independent of hardware problems. For both reconstruction algorithms
%a segment reconstruction efficiency higher than 99\% has been found.
An alternative method to evaluate the segment reconstruction efficiency, almost independent of chamber hardware problems, is described in the following. 
As in the previous case, this method can be used only with no magnetic field in the MS. All segment pairs in two different MDT stations (Inner, Middle or Outer) with a polar angle difference smaller than 7.5 mrad are considered. The segment pairs are fitted to a straight line and this is extrapolated to the third MDT station.
The track is kept if at least three hit tubes are found in the third MDT with a signal charge above the ADC cut and aligned with the track extrapolation within one tube diameter, $\pm$3 cm.
The segment efficiency is then computed as the fraction of selected tracks that have a segment reconstructed with at least 3 hits in the identified drift tubes.
Since the normalization already requires the presence of three hits in the tested MDT station, this segment 
efficiency is almost independent of local hardware problems. A segment reconstruction efficiency higher than 0.99 is found in all MDT stations. 
\end{sloppypar}

%The {\it fake} segment rates have been studied in random triggered
%events. An average rate of about 6$\%$ of fake segments per event
%has been found. This rate is expected to be strongly reduced in
%collision conditions once the segment reconstruction requirements
%will be more stringent.
%tml - Which "other" algorithm?
%For the other muon tracking algorithm studied in this paper more
%stringent segment reconstruction cuts were chosen resulting in a
%fake segment rate, obtained with the same method, of about 2 per
%mill.
The rate of fake segments was studied with a random trigger. An average rate of 0.06 fake segments per event was found with the relaxed hit association criteria used for cosmic muons. 
This rate is expected to be strongly reduced to about $2\times10^{-3}$ if the segment reconstruction requirements are made to be more stringent as shown by using an alternative muon tracking algorithm.

\section{Track reconstruction}
\label{Tracks}
%{\bf Editor: Niels (or Peter ?), Ahmimed - 2x1.5 pages}\\
%{\it Co-editors: Alberto, Eve, Egge, Peter, Niels, Illinois ...}\\
%%%%%%%%%%%%%%% Beginning Text From Ahmimed %%%%%%%%%%%%%%%%%%%%%
The MOORE and Muonboy programs have been optimized to
reconstruct muon tracks originating from  the IP. To cope with the
different topology of cosmic ray muons they have been slightly
modified as explained in Section~\ref{Samples}. To mimic muons in
collision events, the tracks are split at their perigee (point of
closest approach to the beam axis), giving, usually, two
reconstructed tracks: one in the upper part of the MS and one in the
lower part. 
%Moreover, only events with at least one track reconstructed in the
%ID were considered in the track reconstruction efficiency studies.
%tml - First mention of TRT, SCT and Pixels here (I think), so
% references should be added.
Events with at least one ID track satisfying the following
criteria were selected: 
\begin{sloppypar}
\begin{itemize}
\item at least 20 hits in the Transition Radiation Tracker;
\item the number of hits summed over the SemiConductor Tracker (SCT) and the Pixel detector greater than 4  ;
\item the distance of closest approach in the transverse 
plane $|d_0|$ and along the $z$ axis $|z_0|$ smaller than 1 m; ;
\item  the value of the muon track  $\chi^{2}$/ndf $< 3$ ;
\item the value of the reconstructed pseudorapidity $| \eta | < 1$ ;
\item reconstructed momentum greater than 5 GeV.
\end{itemize}
\end{sloppypar}

%{\it i)} at least 20 hits in the TRT (Transition Radiation Tracker) detector; {\it ii)} the number of hits summed over the SCT (Silicon %Tracker) and the Pixel detector greater than 4; {\it iii)} the distance of closest approach in the transverse 
%plane $|d_0|$ and along the $z$ axis $|z_0|$ smaller than 1 m; {\it iv)} the value of the muon track  $\chi^{2}$/ndf $< 3$; {\it v)} the %value of the reconstructed pseudorapidity $| \eta | < 1$; and {\it vi)} reconstructed momentum greater than 5 GeV.
 This selection has been applied for all the studies reported in this Section with the exception of the momentum resolution results. 
\subsection{Resolution}
\begin{figure}[htb]
  \begin{center}
    \includegraphics[width=8cm]{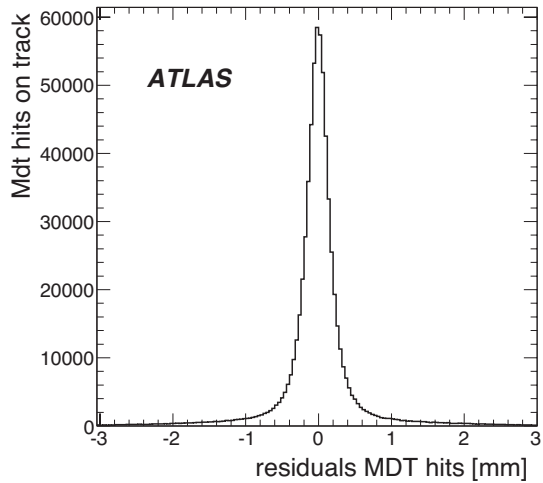}
  \end{center}
  \caption{Distribution of residuals for MDT hits associated to a track. The residuals have been fitted with a double-Gaussian function with common mean. The mean value is 6 $\mu m$, the standard deviation of the narrow Gaussian is about 150 $\mu m$ and the one of the wide Gaussian is about 700 $\mu m$.}
 \label{fig:residualtrack}
\end{figure}

The distribution of residuals for MDT hits associated to a track is
shown in Figure~\ref{fig:residualtrack}. The hit residual is defined
as the difference between the drift radius measured in a tube and
the distance of the track to the tube wire. The distribution refers
only to tracks with MDT hits in at least three different muon
stations (Inner, Middle and Outer) 
because these tracks have well constrained parameters and individual 
hits give a small contribution
to the track parameters. 
The distribution was fitted to a double Gaussian with common mean value. 
%A fit to the distribution using two
%Gaussians with a common mean value gives 6$\mum$ for the mean value,
%tml - The uncertainty on the mean should be included if we want to claim
%compatibility with zero.
%compatible with zero. 
The mean of the distribution was 6$\mum$ and  the RMS widths was 150$\mum$ for the narrow
Gaussian, accounting for 75\% of the distribution, and 700$\mum$ for
the other.
%The residuals have been fitted with a double-Gaussian function with common mean. The mean value
%compatible with zero within 6 $\mu m$ has been found.
%A narrow standard deviation of about 150 $\mu m$ and a wide one of about 700 $\mu m$ are found.
When compared to the distribution of the segment residuals shown in
Section~\ref{MDTperf} two additional effects contribute to the
broadening of this distribution: the misalignment between stations
and multiple scattering in the MS material.

\subsection{Efficiency}
\begin{figure}[!htb]
  \begin{center}
 \includegraphics[width=\columnwidth]{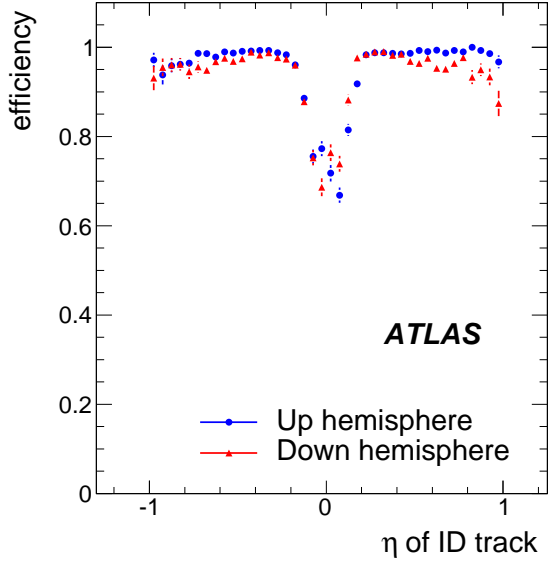}
  \end{center}
  \caption{Track reconstruction efficiency as a function of pseudorapidity.  The loss in efficiency in the region near $| \eta |=0$ is due to the loss of acceptance for detector services.  The presence of a track measured in the ID with $| \eta |<1$ is required.}
  \label{fig:mooreeff}
\end{figure}

The track reconstruction efficiency is computed as the fraction of
events where a track is reconstructed in the MS top or bottom
hemisphere once an ID track was found satisfying the selection
criteria described above. In this case also tracks with hits in only
two out of three MDT stations (Inner, Middle or Outer) are accepted,
even if these tracks have a worse momentum resolution than tracks
reconstructed in three stations. About 15\% of the selected tracks
are in this category. In addition, to compute the track efficiency
in the top (bottom) hemisphere, a momentum cut of 5 GeV (9 GeV) on
the ID track is applied. The result is shown in
Figure~\ref{fig:mooreeff} as a function of the pseudorapidity of the
ID track, for the top and bottom hemisphere separately. 
%The interval
%of low efficiency, for $|\eta | < 0.2$, corresponds to the
%uninstrumented region of the MS due to the gap for the services of
%the internal detectors.

The value of the efficiency, integrated over the $\eta$ acceptance,
is 94.9$\%$ for the top and 93.7$\%$ for the bottom hemisphere
respectively. If the four central bins are removed in
Figure~\ref{fig:mooreeff} the efficiency increases to 98.3$\%$ and
96.3$\%$ respectively. The statistical error on these values is
below 0.1$\%$. The lower efficiency in the central detector region,
around $| \eta | = 0$,  is due to the presence of the main ATLAS
service gap while lower efficiency in the Bottom hemisphere is
explained by the uninstrumented regions occupied by the support 
structure of the ATLAS barrel.

\begin{figure}[htb]
  \begin{center}
        \includegraphics[width=7cm]{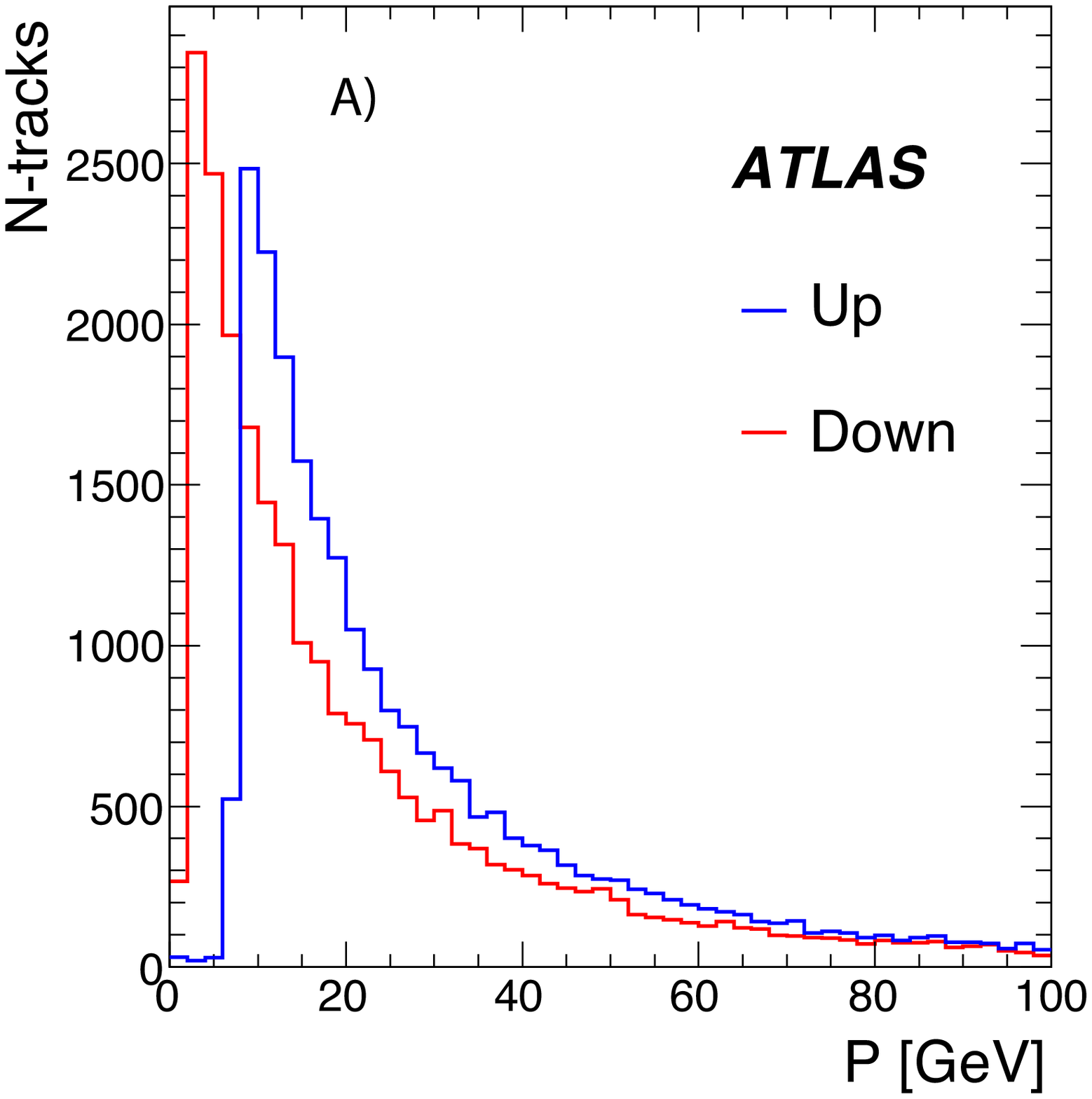}
    \includegraphics[width=7cm]{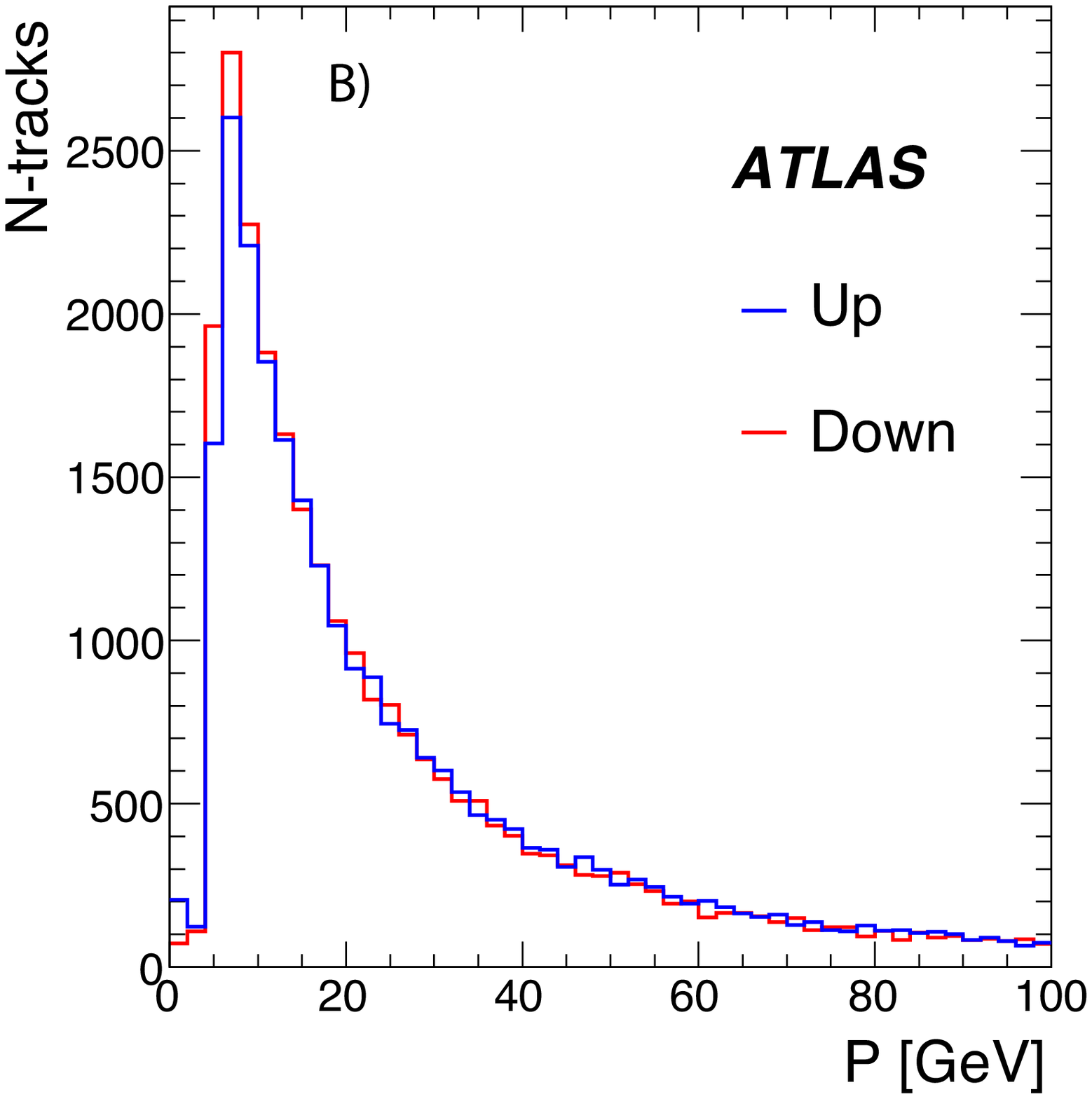}
  \end{center}
  \caption{A): Distribution of momentum  of cosmic muons as measured at the MS entrance for the upper 
and lower hemispheres. The difference between the two distributions is due to the ID track momentum
  cut of 5 GeV. B): same distributions with track momentum extrapolated to the IP.}
 \label{fig:pspectra}
\end{figure}
%When the magnetic field was on (in runs 91890 and 121080) the
%distribution of cosmic muons momentum was measured. 
\subsection{Momentum measurement}
The momentum of cosmic muons was measured in runs with magnetic field. 
The momentum measurement can be defined at the MS entrance or at the point of
closest approach to the IP. In the second case, for tracks
crossing the ID, a correction was made for the energy loss in the
calorimeters. This correction is based on the average energy loss
computed by the track extrapolation algorithm and is on average  3.1 GeV for
muons pointing to the interaction region with a distance of closest approach  of $|d_0| < 1$ m and $|z_0| < 2$ m.

The distribution of momentum at the MS entrance is shown in Figure~\ref{fig:pspectra}-left for the top and bottom
hemispheres separately. The difference between the two distributions
is due to the ID track momentum cut of 5 GeV that translates in a
different momentum cut-off in the two MS hemispheres, since cosmic
muons are directed downwards. 
The same distribution extrapolated to the perigee is shown on the right side of Figure~\ref{fig:pspectra}, demonstrating that the correction for the energy loss in the calorimeter removes the offset.
%The same distribution is shown in
%Figure~\ref{fig:pspectra} right extrapolated at the perigee, the correction
%for the energy loss in the calorimeter removes the offset. 
\begin{figure}[!htbp]
  \begin{center}
        \includegraphics[width=7cm]{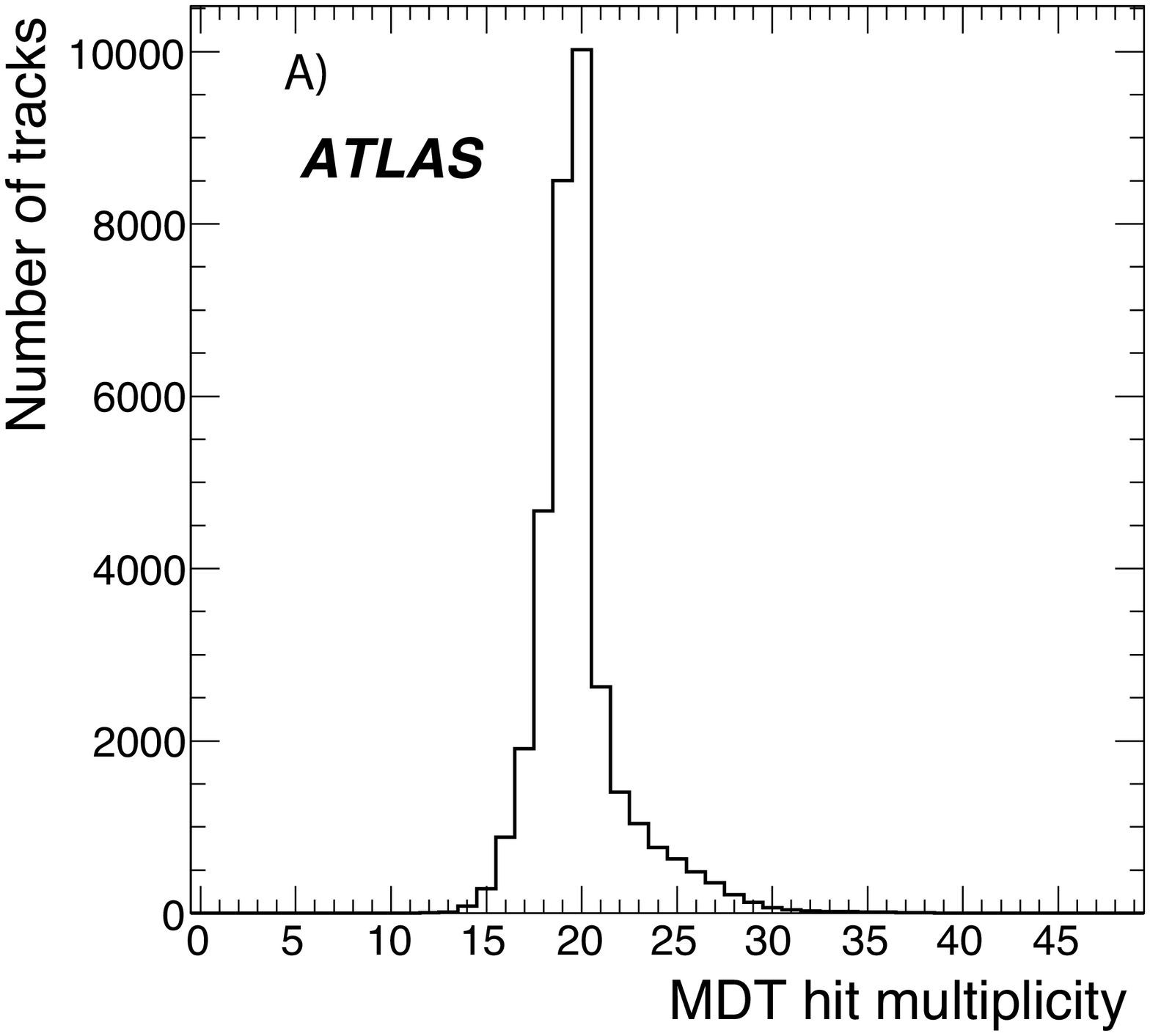}
    \includegraphics[width=7cm]{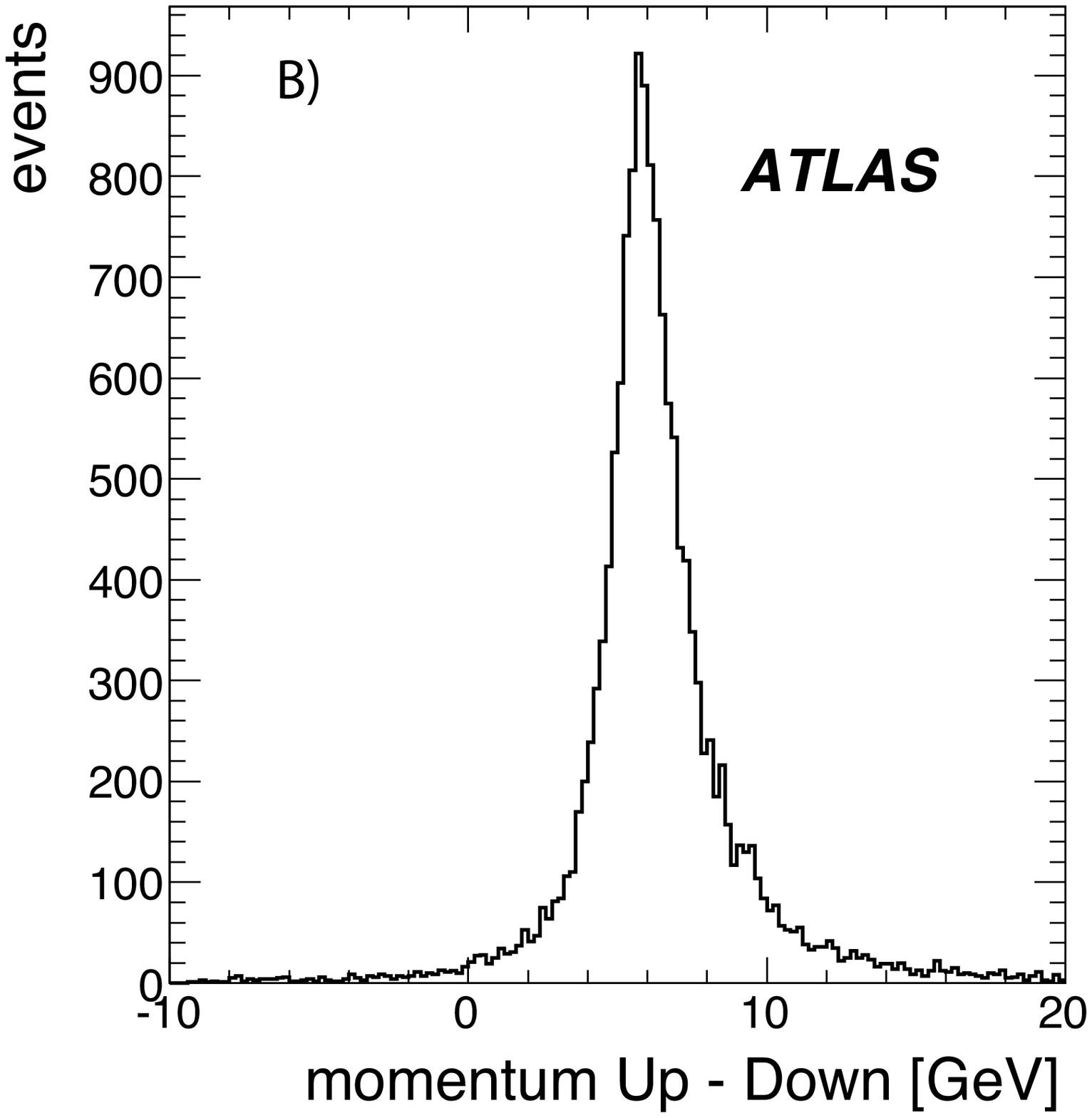}
  \end{center}
  \caption{A): Number of MDT hits on track. B): momentum difference between momenta measured by the MS in the top and bottom hemispheres for cosmic muons.
The momenta are expressed at the MS entrance and only tracks with momenta bigger than 15 GeV are considered. The mean value of 6.3 GeV is due to the energy loss in the calorimeter material.}
 \label{fig:hitontrack}
\end{figure}

%\begin{figure}[!htbp]
% \begin{center}
%  \includegraphics[width=7cm]{plots/MDThitontrack.eps}
% \end{center}
% \caption{Distribution of the number of MDT hits associated to tracks. }
%\label{fig:hitontrack}
%\end{figure}

\begin{sloppypar}
The distribution of the number of MDT hits associated with a track is
shown on the left side of Figure~\ref{fig:hitontrack}. For this plot tracks measured in three MDT stations have been selected.  A clear peak around 20 hits is visible (8 tubes
in the Inner stations, 6 in the Middle and Outer stations).
\end{sloppypar}

In events with tracks that cross the whole MS, the track is split at
the perigee and the two independent momentum measurements, in the
top and bottom hemisphere, can be compared.
Figure~\ref{fig:hitontrack} (right side) shows the distribution of the
difference of the two momentum values, top--bottom measured at the
MS entrance, for tracks with momenta greater than 15 GeV. In this
case the muons cross the calorimeter twice and the energy loss is
twice the value quoted above, in good agreement with the 6.3 GeV
mean value of the distribution.

%{\bf New Part with Niels Corrections:\\
The MS momentum resolution has been estimated by comparing for each
cosmic muon the two independent measurements in the top and bottom
hemispheres. In order to increase the available statistics no requirements on 
the presence of ID tracks were applied in this study.
%The pointing requirement for the ID tracks was tightened to 
%$|d_0|<$ 0.35 m and  $|z_0|<$ 0.5 m.
%The pointing requirement of the ID tracks was tightened to 0.35 m and 0.5 m 
%for the absolute values of the $d0$ and $z0$ parameters respectively. 
Only events with at least two reconstructed tracks in the MS are considered. 
Each track is required to have:
\begin{itemize}
\item at least 17 MDT hits, of which at least 7 in the Inner and 5 in the Middle and Outer stations of the same $\phi$ sector;
\item at least 2 different layers of RPC with a hit in the $\phi$ projection;
\item  polar angle $65^{\circ} < \theta < 115^{\circ}$;
\item distance of closest approach to the IP  $|d_0| < 1$ m and $|z_0| < 2$ m ;
\item polar and azimuthal  angles of the MS track pair agree within 10~$^{\circ}$.
\end{itemize}

% at least 17 MDT hits, of which at 
%least 7 in the Inner and 5 in the Middle and Outer stations of the same $\phi$ sector; 
%at least 2 different layers of RPC with a hit in the $\phi$ projection;
%polar angle $65^{\circ} < \theta < 115^{\circ}$; distance of closest approach to the IP 
%$|d_0| < 1$ m and $|z_0| < 2$ m. Moreover it was required that the polar and azimuthal 
%angles of the MS track pair agree within 10~$^{\circ}$. 
% The $\theta$ cut is designed to avoid using tracks with angles close to 90$\pm$30 degrees, for which the
% G$t_{0}$--refit is bound to fail.
%All these additional cuts ensure that cosmic muons are similar to muons produced in collisions. 
%}

%These additional cuts ensure that only cosmic muons similar to collision
%tracks are selected. 
About 19K top--bottom track pairs were selected in this way. 
For each track the value of transverse momentum was evaluated at the
IP. The difference between the two values divided by their average
\[ \frac{\Delta p_T}{p_T} = 2\ \frac{p_{Tup} - p_{Tdown}}{p_{Tup} + p_{Tdown}} \]
was measured in eleven bins of $p_T$. Since the cosmic muon
momentum distribution is a steep function (see
Figure~\ref{fig:pspectra}), the $p_T$ value of each bin was taken as
the mean value of the distribution in that bin.

The distribution of $\Delta p_T/p_T$ was fitted in each bin with a
double-Gaussian function with common mean value. The narrow Gaussian
was convoluted with a Landau function to account for the distribution of energy loss 
in the calorimeter. For $p_T <$ 10 GeV the normalizations of the two Gaussians were 
constrained such that 95\% of the events are in the narrow Gaussian. Above 10 GeV 
this constraint was lowered to 70\%. 
%In this way we fit the $\Delta p_T/p_T$ distribution with five parameters.
The mean value is representative of the difference in the transverse
momentum scale between the two MS hemispheres. The RMS of the narrow
Gaussian plus the width of the Landau, divided by $\sqrt{2}$, is
taken as an estimate of the transverse momentum resolution for each
$p_T$ bin. The Landau width is added linearly to the narrow Gaussian
RMS since the two quantities are strongly correlated. 

\begin{figure*}[!htbp]
  \begin{center}
  \includegraphics[width=14cm]{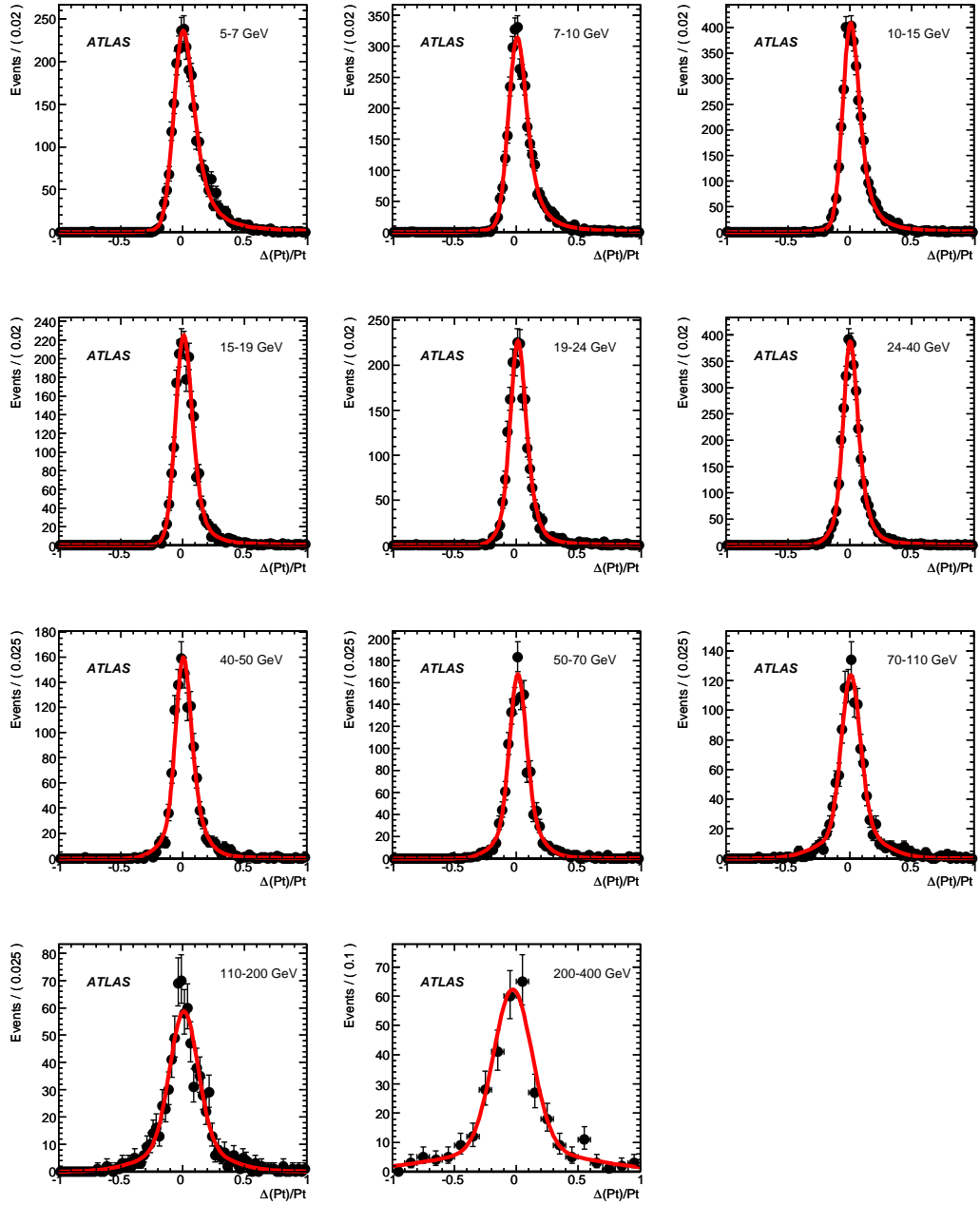}
  \end{center}
  \caption{Distributions of $\Delta p_T/p_T$ in the eleven $p_T$ bins. Fits to the function described
  in the text are superimposed.}
\label{fig:bins}
\end{figure*}

The distribution of $\Delta p_T/p_T$ is shown in Figure~\ref{fig:bins} for all $p_T$ 
bins together with the fitted function. For the eleven bins the fit probability is in the 
range between 45\% and 99\%, showing that the chosen parametrization is a good 
representation of the data distribution. \par
  Different fits have been done to study the 
systematics of the mean and RMS value. 
{\it i)}  The constraint between the two Gaussian areas has been changed by $\pm$10\%.
{\it ii)} A double Gaussian with common mean and asymmetric fit range, with the fit range 
reduced to two standard deviations on the positive side to avoid the energy loss tail. 
{\it iii)} A fit with two independent Gaussians with no range constraint. The result is that 
the estimated resolution is quite independent of the fit assumptions. The variation of the 
fit resolution ranges between 0.5\% at low $p_T$ up to a maximum of 1\% in the highest 
momentum bin.

% \begin{figure}[htb]
%   \begin{center}
% % %  \includegraphics[width=4.cm]{BOX.pdf}
%     \includegraphics[width=14cm]{plots/mean.eps}
%    \end{center}
%    \caption{Mean value of the $\Delta p_T/p_T$ distributions of Figure~\ref{fig:bins} for the 8  $p_T$ bins.
%    {\bf remove grid, shrink y-axis in $\pm 0.04$ to decrease empty spaces, use x- and y-axis labels as in the text, triangles are really ugly, can change?}}
%    \label{fig:scale}
%  \end{figure}

\begin{figure*}[!htb]
  \begin{center}
    \includegraphics[width=15cm]{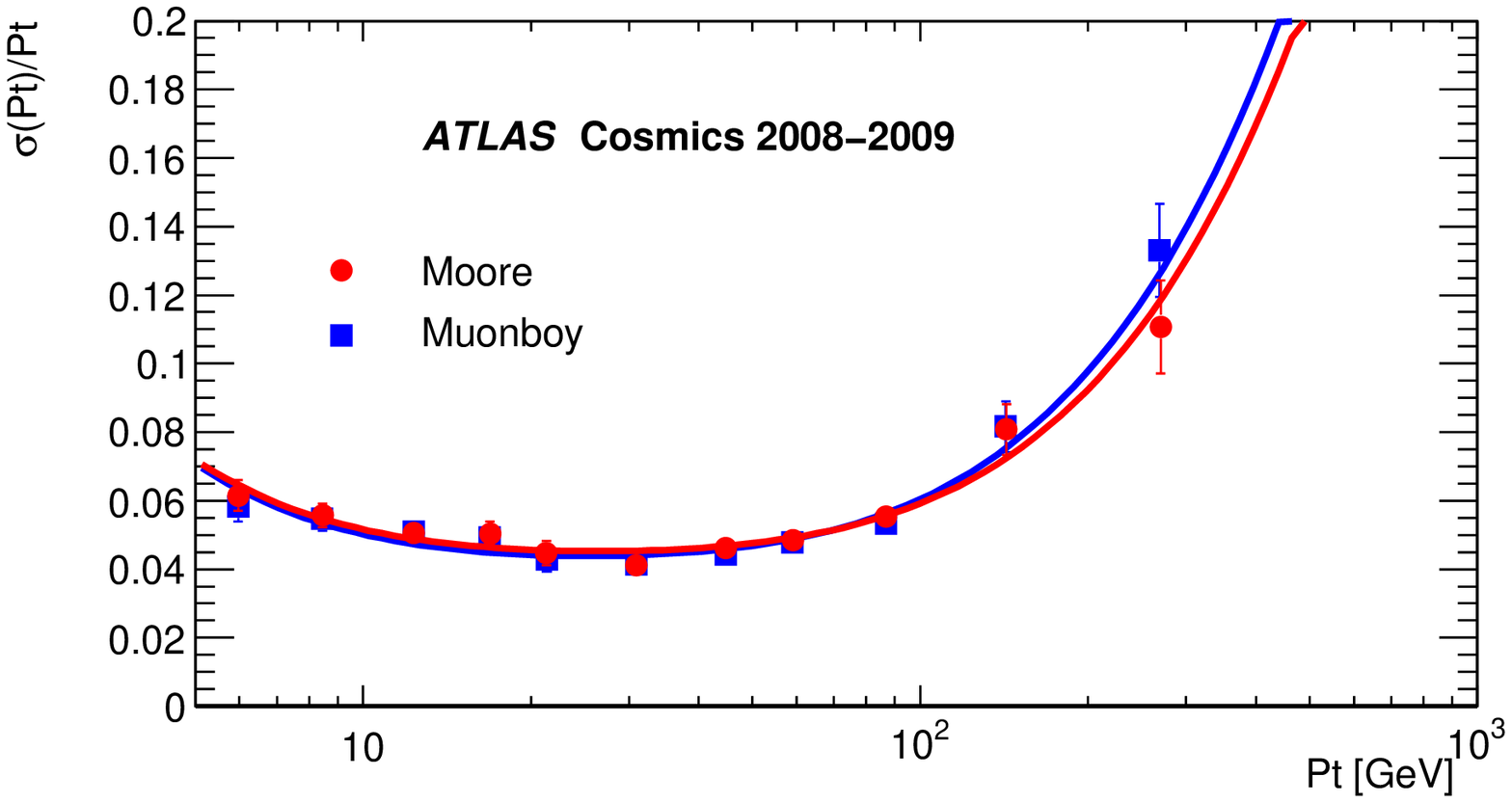}
  \end{center}
  \caption{Transverse momentum resolution evaluated with the top--bottom method explained in the text as a function of $p_T$, barrel region only  ($ | \eta  | < 1.1 $). The fit to the three resolution parameters as described in the text is superimposed.}
 \label{fig:resol}
\end{figure*}

The fit mean values indicate that the $p_T$ scales in the two MS hemispheres are in 
agreement within 1\%, or better.
The relative $p_T$ resolution, $\sigma_{p_T}/p_T = \sigma(\Delta p_T/p_T)/\sqrt{2}$, is shown in Figure~\ref{fig:resol}, for the two
main muon reconstruction algorithms~\cite{MOORE,Mboy}, as a function of the transverse momentum.
The two results are consistent taking into account the independent statistical uncertainties.

The resolution function can be fitted with the sum in quadrature of
three terms, the uncertainty on the energy loss corrections $P_{0}$,
the multiple scattering term $P_{1}$, and  the intrinsic resolution
$P_{2}$.
\begin{equation}
  \frac{\sigma_{p_T}} {p_T} = \frac{P_{0}}{p_T} \ \oplus \ P_{1} \ \oplus \ P_{2}\times p_T
  \nonumber
\end{equation}
The result of the fit is shown in Figure~\ref{fig:resol}. The values
of the parameters 
%obtained with one of the two reconstruction algorithms 
are: $P_{0} = 0.29 \pm 0.03 \pm 0.01$ GeV, $P_{1} = 0.043
\pm 0.002 \pm0.002$, $P_{2} = (4.1 \pm 0.4 \pm0.6)\times10^{-4}$
GeV$^{-1}$. 
The second uncertainty, due the systematics of the bin-by-bin fit method, was evaluated 
by changing the fitting assumptions as explained above. The expected values for 
these parameters were computed in reference~\cite{CSCpaper} on the basis of an 
analytic calculation of the $p_T$ resolution that takes into account the detailed 
description of the material in the MS, the single tube resolution, the alignment accuracy
and the magnetic field map.
%The second uncertainty listed represents an estimate of the systematic error of the 
%bin-by-bin fit method that has been obtained by changing the resolution fitting assumptions 
%(Narrow to Wide Gaussian fraction, two Gaussian fit with asymmetric fit interval, two 
%Gaussian fit with independent means, ...). The expected values for these parameters 
%have been computed in~\cite{CSCpaper} on the basis of an analytic calculation of the
%momentum resolution that took into account detailed material distribution in teRMS of 
%radiation lengths, average single tube resolution, average alignment accuracy in sagitta 
%direction, and the magnetic field map. 
The values obtained for the barrel MS were:
$P_{0} = 0.35$ GeV, $P_{1} = 0.035$ and $P_{2} = 1.2 \times 10^{-4}$
GeV$^{-1}$. The result is in fair agreement with the expected values
for the first two terms, while the intrinsic term is  worse.
%tml - Not sure what this sentence is trying to say.  Is the difference under
%investigation?  If it "has been investigated" then what's the source of the difference?
%Or is this a prelude to the next paragraphs?
The difference has been investigated to trace the effects that
contribute to worsen the resolution as determined with cosmic muons.

First, more than 70\% of the track pairs considered in the analysis are in the Large 
 sectors 5--13. At high $p_T$ the momentum resolution in the barrel Large sectors is 
 worse than in Small sectors because the field integral is smaller (see Figure~\ref{fig:muon-spect}). 
 Instead, in the low $p_T$ region dominated by multiple scattering the resolution in 
 Large sectors is better since the magnet coils are in the Small sectors.
%First, more than 70\% of the track pairs considered in the analysis are in the sector 
%pair 5--13. In the barrel at high $p_T$ the momentum resolution in Large sectors is 
%expected to be worse than in Small sectors because the field integral is smaller (see
%Figure~\ref{fig:muon-spect}). In the $p_T$  region dominated by the multiple scattering 
%the Large sector resolution is instead expected to be better due to the higher amount 
%of material, magnet coils, present in the Small sectors between Inner and Outer chambers.
%In fact, a higher value for the intrinsic term, $P_{2} = 1.8 \times 10^{-4}$ GeV$^{-1}$, results for Large sectors from Monte Carlo simulations done with ''nominal'' MDT spatial resolution and alignment.

Second, the single tube resolution is affected by imperfect
calibrations and  the additional time jitter is not completely
recovered by the G$t_{0}$--refit (see Figure~\ref{resolBML2A03}).
Part of the tracks in the sample contain segments with a badly
converging G$t_{0}$--refit. As a cross-check, all the tracks with
bad convergence were removed and the analysis was repeated. The
intrinsic term decreased by about 30$\%$.
%4.1$\times$10$^{-4}$ to 2.8$\times$10$^{-4}$ GeV$^{-1}$.

Third, the alignment
%though in some sectors almost reaching the required specifications,
in many sectors of the MS is still not at
the required level due to the limited statistics of straight tracks in the cosmic ray 
data sample.
Last, several other effects that contribute to resolution have not been removed, such as chamber deformations (due to temperature effects), wire sagging (particularly important in large chambers), single chamber geometrical defects.
Each of these effects contribute to
worsening the resolution and can be removed with dedicated
software tools. At the present stage of commissioning, the momentum
resolution is close to the design value for $p_T < 50$ GeV, but is
not as good for higher momenta. 
%It has also to be stressed that due to the rarity of cosmic muons with 
%transverse momenta above 200 GeV, the multiple scattering term, $P_{1}$, 
%and the intrinsic term, $P_{2}$, fitted in the data are strongly correlated.
%More work, more statistics and, in particular,  muons produced in beam collisions with the correct
%synchronization and in pointing geometry are needed to improve the MS performance at higher momenta.
%%%%%%%%%%%%%%% End Text From Ahmimed %%%%%%%%%%%%%%%%%%%%%

\section{Summary}
\label{Conclusions}
%{\bf Editor: Fabio - 0.5 pages}
The data collected in several months during the 2008-2009 cosmic ray
runs have been analyzed to assess the performance of the Muon
Spectrometer after its installation in the ATLAS experiment. Parts
of the detector, the Small Wheels in front of the end-cap toroids,
were installed during the runs and the commissioning of the many
detectors was proceeding while debugging the data acquisition and
the data control systems. The detector coverage during most of the
run period was higher than 99$\%$, with the exception of the RPC
chambers which were still under commissioning. For this detector subsystem 
the coverage steadily improved during the commissioning runs reaching more than
95\% in Spring 2009.
Results on several aspects of the Muon Spectrometer performance have
been presented. These include detector coverage, efficiency, 
resolution and relative timing of trigger and precision tracking chambers, 
track reconstruction, calibration, alignment and data quality.

%These include the detector coverage, the trigger and
%precision tracking chambers, their relative timing, calibration and
%alignment, survey of data quality, and track reconstruction.
Finally, with data collected when the magnetic field was on, a first
estimate of the spectrometer momentum resolution was obtained.
Efficiency and resolution of single elements have been measured
%The basic performance in teRMS of single-element efficiency and
%resolution have been measured 
for MDT, RPC and TGC chambers and were
found in agreement with results obtained previously  with high-momentum muon
beams. The trigger chamber timing has been adjusted with enough
precision to guarantee that the interaction bunch crossing can be
identified with a minimal number of failures. The muon trigger
logic, based on fast tracking of pointing muons has been extensively
tested in the regions of the detector with good cosmic ray
illumination. A slight deterioration of the MDT spatial resolution,  compared to test beam results,
was observed, which can be understood in terms of an additional time jitter
due to the asynchronous timing of cosmic muons and to their
non-pointing geometry. These effects were partially removed,
modifying the track reconstruction programs with dedicated
algorithms. Allowing for an increase of the single hit resolution,
to cope with these effects, the track segment efficiency in
individual chambers was found to be satisfactory and uniform over
the large number of chambers.

The performance of the end-cap and barrel optical alignment systems
have been measured using cosmic muon tracks with no magnetic field.
The results demonstrate that the end-cap optical system
is able to provide the required precision for 
chamber alignment. The design of the alignment system in the barrel
requires additional constraints provided by straight tracks. The
method has been tested with good results, but is limited by the
statistics of high-momentum muons with the required pointing
geometry.

\begin{sloppypar}
With the geometry corrections provided by the alignment system,
tracks in projective geometry were reconstructed in the barrel
showing that the reconstruction efficiency is uniform over the
entire acceptance and that the sagitta error is in agreement with
the  detector resolution, the alignment precision and the
effect of multiple coulomb scattering.
\end{sloppypar}

Finally with magnetic field, tracks crossing the whole spectrometer
were used to obtain two independent measurements of the momentum.
The momentum resolution was evaluated using the two values in the top and bottom part of the detector   and the
results were analyzed, fitting the distribution of the difference  as function of the
momentum. Taking into account the momentum spectrum, the multiple
scattering in the spectrometer and the energy loss in traversing the
calorimeters, the momentum resolution is in good agreement with
results from simulation for transverse momenta smaller than 50 GeV.
The statistics of high-momentum pointing tracks limits the
accuracy of the individual chamber calibration and the precision of
the alignment. At higher momenta, these limitations result in  degraded
momentum resolution.

%The cosmic rays collected by the ATLAS experiment in 2008 and 2009 runs have been used to commission 
%the Muon Spectrometer. 
%Excellent results have been achieved in teRMS of detector coverage, timing, single-hit performance, calibration, 
%alignment and momentum resolution.
%The design performance of the MS have been reached for muons with transverse momenta up to about 50 GeV. 
%The first collision data, as it was planned, will be used to complete the MS commissioning in teRMS of trigger timing, 
%precision chambers calibration and alignment and reconstruction software optimization.

During the long period of commissioning with cosmic rays it was possible to optimise the performance of the various hardware and software elements 
and to reach a level of understanding, such that we can consider  the Muon Spectrometer to be ready to efficiently detect 
muons produced in high-energy proton-proton collisions.

\section{Acknowledgements}

We are greatly indebted to all CERN's departments and to the LHC project for their immense efforts not only in building the LHC, but also for their direct contributions to the construction and installation of the ATLAS detector and its infrastructure. 
We acknowledge equally warmly all our technical colleagues in the collaborating Institutions without whom the ATLAS detector could not have been built. Furthermore we are grateful to all the funding agencies which supported generously the construction and the commissioning of the ATLAS detector and also provided the computing infrastructure.

The ATLAS detector design and construction has taken about fifteen years, and our thoughts are with all our colleagues who sadly could not see its final realisation.

\begin{sloppypar}
We acknowledge the support of ANPCyT, Argentina; Yerevan Physics Institute, Armenia; ARC and DEST, Australia; Bundesministerium f\"ur Wissenschaft und Forschung, Austria; National Academy of Sciences of Azerbaijan; State Committee on Science \& Technologies of the Republic of Belarus; CNPq and FINEP, Brazil; NSERC, NRC, and CFI, Canada; CERN;
CONICYT, Chile;
 NSFC, China; 
 COLCIENCIAS, Colombia;
 Ministry of Education, Youth and Sports of the Czech Republic, Ministry of Industry and Trade of the Czech Republic, and Committee for Collaboration of the Czech Republic with CERN; Danish Natural Science Research Council and the Lundbeck Foundation; European Commission, through the ARTEMIS Research Training Network; IN2P3-CNRS and Dapnia-CEA, France; Georgian Academy of Sciences; BMBF, HGF, DFG and MPG, Germany; Ministry of Education and Religion, through the EPEAEK program PYTHAGORAS II and GSRT, Greece; ISF, MINERVA, GIF, DIP, and Benoziyo Center, Israel; INFN, Italy; MEXT, Japan; CNRST, Morocco; FOM and NWO, Netherlands; The Research Council of Norway; Ministry of Science and Higher Education, Poland; GRICES and FCT, Portugal; Ministry of Education and Research, Romania; Ministry of Education and Science of the Russian Federation and State Atomic Energy Corporation ``Rosatom''; JINR; Ministry of Science, Serbia; Department of International Science and Technology Cooperation, Ministry of Education of the Slovak Republic; Slovenian Research Agency, Ministry of Higher Education, Science and Technology, Slovenia; Ministerio de Educaci\'{o}n y Ciencia, Spain; The Swedish Research Council, The Knut and Alice Wallenberg Foundation, Sweden; State Secretariat for Education and Science, Swiss National Science Foundation, and Cantons of Bern and Geneva, Switzerland; National Science Council, Taiwan; TAEK, Turkey; The Science and Technology Facilities Council and The Leverhulme Trust, United Kingdom; DOE and NSF, United States of America. 
\end{sloppypar}

\end{document}